\documentclass[12pt,pdftex]{article}
\pdfoutput=1

\usepackage[a4paper,text={16.8cm,22.4cm}]{geometry}
\usepackage{amsmath,amssymb,bm,graphicx,color,float}
\allowdisplaybreaks 
\newcommand{\degree}{\ensuremath{^\circ}}
\newcommand{\mcdot}{\!\cdot\!}

\newcommand{\bra}[1]{\langle #1|}
\newcommand{\ket}[1]{|#1\rangle}
\newcommand{\braket}[2]{\langle #1|#2\rangle}

\newcommand{\e}{\mathrm{e}}

\newcommand{\eq}[1]{Eq.~\eqref{#1}}

\def\eps{\varepsilon}

\def\SCETG{${\rm SCET}_{\rm G}\,$}

\newcommand{\be}{\begin{equation}}
\newcommand{\ee}{\end{equation}}

\def\OMIT#1{{}}

\newcommand{\vc}[1]{{\bf{#1}}}

\def\nslash{n\!\!\!\slash}
\def\bnslash{{\bar{n}}\!\!\!\slash}

\newcommand{\Tr}{\mathrm{Tr}}

\newcommand{\Eqs}[2]{Eqs.~(\ref{#1}) and (\ref{#2})}

\def\SCETG{${\rm SCET}_{\rm G}\,$}

\begin{document}
\begin{titlepage}

\begin{flushright}
    MITP/13-025\\
    April 15, 2013
  \end{flushright}

\begin{flushright}
\end{flushright}

\vspace{0.2cm}
\begin{center}
\LARGE\bf
Angular distributions \\ of higher order splitting functions \\ in the vacuum and in dense QCD matter
\end{center}

\vspace{0.2cm}
\begin{center}
Michael Fickinger$^{a}$, Grigory Ovanesyan$^b$, and Ivan Vitev$^b$

\vspace{0.6cm}
{\sl 
${}^a$\, PRISMA Cluster of Excellence \\
 Mainz Institute for Theoretical Physics\\
Johannes Gutenberg University, 55099 Mainz, Germany.\\[2mm]
}

\vspace{0.6cm}
{\sl 
${}^b$\, Theoretical Division, Los Alamos National Laboratory \\
MS B283, Los Alamos, NM 87545, U.S.A.\\[2mm]
}
\end{center}

\vspace{0.2cm}
\begin{abstract}
\vspace{0.2cm}
\noindent 

We study the collinear splitting functions needed for next-to-next-to-leading order calculations of jet production in the vacuum and in dense QCD matter. These splitting functions describe the probability of a parton to evolve into three-parton final state and are generalizations of the traditional DGLAP splitting kernels to a higher perturbative order. Of particular interest are the angular distributions of such splitting functions, which can elucidate the significance of multiple parton branching for jet observables and guide the construction of parton shower Monte Carlo generators. We find that to ${\cal O}(\alpha_s^2)$ both the vacuum and  the in-medium collinear splitting functions are neither angular ordered nor anti-angular ordered. Specifically, in dense QCD matter they retain the characteristic broad angular distribution already found in the ${\cal O}(\alpha_s)$  result.

\end{abstract}

\vfil

\end{titlepage}

 


\section{Introduction}
To understand the various aspects of an experimental measurement, the LHC program relies on simulations from parton shower Monte Carlo codes like PYTHIA \cite{Sjostrand:2006za} and HERWIG \cite{Bahr:2008pv}. One of the ingredients of such event generators are the collinear Dokshitzer-Gribov-Lipatov-Altarelli-Parisi (DGLAP) splitting 
functions~\cite{Gribov:1972rt,Dokshitzer:1977sg,Altarelli:1977zs}. These functions describe to leading order ${\cal O}(\alpha_s)$  collinear radiation on a distance scales of the order of the inverse transverse size of a typical jet. Coherent branching effects~\cite{Marchesini:1983bm,Marchesini:1987cf} are encoded in some of these generators, like HERWIG, in a way that forces subsequent branchings to happen at smaller angles than the previous ones. This property is known as {\it{angular ordering}}. Typical parton showers resum leading order collinear large logarithms, while the angular-ordered parton shower have been argued to include leading order infrared logarithms.

While implementations of  angular ordered parton showers have been phenomenologically successful, they suffer from a conceptual inconsistency. Angular ordering is directly applied to the collinear splitting functions. The coherent branching is correct in the soft gluon emission limit from a hard $N$-parton final state. Conversely, the collinear (DGLAP) splitting functions are derived in the small angle approximation, factorize from the hard scattering and have no knowledge of the global event structure.
In this paper we are interested in characteristics of well-separated and energetic jets.  In order to understand the properties of highly-collimated showers, and in particular the questions of angular ordering, angular anti-ordering, or lack of any ordering, we investigate the higher order  ${\cal O}(\alpha_s^2)$  collinear splitting functions.
In the vacuum, all such $1\rightarrow 3$ parton branchings have already been calculated \cite{Catani:1998nv,Catani:1999ss}. We use Soft Collinear Effective Theory (SCET)~\cite{Bauer:2000ew,Bauer:2000yr,Bauer:2001ct,Bauer:2001yt}  to demonstrate, on the example of the $q\rightarrow ggq$  splitting, that it recovers these results. Our main focus, however, are the medium-induced splitting processes, where the characteristic large-angle radiation pattern, first understood  in the soft gluon approximation limit~\cite{Vitev:2005yg}, is the theoretical basis for interpreting jet production in heavy ion collisions at the LHC. Naturally, the question of what effects, if any, multiple branchings may have on the angular distribution of an in-medium shower is a very important one. So far, possible qualitative features of the gluon bremsstrahlung have been discussed on the example of a dipole antenna
 model~\cite{MehtarTani:2010ma,MehtarTani:2011jw}. Actual calculations of  $1 \rightarrow 3$ parton branching in dense QCD matter are absent in the literature.
We use Soft Collinear Effective Theory with Glauber gluons~\cite{Idilbi:2008vm,D'Eramo:2010ak,Bauer:2010cc,Ovanesyan:2011xy,Ovanesyan:2011kn,Benzke:2012sz,Ovanesyan:2012fr} (\SCETG) to derive the splitting function of a quark to emit two gluons and study the angular distributions in such higher order medium-induced splitting.  This information might lead to important insights as to what corrections arise to fixed order and/or resummed calculations~\cite{Vitev:2009rd,Neufeld:2010fj,He:2011pd,He:2011sg,Neufeld:2012df,Kang:2012zr,Stavreva:2012aa,Dai:2012am,Kang:2013wca} 
and Monte Carlo simulations~\cite{ColemanSmith:2012vr,Qin:2012gp,Renk:2012ve,Wang:2013cia,Ma:2013bia} of jet observables in heavy ion reactions. It will also help interpret the exciting experimental measurements with jet final states in heavy ion collisions at the LHC, see for example~\cite{Aad:2010bu,Chatrchyan:2012vq,CMS:2012wxa,Aad:2012vca}.

Our paper is organized as follows. In section \ref{sec:vacuumsplitting} we derive the $q\rightarrow ggq$ splitting function using SCET. In section \ref{sec:SCETG} we review the basics of Soft Collinear Effective Theory with Glauber gluons. In section \ref{sec:mediumsplitting} use \SCETG to derive the first order in opacity splitting function of $q\rightarrow ggq$. In section \ref{sec:angulardistribs} we study the angular distributions of vacuum and medium-induced $1\rightarrow 3$ splitting functions. We conclude our paper in section \ref{sec:conclusions}. 


\section{The splitting function $q\rightarrow ggq$ in the vacuum}\label{sec:vacuumsplitting}
In this section we calculate the vacuum splitting function for $q\rightarrow ggq$.\footnote{Note that to ${\cal O}(\alpha_s^2)$ the kinematic variables, such as the light cone momentum fractions and transverse momenta do not factorize. 
}  We use SCET and show, that our SCET calculation yields the same result as obtained in the collinear approximation of massless QCD, Ref. \cite{Catani:1998nv}. 

We follow the method used in Ref. \cite{Ovanesyan:2011xy} for radiative energy loss. We use the light-cone gauge which allows us to work directly with physical transverse polarization vectors. For aThe gluon polarization vector $\eps$ in light-cone components is
\begin{eqnarray}
\left[\bar{n}\mcdot \eps, n\mcdot\eps, \vc{\eps}_{\perp}\right]=\left[0,\frac{2\vc{p}_{\perp}\mcdot \vc{\eps}_{\perp}}{\bar{n}\mcdot p}, \vc{\eps}_{\perp}\right], \label{eq:lightconevec}
\end{eqnarray}
where we have used the gauge conditions $\eps(p)\mcdot p=0, \bar{n}\mcdot \eps=0$.
The spin sum matrix of these transverse polarization vectors is
\begin{eqnarray}
\sum_{\text{gluon polarizations}}\vc{\eps}^{i}_{\perp}  \vc{\eps}^{i'}_{\perp} =\delta^{i i'}.\label{eq:gluonspinsum}
\end{eqnarray}
$i$ and $i'$ are indices in Euclidean three-space.

\subsection{Notation and kinematics}
\begin{figure*}[!t]
\includegraphics[width=400pt]{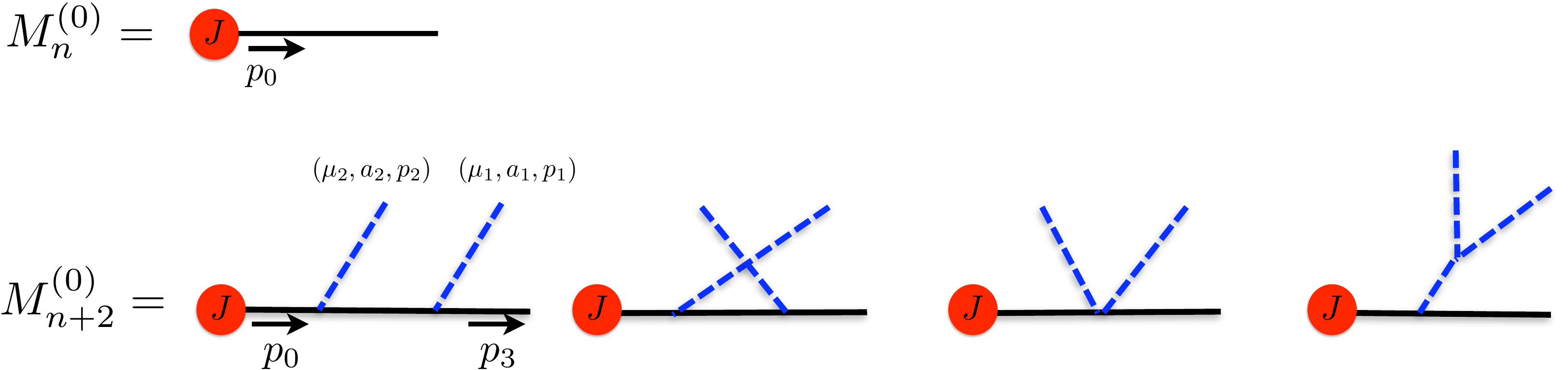}
\caption{Definition of the matrix elements $M_n^{(0)}$ and $M_{n+2}^{(0)}$ and the Feynman diagrams in SCET that contribute to the splitting $q\rightarrow ggq$.}
\label{fig:MnMnplus2}
\end{figure*}

We consider a general hard scattering amplitude $J$ that creates, apart from other partons, the collinear parent quark  with momentum $p_0$. This parton subsequently emits  two gluons with momenta $p_1$ and $p_2$, and the final-state  quark emerges with momentum $p_3$. The Feynman diagrams for matrix elements\footnote{Subscripts $^{(0)}$ refer to the fact that we deal with vacuum splittings in this subsection.} ${\mathcal M}_{n}^{(0)}$ and ${\mathcal M}_{n+2}^{(0)}$ in SCET are shown in Figure \ref{fig:MnMnplus2}. These matrix elements equal to
\begin{eqnarray}
{\mathcal M}_{n}^{(0)}&=&\bar{\chi}_{n,p_0}J,\\
{\mathcal M}_{n+2}^{(0)}&=&g^2\,\vc{\eps}^{i_1}_{1\perp}\,\vc{\eps}^{i_2}_{2\perp}\,\bar{\chi}_{n,p_3}\,\Gamma_{\text{eff}}^{i_1 i_2}J,\label{eq:mnplus2def}
\end{eqnarray}
where $\bar\chi_{n}$ is the gauge invariant collinear quark field of SCET and the expression for $\Gamma_{\text{eff}}$ follows directly from the Feynman rules of SCET.
The squared matrix elements can be written as
\begin{eqnarray}
&&\sum_{\text{spin}, \text{color}}\left|{\cal M}_{n}^{(0)}\right|^2 =\text{Tr}\left(\frac{\nslash}{2}\bar{n}\mcdot p_0\, J(p_0)\bar{J}(p_0)\right),\label{eq:mn}\\
&&\sum_{\text{spin}, \text{color}}\left|{\cal M}_{n+2}^{(0)}\right|^2 =g^4\,\text{Tr}\left(\frac{\nslash}{2}\bar{n}\mcdot p_3\, J(p_0)\bar{J}(p_0)\,\rho_0\right), \text{\, where}\label{eq:mnplus2}\\
&&\rho_0 =\sum_{i_1,i_2=1}^2\gamma^0\,(\Gamma_{\text{eff}}^{i_1 i_2})^{\dagger}\,\gamma^0\,\Gamma^{i_1 i_2}_{\text{eff}}.\label{eq:rhodef}
\end{eqnarray}
The trace in the equations above is over spin and color indices. To obtain \eq{eq:mnplus2} we have used \eq{eq:gluonspinsum} to sum over the gluon polarizations.
If
\begin{eqnarray}
\rho_0=\psi_0\left(\mathbb{I}_{\text{Dirac}}\right)\left(\mathbb{I}_{\text{color}}\right),\label{eq:rhoreduced}
\end{eqnarray}
with $\psi_0$ a real number\footnote{This is the case if the jet has been created by a pure QCD interaction.}, the $n+2$ parton matrix element and the $n$ parton matrix element can be related
\begin{eqnarray}
\sum\left|{\cal M}_{n+2}^{(0)}\right|^2 &= \frac{4g^4}{s_{123}^2}\langle\hat{P}_{q\rightarrow ggq}\rangle\,\sum\left|{\cal M}_{n}^{(0)}\right|^2\label{eq:Pdef}.
\end{eqnarray}
The sums are over spin and color.
The splitting function\footnote{The brackets in this notation indicate that we sum over the initial quark polarization. In all splittings that we consider in this paper we do the same. Spin correlations in the vacuum splittings have been studied in Refs. \cite{Catani:1998nv,Catani:1999ss}.} for $q\rightarrow ggq$
\begin{eqnarray}
\langle\hat{P}_{q\rightarrow ggq}\rangle= \frac{z_3 s_{123}^2}{4 g^4}\psi_0,\label{eq:Pmasterformula}
\end{eqnarray}
where $z_i=\bar{n}\mcdot p_i/\bar{n}\mcdot (p_1+p_2+p_3)=\bar{n}\mcdot p_i/\bar{n}\mcdot p_0$, and $s_{123}=(p_1+p_2+p_3)^2$, is calculated by substituting \eq{eq:rhoreduced} into \eq{eq:mnplus2} and comparing with \eq{eq:mn} and \eq{eq:Pdef}.
The factorization formula \eq{eq:Pdef} and the splitting function \eq{eq:Pmasterformula} were first derived in Ref.~\cite{Catani:1998nv}. Note that if not for \eq{eq:Pdef}, a process independent splitting function could not be defined.

In section~\ref{subseq:vac} we use the vectors
\begin{eqnarray}
\vc{U}_{Q_1, Q_2}^{j}=\bar{n}\,\mcdot \,p_0\left(\frac{\vc{Q}^j_{1\perp}}{\bar{n}\,\mcdot \,Q_1}-\frac{\vc{Q}^j_{2\perp}}{\bar{n}\,\mcdot \,Q_2}\right)=\frac{\vc{Q}^j_{1\perp}}{z_{Q_1}}-\frac{\vc{Q}^j_{2\perp}}{z_{Q_2}},
\end{eqnarray}
where the four-vectors $Q_1$ and $Q_2$ are linear combinations of $p_1, p_2, p_3$.
They are related to $s_{ij}\equiv (p_i+p_j)^2$ via
\begin{eqnarray}
s_{13}=z_1 z_3\,\vc{U}^2_{p_1,p_3},\qquad s_{23}=z_2 z_3\,\vc{U}^2_{p_2,p_3}, \qquad s_{12}=z_1 z_2\,\vc{U}^2_{p_1,p_2}.
\end{eqnarray}
These relations are needed to compare our results to Ref.~\cite{Catani:1998nv}.
Note that out of the six transverse vectors that appear in vacuum Feynman diagrams (see section~\ref{subseq:vac}), $\vc{U}_{p_1,p_3}$, $\vc{U}_{p_2,p_3}$, $\vc{U}_{p_1,p_2}$, $\vc{U}_{p_2,p_1+p_3}$, $\vc{U}_{p_1,p_2+p_3}$, $\vc{U}_{p_1+p_2,p_3}$,  only two are linearly independent.\footnote{Due to boost invariance along the collinear direction only two of the three momenta $p_1, p_2, p_3$ are independent.} They all can be written as linear combinations of $\vc{U}_{p_1,p_3}$, $\vc{U}_{p_2,p_3}$ and, moreover, any product of these six vectors can be written as combination of $s_{13}$, $s_{23}$, $s_{12}$ with coefficients that depend on $z_1$, $z_2$, $z_3$.

\subsection{Individual contributions of diagrams}\label{subseq:vac}
In this subsection we calculate the diagrams that contribute to the effective vertex $\Gamma_{\text{eff}}$, shown in Figure~\ref{fig:MnMnplus2}. We use the SCET Feynman rules, the polarization vectors in the light-cone gauge given in \eq{eq:lightconevec}, and the relation
\begin{eqnarray}
&&\gamma_{\perp}^{i} \gamma_{\perp}^{j}=-\delta^{ij}-i \eps^{ij3} \Sigma^3, \text{ where } i, j = 1,2, \text{ and\,\,\,}
\Sigma^3 = \left( \begin{array}{cc}
\sigma^3 & 0  \\
0 & \sigma^3 \\
 \end{array} \right).
\end{eqnarray}
The  explicit form of the matrix $\Sigma_3$ is only valid in the Weyl representation. Its properties, $\Sigma^{3\dagger}=\Sigma^3$ and $(\Sigma^3)^2=1$, which we use in section~\ref{sec:vacuumresult}, however, are independent of the representation. For the contributions from each of the four diagrams to $\Gamma_{\text{eff}}$ we obtain $C_i \,\Gamma_{i}^{i_1 i_2}$, with
\begin{eqnarray}
&&C_1=\frac{1}{s_{123}},\qquad\qquad\,\,\,\,\,\,\,\,\Gamma_1^{i_1 i_2}=\left(T^{a_1}T^{a_2}\,O_{1a}^{i_1i_2}+T^{a_2}T^{a_1}\,O_{1b}^{i_1 i_2}\right),\label{eq:Gamma1}\\
&&C_2=\frac{1}{s_{123} s_{13}},\qquad\qquad\Gamma_2^{i_1  i_2}=\sum_{j_1, j_2} T^{a_1}T^{a_2}\, \vc{U}_{p_1,p_3}^{j_1}\, \vc{U}_{p_2,p_1+p_3}^{j_2}\,O_2^{i_1 i_2 j_1 j_2},\label{eq:Gamma2}\\
&&C_3=\frac{1}{s_{123} s_{23}},\qquad\qquad\Gamma_3^{i_1  i_2}=\sum_{j_1, j_2} T^{a_2}T^{a_1}\, \vc{U}_{p_2,p_3}^{j_1}\, \vc{U}_{p_1,p_2+p_3}^{j_2}\,O_3^{i_1 i_2 j_1 j_2},\label{eq:Gamma3}\\
&&C_4=\frac{1}{s_{123} s_{12}},\qquad\qquad\Gamma_4^{i_1  i_2}=\sum_{j_1, j_2} [T^{a_1},T^{a_2}]\, \vc{U}_{p_1,p_2}^{j_1}\, \vc{U}_{p_1+p_2,p_3}^{j_2}\,O_4^{i_1 i_2 j_1 j_2}.\label{eq:Gamma4}
\end{eqnarray}
The operators $O_j$ depend only on $z_1, z_2, z_3$ and $\Sigma_3$ and are defined as
\begin{eqnarray}
&&O=\left(
\begin{array}{ccccc}
O^{i_1 i_2}_{1a}; & O^{i_1 i_2}_{1b}; & O^{i_1 i_2 j_1 j_2}_2; & O^{i_1 i_2 j_1 j_2}_3; & O^{i_1 i_2 j_1 j_2}_4
\end{array}
\right),\nonumber\\
&&Q=\left(
\begin{array}{cccccccc}
\delta^{i_1 i_2}; & \eps^{i_1 i_2 3}; &  \delta^{i_1 j_1} \delta^{i_2 j_2}; & \delta^{i_1 j_2} \delta^{i_2 j_1};  & \delta^{i_1 i_2} \delta^{j_1 j_2}; & \delta^{i_1 j_1} \eps^{i_2 j_2 3}; & \delta^{i_2 j_2} \eps^{i_1 j_1 3}; & \delta^{i_1 i_2} \eps^{j_1 j_2 3}
\end{array}
\right),\nonumber\\
&&O_i=\sum_{j} Q_{j} \,M_{ji}.\label{eq:Odef}
\end{eqnarray}
The 8x5 matrix $M$ is equal to
\footnotesize
\begin{eqnarray}
&&  M_{ji}=  \label{eq:Mdef}\\ 
&&\left(
\begin{array}{ccccc}
 \frac{1}{1-z_2}-\frac{2 (z_1-z_2)}{(z_1+z_2)^2} & \frac{1}{1-z_1}+\frac{2 (z_1-z_2)}{(z_1+z_2)^2}
& 0 & 0 & 0 \\
 -\frac{i \Sigma_3}{-1+z_2} & \frac{i \Sigma_3}{-1+z_1} & 0 & 0 & 0 \\
 0 & 0 & (2-z_2) (z_1+2 z_3) & z_1 z_2 & -2 z_2 (1+z_3) \\
 0 & 0 & z_1 z_2 & (2-z_1) (z_2+2 z_3) & -2 z_1 (1+z_3) \\
 0 & 0 & -z_1 z_2 & -z_1 z_2 & \frac{2 z_1 z_2 (1+z_3)}{z_1+z_2} \\
 0 & 0 & -i z_2 (z_1+2 z_3) \Sigma_3 & i (-2+z_1) z_2 \Sigma_3 & 2 i z_2 (z_1+z_2)
\Sigma_3 \\
 0 & 0 & i z_1 (-2+z_2) \Sigma_3 & -i z_1 (z_2+2 z_3) \Sigma_3 & 2 i z_1 (z_1+z_2)
\Sigma_3 \\
 0 & 0 & 0 & -2 i ((-1+z_1) z_2+z_1 z_3) \Sigma_3 & 2 i z_1^2 \Sigma_3
\end{array}
\right)_{ji} .\nonumber
\end{eqnarray}
\normalsize

Note that \eq{eq:Gamma1} and \eq{eq:Gamma4} do not coincide exactly with the corresponding diagrams 1 and 4 in Figure~\ref{fig:MnMnplus2} since we have rearranged terms in these equations. The terms proportional to ${2(z_1-z_2)}/{(z_1+z_2)^2}$ in the first row of $M_{ji}$ are included into operators $O_{1a}$ and $O_{1b}$ and therefore contribute to diagram 1. In Figure~\ref{fig:MnMnplus2}, these
terms are contained in diagram 4. This rearrangement does not effect the sum of diagrams 1 and 4 and is especially convenient for the in-midium calculation.

\subsection{Result for the vacuum splitting}\label{sec:vacuumresult}
As one can see from the previous subsection, the sum of all four diagrams can be written in the following form,
\begin{eqnarray}
\Gamma_{\text{eff}}^{i_1 i_2}&=&\sum_{k=1}^{4}C_k\,\Gamma_k^{i_1 i_2}=\left[T^{a_1}T^{a_2}\left(\alpha_{1,1}^{i_1i_2}+i\alpha_{2,1}^{i_1 i_2}\,\Sigma^3\right)+T^{a_2}T^{a_1}\left(\alpha_{1,2}^{i_1i_2}+i\alpha_{2,2}^{i_1 i_2}\,\Sigma^3\right)\right].\nonumber\\
&=&\sum_{j}\,\e_{j}^{(0)}\left(\alpha_{1,j}^{i_1 i_2}+\alpha_{2,j}^{i_1 i_2}\,i\Sigma^3\right).
\label{eq:gammafinal}
\end{eqnarray}
In general, $\e_{j}^{(0)}$ is a list of color operators and $\alpha_{1,j}^{i_1 i_2}$, $\alpha_{2,j}^{i_1 i_2}$ are complex numbers. In our case $e_{1}^{{(0)}}=T^{a_1}T^{a_2}$, $e_{2}^{{(0)}}=T^{a_2}T^{a_1}$ and $\alpha_{1,j}$, $\alpha_{2,j}$ are real functions which can be extracted from \eq{eq:Gamma1}$-$\eq{eq:Mdef}.

Squaring $\Gamma_{\text{eff}}$ yields
\begin{eqnarray}
&& \rho_0=\sum_{i_1,i_2=1}^2\gamma^0\,(\Gamma_{\text{eff}}^{i_1 i_2})^{\dagger}\,\gamma^0\,\Gamma^{i_1 i_2}_{\text{eff}}=\sum_{j',j}\langle{e_{j'}^{(0)}}|{e_{j}^{(0)}}\rangle\,\left(\text{Re}\left(\alpha_{1,j'}^*\mcdot\alpha_{1,j}+\alpha_{2,j'}^*\mcdot\alpha_{2,j}\right) \right) \nonumber \\
&& \hspace*{2.5in}   \left. -\Sigma^3\,\text{Im}\left(\alpha_{1,j'}^*\mcdot\alpha_{2,j}-\alpha_{2,j'}^*\mcdot\alpha_{1,j}\right)\right).  \label{eq:GammaeffSquareformula}
\end{eqnarray}
In the equations above, the dot  between $\alpha$'s is a shorthand notation for summing over $i_1,i_2$, for example $\alpha_{1,j'}^*\mcdot\alpha_{1,j}\equiv \sum_{i_1,i_2=1}^2\,\alpha_{1,j'}^{i_1 i_2*}\alpha_{1,j}^{i_1 i_2}$.
Note that The Gram matrix for the two basis color operators $\e_{j}^{(0)}$,
\begin{eqnarray}
\braket{e^{(0)}_{j'}}{ e^{(0)}_j}=
\left[ \begin{array}{cccccc}
C_F^2& C_F\left(C_F-\frac{C_A}{2}\right) \\
C_F\left(C_F-\frac{C_A}{2}\right) & C_F^2 \end{array} \right]\times \mathbb{I}_{\text{color}},\label{gram0}
\end{eqnarray}
is symmetric\footnote{We used that the Gram matrix $\langle{e_{j'}^{(0)}}|{e_{j}^{(0)}}\rangle$ is a symmetric matrix to arrive at \eq{eq:GammaeffSquareformula}.} and proportional to unity in color space. Since the tensors $\alpha$ are real the imaginary part in \eq{eq:GammaeffSquareformula} vanishes and the Dirac part is also proportional to unity. As an immediate result we obtain that \eq{eq:rhoreduced} holds,
\begin{eqnarray}
&&\rho_0=\psi_0 \left(\mathbb{I}_{\text{Dirac}}\right)\left(\mathbb{I}_{\text{color}}\right),\, \text{ where} \\
&&\psi_0=C_F\left[C_F\left((\alpha_{1,1}+\alpha_{2,1})^2+(\alpha_{1,2}+\alpha_{2,2})^2\right)-C_A\left(\alpha_{1,1}\mcdot \alpha_{2,1}+\alpha_{1,2}\mcdot \alpha_{2,2}\right)\right].\label{eq:rho0final}
\end{eqnarray}
As before, it is understood that a square or a product contains summation over $i_1, i_2$.

Substituting \eq{eq:rho0final} into \eq{eq:Pmasterformula} yields
\begin{eqnarray}
\langle\hat{P}_{g_1 g_2 q_3}\rangle&=& C_F^2 \langle\hat{P}^{\text{(ab)}}_{g_1 g_2 q_3}\rangle+ C_F C_A \langle\hat{P}^{\text{(nab)}}_{g_1 g_2 q_3}\rangle, \text{ where}\\
\langle\hat{P}^{\text{(ab)}}_{g_1 g_2 q_3}\rangle&=&\frac{z_3 s_{123}^2}{4}\,\left((\alpha_{1,1}+\alpha_{2,1})^2+(\alpha_{1,2}+\alpha_{2,2})^2\right),\\
\langle\hat{P}^{\text{(nab)}}_{g_1 g_2 q_3}\rangle&=&\frac{z_3 s_{123}^2}{4}\,\left(-\alpha_{1,1}\mcdot \alpha_{2,1}-\alpha_{1,2}\mcdot \alpha_{2,2}\right).
\end{eqnarray}
In the abelian part, the contribution from $\Gamma_4$ cancels exactly in the sums $\alpha_{1,1}+\alpha_{2,1}$ and $\alpha_{1,2}+\alpha_{2,2}$ because of the color commutator in \eq{eq:Gamma4}. The remaining three diagrams result in
\begin{eqnarray}
\langle\hat{P}^{\text{(ab)}}_{g_1 g_2 q_3}\rangle&=&\frac{s_{123}^2}{2s_{13}s_{23}}\frac{z_3(1+z_3^2)}{z_1 z_2}+\frac{s_{123}}{s_{13}}\frac{z_{3}(1-z_1)+(1-z_2)^3}{z_1 z_2}-\frac{s_{23}}{s_{13}}+(1\leftrightarrow 2).\label{finalab}
\end{eqnarray}
The non-abelian part reduces to
\begin{eqnarray}
&&\langle\hat{P}^{\text{(nab)}}_{g_1 g_2 q_3}\rangle=\frac{\left[2(z_1 s_{23}-z_2 s_{13})+(z_1-z_2)s_{12}\right]^2}{4(z_1+z_2)^2s_{12}^2}+\frac{1}{4}+\frac{s_{123}^2}{2s_{12}s_{13}}\Bigg(\frac{1+z_3^2}{z_2}+\frac{1+(1-z_2)^2}{1-z_3}\Bigg)\nonumber\\
&&\qquad\qquad-\frac{s_{123}^2}{4s_{13}s_{23}}\frac{z_3(1+z_3^2)}{z_1 z_2}+\frac{s_{123}}{2s_{12}}\Bigg(\frac{z_1(2-2z_1+z_1^2)-z_2(6-6z_2+z_2^2)}{z_2(1-z_3)}\Bigg)\nonumber\\
&&\qquad\qquad+\frac{s_{123}}{2s_{13}}\Bigg(\frac{(1-z_2)^3+z_3^2-z_2}{z_2(1-z_3)}-\frac{z_3(1-z_1)+(1-z_2)^3}{z_1 z_2}\Bigg)+(1\leftrightarrow 2).\label{finalnab}
\end{eqnarray}
Both \eq{finalab} and \eq{finalnab} are in agreement with Ref.~\cite{Catani:1998nv} for $\eps=0$.

\subsection{Cascade approximation for the two gluon splitting function}
In this section we derive an approximation for the $q\rightarrow ggq$ splitting function based solely on our knowledge of  $1\rightarrow 2$ splitting functions. We refer to this approximation as "cascade", since it is closely related to how parton shower generators would approximate such higher order collinear splitting function.
 We start from the definition of arbitrary $1\rightarrow 2$ and $1\rightarrow 3$ splittings
\begin{eqnarray}
&&\sum_{\text{spin,color}}\left|{\cal M}^{(0)}_{n+1}\right|^2 = \frac{2g^2}{s_{jk}}\langle P^{(0)}_{i\rightarrow jk}[p_j,p_k]\rangle\sum_{\text{spin,color}}\left|{\cal M}^{(0)}_{n}\right|^2,\label{eq:split1to2def0}\\
&&\sum_{\text{spin,color}}\left|{\cal M}^{(0)}_{n+2}\right|^2 = \frac{4g^4}{s_{jkl}^2}\langle P^{(0)}_{i\rightarrow jkl}[p_j,p_k,p_l]\rangle\sum_{\text{spin,color}}\left|{\cal M}^{(0)}_{n}\right|^2.\label{split1to3def0}
\end{eqnarray}
Next we calculate an expression for $\left|{\cal M}^{(0)}_{n+2}\right|^2$ by iteratively applying \eq{eq:split1to2def0} twice and summing over all possible branching sequences that produce two gluons. Comparing this expression to \eq{split1to3def0}, we obtain the cascade $1\rightarrow 3$ splitting function
\begin{eqnarray}
&&\langle P_{q\rightarrow ggq}^{\text{casc}}[p_1,p_2,p_3]\rangle^{(0)}= \nonumber \\
&&  \qquad\qquad\qquad s_{123}\Bigg(\frac{\langle P^{(0)}_{q\rightarrow gq}[p_2,p_3+p_1]\rangle \langle P^{(0)}_{q\rightarrow gq}[p_1,p_3]\rangle}{s_{13}}+\frac{\langle P^{(0)}_{q\rightarrow gq}[p_1,p_3+p_2]\rangle \langle P^{(0)}_{q\rightarrow gq}[p_2,p_3]\rangle}{s_{23}}\nonumber\\
&&\qquad\qquad\qquad\qquad\quad+\frac{\langle P^{(0)}_{q\rightarrow gq}[p_1+p_2,p_3]\rangle \langle P^{(0)}_{g\rightarrow gg}[p_1,p_2]\rangle}{s_{12}}\Bigg).\label{cascade}
\end{eqnarray}
In section~\ref{subsec:AOvac} we compare the cascade approximation to the full splitting. The cascade formula omits certain interference terms, which are contained in the full $1\rightarrow 3$ splitting.

\section{Soft Collinear Effective Theory with Glauber Gluons}\label{sec:SCETG}

When a highly energetic parton traverses  dense QCD matter, the perturbative QCD approach can be used to describe its elastic and inelastic interactions and the formation of an in-medium parton shower. In this approach, the medium can be modeled as consisting of effective scattering centers that provide a color-screened Coulomb potential, which serves as a background field for the partons that travel through the medium~\cite{Gyulassy:1993hr}. Consequently, the processes that characterize the evolution of a parton shower in strongly-interacting matter, which can be cold nuclear matter or a Quark-Gluon-Plasma (QGP), can be divided into two categories. In the first category the familiar soft and collinear splittings appear at leading order at high energies, analogously to  the vacuum case. The second category involves elastic scattering with the medium quasi-particles. The first type of processes are described by the known Soft Collinear Effective Theory. For example, small angle collinear radiation in the parton shower is correctly captured by the SCET Lagrangian. However, the elastic scattering off of medium quasi-particles forces us to go beyond traditional SCET.
 
Soft Collinear Effective Theory with Glauber gluons (\SCETG) is an effective theory appropriate for describing parton shower formation in the ambiance of dense QCD matter and the corresponding jet observables in heavy ion collisions. In addition to the  interactions of SCET, it has interactions of collinear quarks~\cite{Idilbi:2008vm,Ovanesyan:2011xy} and collinear gluons~\cite{Ovanesyan:2011xy} with $t-$ channel off-shell gluons with momentum scaling\footnote{This statement is correct for the static and collinear sources. However for the soft source the correct mode is $(\lambda, \lambda^2,\lambda)$ \cite{Ovanesyan:2011xy}.} $(\lambda^2,\lambda^2,\lambda)$, which are usually called Glauber gluons. So far, the soft gluons have been neglected and \SCETG contains only interactions of collinear fields with Glauber gluons. Because this mode is off-shell, the proper description for it is to treat the source field and the Glauber gluon as a background field. Thus, based on the assumptions for the momentum scaling of the source, as well as on the gauge fixing, one can derive the scaling of the background field created by the source. With this scaling at hand, it is a matter of putting this background field into the covariant derivative of the SCET Lagrangian and extracting the Feynman rules of~\SCETG. The resulting Lagrangian of \SCETG is~\cite{Ovanesyan:2011xy}:
\begin{eqnarray}
&&\mathcal{L}_{\rm{SCETG}}(\xi_n,A_n,A_G)=\mathcal{L}_{\rm{SCET}}(\xi_n,A_n)\label{scetglagrangian}\\
&&\qquad\qquad\qquad\qquad\qquad\! +g\sum_{p,p'}\,\e^{-i(p-p')x}\left(\bar{\xi}_{n,p'}T^a\frac{\bnslash}{2}\xi_{n,p}-if^{abc}A^{\lambda c}_{n,p'}A^{\nu b }_{n,p}\,g_{\nu\lambda}^{\perp}\,\bar{n}\mcdot p\right)\,n\mcdot A^a_{G}(x).\nonumber
\end{eqnarray}
The details of the Lagrangian depend on the type of the source and the gauge fixing condition. In \cite{Ovanesyan:2011xy} different types of such choices have been considered. The Lagrangian above corresponds to the static source with the momentum scaling  $p_{\text{source}}=M v+k$, where the mass of the source particle $M\rightarrow \infty$ and $k\sim (\lambda,\lambda,\lambda)$. As for the gauge choice, it is the hybrid gauge, when the collinear gluons are quantized in the light-cone gauge and the potential off-shell Glauber gluons are quantized in the covariant gauge. This choice is simple for two reasons. First, the number of Feynman rules and their structure is minimal with this choice. For example, one can compare the two terms in \eq{scetglagrangian} with similar Lagrangians of \SCETG derived in \cite{Ovanesyan:2011xy} using covariant or light-cone gauge. Second, the two Wilson lines that appear in matrix elements of the effective theory, the collinear Wilson line and the transverse gauge link, do not add additional diagrams\footnote{This was derived specifically for the static source.} \cite{Ovanesyan:2011xy}.

The Lagrangian in~\eq{scetglagrangian} contains the background field in position space. The Feynman rules of such a Lagrangian contain the Fourier transform of the vector potential $n\mcdot A_G(x)\rightarrow \int {\rm d}^4 q/(2\pi)^4\,\e^{iq x} v(q)$. For the static source $v(q)=2\pi\delta(q^{0})\tilde{v}(\vc{q})$ due to the fact that the recoil energy is negligible. This formally follows from time independence of the background field $A_G(x)$.

Since every appearance of the Glauber gluon interaction leads to the integral over the Glauber gluon momentum, the following notation will be useful:
\begin{eqnarray}
{\rm d}\Phi_i=\frac{{\rm d}^4 q_i}{(2\pi)^4}\,\e^{iq_i \delta x_i}\, v(q_i), \qquad{\rm d}\vc{\Phi}_{i\perp}=\frac{{\rm d}^2 \vc{q}_{i\perp}}{(2\pi)^2}\,\e^{-i\vc{q}_{i\perp} \delta \vc{x}_i}\, \tilde{v}(\vc{q}_{i\perp}),
\end{eqnarray}
where $\delta x_i=x_i-x_0$,  $x_0$ is the space-time position where the jet was created, and $x_i$ is the space-time position of the interaction with the medium quasi-particle $i$. The transverse part of the four-vector $\delta x_i$ is defined as $\delta \vc{x}_i$. The relation between these two definition is simple:
\begin{eqnarray}
{\rm d}\Phi_i={\rm d}\vc{\Phi}_{i\perp}\,\frac{{\rm d}\, (n\mcdot q_i)}{2\pi}\,\e^{i (n\,\mcdot \,q_i)\delta z_i},
\end{eqnarray}
where $\delta z_i=\delta x_i^{3}$. Finally, in order to relate the cross section to physical observables as elastic scattering length and cross sections, we use:
\begin{eqnarray}
\frac{{\rm d}\sigma_{\text{el}}}{{\rm d}^2\vc{q}_{\perp}}(R,T)=\frac{C_2 ( R )\, C_2(T)}{d_A}\,\frac{|\tilde{v}(\vc{q}_{\perp})|^2}{(2\pi)^2}  =\frac{4\,\alpha_s^2}{d_A(\vc{q}_{\perp}^2+\mu^2)^2}\mcdot\left\{ \begin{array}{c}
C_F^2, \text{\,\,\,for\,\,\, } qq\rightarrow qq  \\
C_A^2, \text{\,\,\,for\,\,\, } gg\rightarrow gg  \\
C_F C_A, \text{\,\,\,for\,\,\, } qg\rightarrow qg\\ \end{array} \right\}.
\end{eqnarray}
In the equation above, $C_2(R)$ and $C_2(T)$ are the quadratic Casimirs of the incident parton and target (source) representations. $d_A = 8$ is the dimension of the adjoint representation. The formula above is valid in the high energy limit and neglecting the masses of the partons. As a result one can read out the value of $\tilde{v}(\vc{q}_{\perp})=4\pi\alpha_s/(\mu^2+\vc{q}^2_{\perp})$.

\section{The  $q\rightarrow ggq$ splitting function in  dense QCD matter}\label{sec:mediumsplitting}
In this section we calculate the $q\rightarrow ggq$ splitting function in the medium to first order in opacity,  using \SCETG and keeping the full $z_1, z_2, z_3$ dependence. 
First order of opacity contains single Born diagrams, representing interactions of the propagating system at longitudinal positions $x_i^3$. It also contains double Born diagrams, which can be viewed as  the contact limit $x_j^3 \rightarrow x_i^3$ of 2 interactions. In the first type of interactions one Glauber gluon is exchanged in both the matrix element and the complex conjugate of the matrix element. In the second type two Glauber gluons at the same point are exchanged either only in the matrix element or only in the complex conjugate of the matrix element. The organization of the opacity series is independent of the propagating system, for more details see Refs.~\cite{Gyulassy:2000er,Vitev:2007ve}.

The calculation in medium is very similar to the one in vacuum and we use many definitions of section~\ref{sec:vacuumsplitting} in this section. In particular, since Glauber gluons do not carry  large momenta, the entire part that depends solely on $z_1, z_2, z_3$ is identical in the vacuum and medium calculations. Thus, we use the same operators $O_j$ given in \eq{eq:Odef} as well as the same  matrix given in \eq{eq:Mdef}. 
\begin{figure*}[!t]
\center
\includegraphics[width=385pt]{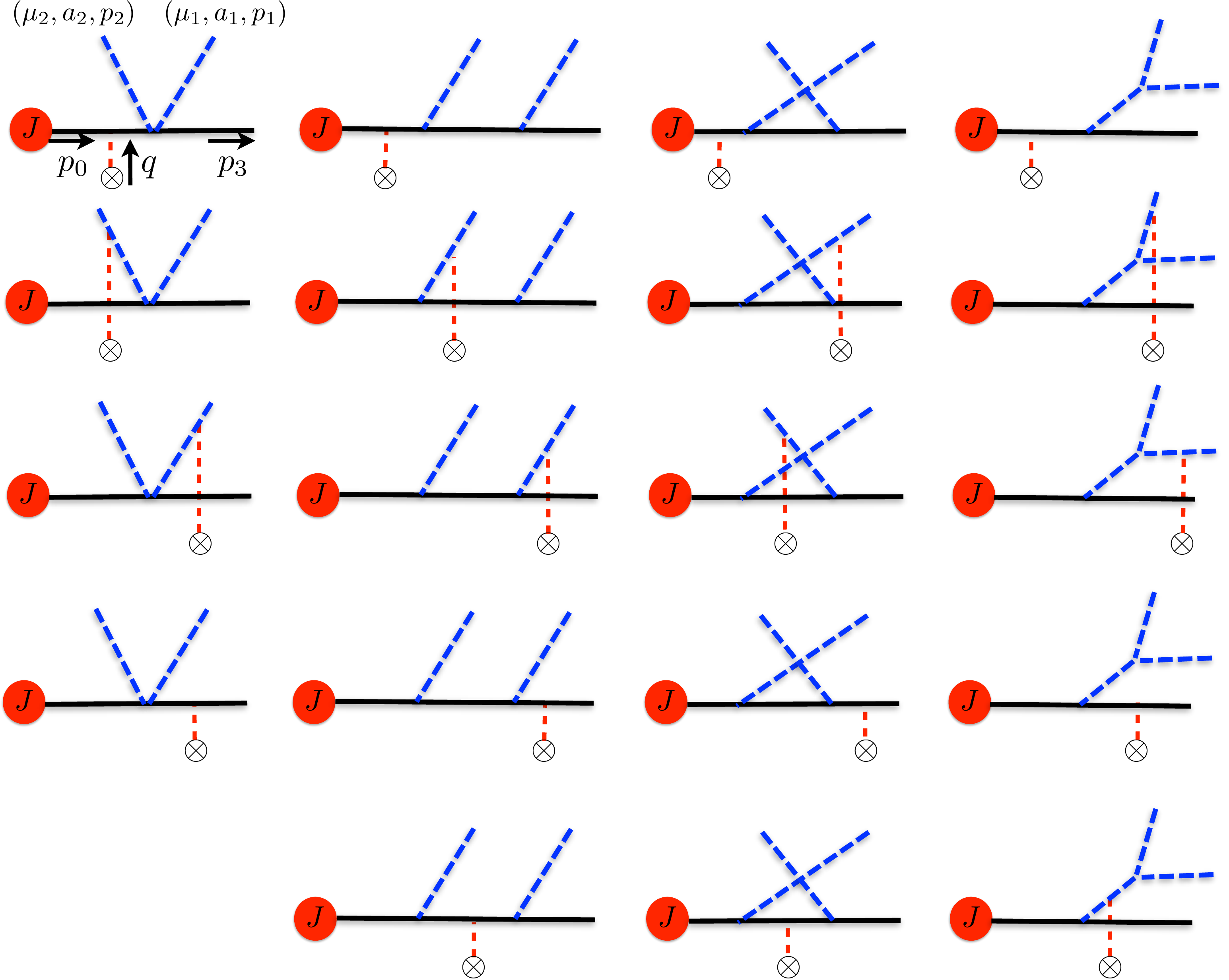}
\caption{Single Born diagrams. See text for explanation of the different topologies.}
\label{fig:singleBdiagrams1}
\end{figure*}

\subsection{Single Born diagrams}\label{subseq:singleB}
All single Born graphs are shown in Figure \ref{fig:singleBdiagrams1}. Graphs 1-4 have the same collinear structure as the vacuum graph 1 in Figure~\ref{fig:MnMnplus2}, and we refer to them as of topology 1. Graphs 5-9 are of topology 2, graphs 10-14 are of topology 3, and graphs 15-19 are of topology 4. The amplitude of an arbitrary single Born graph $k$ with $1\le k \le 19$ looks like
\begin{eqnarray}
\mathcal{M}_k^{(1)}=-g^2\,\vc{\eps}_1^{i_1}\,\vc{\eps}_2^{i_2}\,\bar{\chi}_{n,p}\left(\int d{\bf{\Phi}_{\perp}}\,C_k\,\Gamma_k^{i_1 i_2}I_k^{(1)} \right)J.\label{eq:singlebornmaster}
\end{eqnarray}
The minus sign in \eq{eq:singlebornmaster} cancels when squaring the matrix element. It is chosen for convenience, since it leads to a color operator matrix with more positive than negative numbers, see \eq{eq:singleborncoloroperators}.
In the remainder of this subsection we give detailed expressions for the longitudinal integrals $I_{k}^{(1)}$, the factors $C_k$ and effective vertices $\Gamma_{k}^{i_1i_2}$.

The longitudinal integrals are defined as
\begin{eqnarray}
I_k^{(1)}=\int\frac{\text{d}q^{-}}{2\pi} \e^{i q^{-}\delta z}\,\Delta_g(Q_{1},q)\,...\,\Delta_g(Q_{N_k},q),\label{eq:DefLISB}
\end{eqnarray}
where the integrand contains the product of all propagators with momentum $Q_{i}-q$, that depend on the medium transfer momentum $q$
\begin{eqnarray}
&&\Delta_g(p,q)=\left[\Omega(p,\vc{q}_{\perp})-q^{-}+i\eps/\bar{n}\mcdot p\right]^{-1},\\
&&\Omega(p,\vc{q}_{\perp})=p^{-}-\frac{(\vc{p}_{\perp}-\vc{q}_{\perp})^2}{p^+}.
\end{eqnarray}
For diagram $k$ the number of $q$ dependent propagators is $N_k$. For single Born diagrams $N_k$ is at least $N_{\text{min}}=1$ and at most $N_{\text{max}}=2$ for topology $1$ and $N_{\text{max}}=3$ for the other topologies $2,3,4$.

Performing the integrals yields
\begin{eqnarray}
I^{(1)}_k=\left\{ \begin{array}{c}
I_1(\Omega_1), \text{\,\,\,if\,\,\, } N_k=N_{\text{min}}  \\
I_2(\Omega_1,\Omega_2), \text{\,\,\,if\,\,\, } N_k=N_{\text{min}}+1  \\
I_3(\Omega_1,\Omega_2,\Omega_3), \text{\,\,\,if\,\,\, } N_k=N_{\text{min}}+2\\ \end{array} \right\},
\end{eqnarray}
where $\Omega_i=\Omega(Q_i,\vc{q}_{\perp})$ and
\begin{eqnarray}
I_{1}(\Omega_1)&=&-i\,\e^{i\Omega_1\delta z},\label{eq:I1}\\
I_{2}(\Omega_1,\Omega_2)&=&i\,\frac{\e^{i\Omega_2\delta z}-\e^{i\Omega_1\delta z}}{\Omega_2-\Omega_1},\label{eq:I2}\\
I_{3}(\Omega_1,\Omega_2, \Omega_3)&=&i\,\left(\frac{\e^{i\Omega_2\delta z}-\e^{i\Omega_1\delta z}}{\Omega_2-\Omega_1}-\frac{\e^{i\Omega_3\delta z}-\e^{i\Omega_1\delta z}}{\Omega_3-\Omega_1}\right)\frac{1}{\Omega_3-\Omega_2}.\label{eq:I3}
\end{eqnarray}
For details on the longitudinal integrals of single and double Born graphs, see Appendix~\ref{appendix:longintegrals}.

The factors $C_k$ and effective vertices $\Gamma_{k}^{i_1i_2}$ are
\begin{eqnarray}
&&\Gamma_k^{i_1 i_2}=e_k^{(a)}\,O_{1a}^{i_1 i_2}+e_k^{(b)}\,O_{1b}^{i_1 i_2}, \,\,\,C_k=\left\{ \begin{array}{c}
\frac{1}{s_{123}}, \text{\,\,\,if\,\,\, } N_k=N_{\text{min}}  \label{eq:top1}\\
\frac{1}{\bar{n}\,\mcdot \,p_{0}}, \text{\,\,\,if\,\,\, } N_k=N_{\text{min}}+1  \\ \end{array} \right\},\\
&&\Gamma_k^{i_1 i_2}=e_k\,{\vc{U}^{j_1}_{p_{k_1},p_{k_2}}\vc{U}^{j_2}_{p_{k_3},p_{k_4}}}\,O^{i_1 i_2 j_1 j_2}_{t_k},\,\,\,C_k=\left\{ \begin{array}{c}
\frac{1}{p^2_{k_{12}} p^2_{k_{34}}}, \text{\,\,\,if\,\,\, } N_k=N_{\text{min}}  \\
\frac{1}{p^2_{k_{12}} \bar{n}\,\mcdot \,p_0}, \text{\,\,\,if\,\,\, } N_k=N_{\text{min}}+1  \\
\frac{1}{\bar{n}\,\mcdot\,(p_{k_1}+p_{k_2}) \,\bar{n}\,\mcdot \,p_0}, \text{\,\,\,if\,\,\, } N_k=N_{\text{min}}+2 \end{array} \right\}\nonumber,\\\label{eq:top2}
\end{eqnarray}
where $p_{k_1}, p_{k_2}$ are the two four-vectors that come out of the second collinear splitting and similarly $p_{k_3}, p_{k_4}$ are those coming out of the first splitting. Since $\vc{U}_{Q_1, Q_2}$ is antisymmetric under exchange of its arguments $Q_1\leftrightarrow  Q_2$, we need to define the order of the arguments: For $q\rightarrow gq$ splittings, the gluon momentum is the first argument of $U$ followed by the quark momentum; for $g\rightarrow gg$ splittings, the momentum containing $p_1$ is the first argument of $U$ followed by the momentum containing $p_2$.

The color operator for single Born amplitudes $e_{k}$ is provided in Appendix~\ref{seq:color} in the basis of six elements $e^{(1)}_j$, see section~\ref{seq:squaring}. For topology $k=1$ there are two color operators per diagram, with indices~$^{(a)}$ and~$^{(b)}$, while for the other topologies there is only one color operator per diagram, consistent with the notation in \eq{eq:top1} and \eq{eq:top2}.

Note that \Eqs{eq:top1}{eq:top2} are very similar to the corresponding vacuum equations, \eq{eq:Gamma1}$-$\eq{eq:Gamma4}.\footnote{This similarity is very much expected due to the fact that Glauber exchanges do not change the large momentum fractions $z_1,z_2,z_3$ and, thus, the part of the amplitude that depends only on these fractions is identical to the vacuum case. Hence, the operators $O_{t_k}$ are equivalent to \eq{eq:Odef} and \eq{eq:Mdef}, and $t_k=2,3$ or $4$ depending on the topology of the diagram.} We also have used the same rearrangements between topology 1 and 4 as in the vacuum case, mentioned at the end of section~\ref{subseq:vac}.

Even though we provided all rules necessary to evaluate any single Born graph in this subsection,  we also summarize all values for $C_k, \vc{U}_{p_{k_1},p_{k_2}}, \vc{U}_{p_{k_3},p_{k_4}}, I_k^{(1)}$ in Appendix~\ref{sec:FeynmanGraphs}.

\subsection{Double Born diagrams}\label{subseq:doubleB}
\begin{figure*}[!t]
\center
\includegraphics[width=385pt]{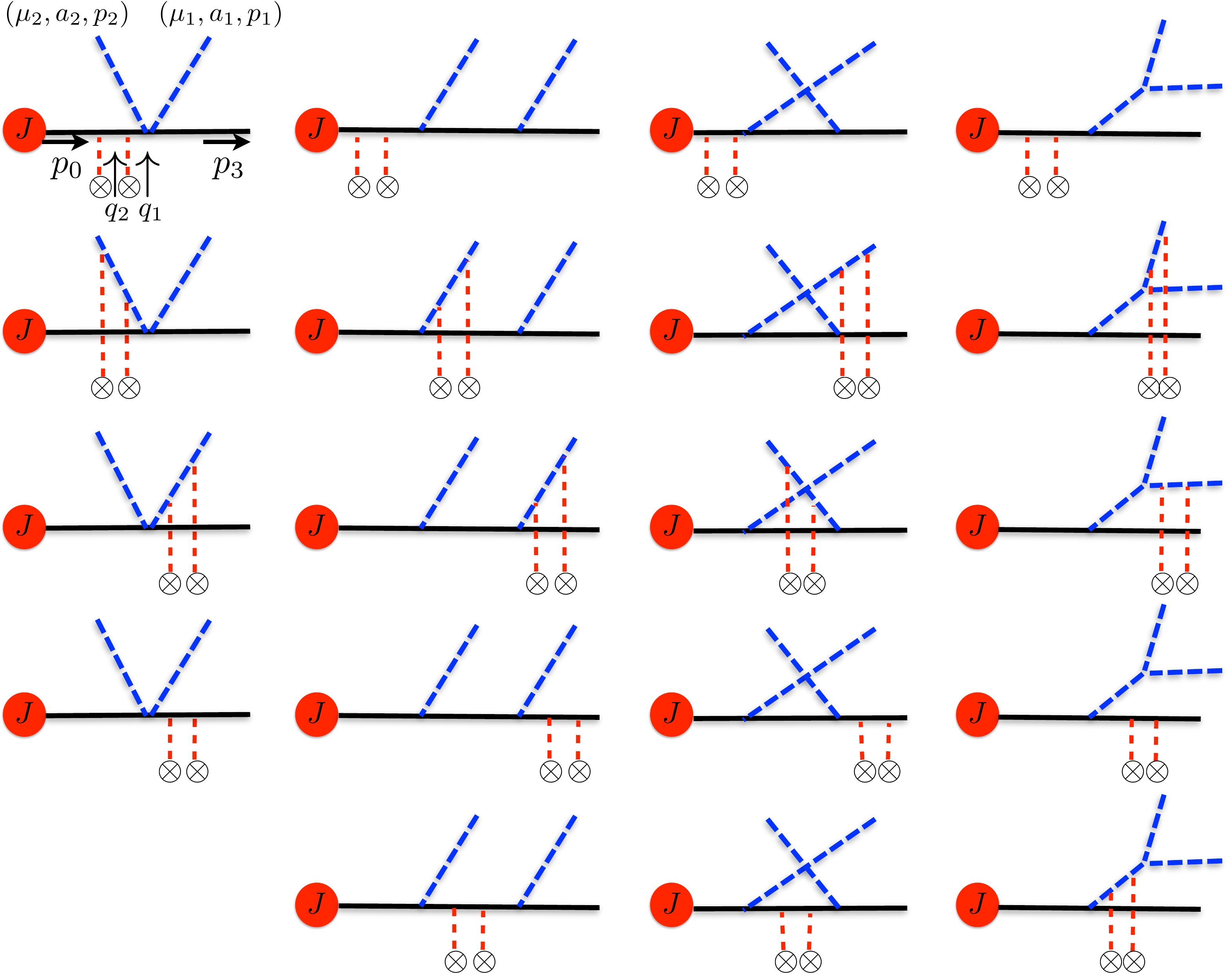}\\
\includegraphics[width=385pt]{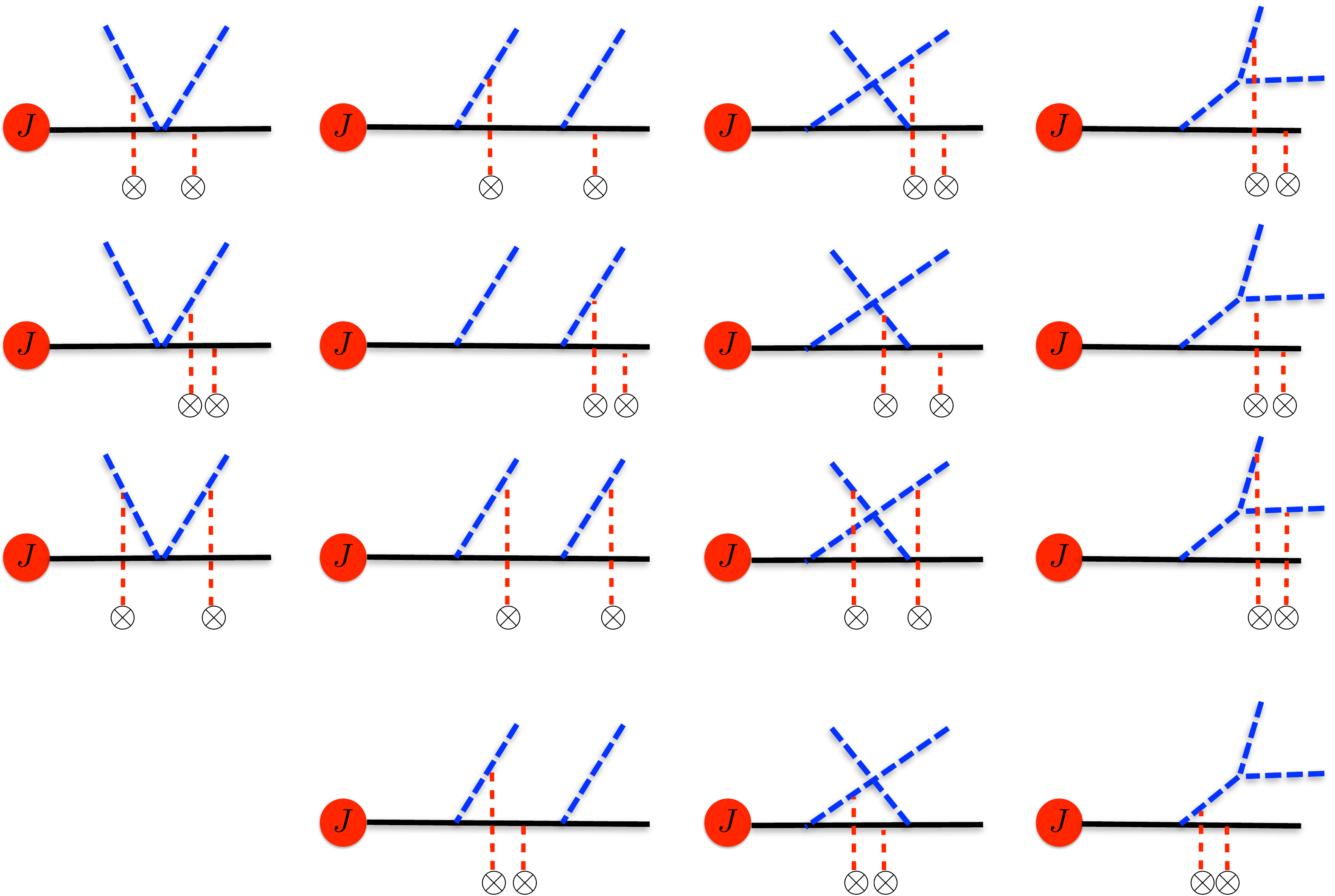}
\caption{Double Born diagrams. See text for explanation of the different topologies}
\label{fig:doubleborngraphs}
\end{figure*}

All 34 double Born graphs are presented in Figure \ref{fig:doubleborngraphs}. Graphs 1-7 are of topology 1, graphs 8-16 are of topology 2, graphs 17-25 are of topology 3, and graphs 26-34 are of topology 4. A general diagram $k$ with $1\le k\le 34$ equals
\begin{eqnarray}
\mathcal{M}_k^{(2c)}=\,g^2\,\vc{\eps}_1^{i_1}\,\vc{\eps}_2^{i_2}\,\bar{\chi}_{n,p}\left(\int d{\bf{\Phi}}_{1\perp}\,d{\bf{\Phi}}_{2\perp}\,C_k\,\Gamma_k^{i_1 i_2}I_k^{(2c)} \right)J,
\end{eqnarray}
where $C_k$ and $\Gamma_k^{i_1 i_2}$ are identical to the single Born case, given in \eq{eq:top1} and \eq{eq:top2}. But for double Born diagrams $N_{\text{min}}$ and $N_{\text{max}}$ are both larger by 1 in comparison to the single Born case. The maximum number of $q-$dependent denominators is still $N_{\text{max}}=N_{\text{min}}+1$ for topology 1 and $N_{\text{max}}=N_{\text{min}}+2$  for the remaining topologies.

The longitudinal integrals for double Born diagrams are defined similarly to \eq{eq:DefLISB},
\begin{eqnarray}
I_k^{(2c)}=\int\frac{{\rm d} q_1^-}{2\pi}\frac{{\rm d} q^-_2}{2\pi}\,\e^{iq_1^- \delta z_1+iq^-_2 \delta z_2}
\,\Delta_g(Q_{1},\tilde{q}_1)\,...\,\Delta_g(Q_{N_k},\tilde{q}_{N_k}),
\end{eqnarray}
where $\tilde{q}_i$ is $q_1$, $q_2$ or $q_1+q_2$ dependent on the diagram.
The results can be expressed through the same functions $I_1, I_2, I_3$ defined in \eq{eq:I1}$-$\eq{eq:I3},
\begin{eqnarray}
I^{(2c)}_k=(-i)\mcdot\left\{ \begin{array}{c}
I_1(\Omega_2)/2, \text{\,\,\,if\,\,\, } n_k=(1,0,1)  \\
I_2(\Omega_2,\Omega_3)/2, \text{\,\,\,if\,\,\, } n_k=(1,0,2)  \\
I_3(\Omega_2,\Omega_3,\Omega_4)/2, \text{\,\,\,if\,\,\, } n_k=(1,0,3)  \\
I_2(\Omega_1+\Omega_2,\Omega_3), \text{\,\,\,if\,\,\, } n_k=(1,1,1)  \\
I_3(\Omega_4,\Omega_1+\Omega_3, \Omega_2+\Omega_3), \text{\,\,\,if\,\,\, } n_k=(2,1,1)  \\
I_3(\Omega_4, \Omega_1+\Omega_2,\Omega_1+\Omega_3), \text{\,\,\,if\,\,\, } n_k=(1,2,1)  \\
I_3(\Omega_1+\Omega_2,\Omega_3, \Omega_4), \text{\,\,\,if\,\,\, } n_k=(1,1,2)  \\ \end{array} \right\},\label{eq:LIDB}
\end{eqnarray}
where $n_{k}=(n_{q_1}, n_{q_2}, n_{q_{12}})$, with $n_{q_1}$ being the number of $q_1$ dependent denominators, etc. Obviously $N_k=n_{q_1}+n_{q_2}+n_{q_{12}}$. For details on how to perform single and double Born longitudinal integrals see Appendix~\ref{appendix:longintegrals}. The $\Omega_i$ in \eq{eq:LIDB} are defined as: $\Omega_1$\ldots$\Omega_{n_{q_1}}$ for $q_1$ dependent propagators, $\Omega_{n_{q_1}+1}$\ldots$\Omega_{n_{q_1}+n_{q_2}}$ for $q_2$ dependent propagators, and  $\Omega_{n_{q_1}+n_{q_2}+1}$\ldots$\Omega_{n_{q_1}+n_{q_2}+n_{q_{12}}}$ for $q_1+q_2$ dependent propagators.

The color operators $e_{k}$ for the double Born amplitudes in the basis of 24 basis elements $e^{(2)}_j$ (see section \ref{seq:squaring}) are provided in Appendix~\ref{seq:color}.

Even though we provided the rules necessary to evaluate any double Born graph in this subsection, we also summarize all values for $C_k, \vc{U}_{p_{k_1},p_{k_2}}, \vc{U}_{p_{k_3},p_{k_4}}, I_k^{(2c)}$ in Appendix~\ref{sec:FeynmanGraphs}.

\subsection{Squaring the matrix element}\label{seq:squaring}
In this subsection we combine vacuum, single and double Born amplitudes and derive a formula for the total squared matrix element averaged over the dense QCD matter. We start from general expressions for vacuum, single and double Born amplitudes
\begin{eqnarray}
&&\mathcal{M}^{(0)}_{n+2}=g^2\,\eps_1^{i_1}\eps_2^{i_2}\,\bar{\chi}_{n,p_3}\,\sum_{j} e_{j}^{(0)}\left(\alpha_{1,j}^{i_1 i_2}+\alpha_{2,j}^{i_1 i_2}\,i\Sigma^3\right)\,J,\\
&&\mathcal{M}^{(1)}_{n+2}=g^2\,\eps_1^{i_1}\eps_2^{i_2}\,\bar{\chi}_{n,p_3}\int \text{d}\vc{\Phi}_{\perp}\,\sum_{j} e_{j}^{(1)}\left(\beta_{1,j}^{i_1 i_2}+\beta_{2,j}^{i_1 i_2}\,i\Sigma^3\right)\,J,\\
&&\mathcal{M}^{(2c)}_{n+2}=g^2\,\eps_1^{i_1}\eps_2^{i_2}\,\bar{\chi}_{n,p_3}\int\text{d}\vc{\Phi}_{1\perp}\,\text{d}\vc{\Phi}_{2\perp}\,\sum_{j} e_{j}^{(2)}\left(\gamma_{1,j}^{i_1 i_2}+\gamma_{2,j}^{i_1 i_2}\,i\Sigma^3\right)\,J.
\end{eqnarray}
The tensors structures $\alpha_{1,2},\beta_{1,2},\gamma_{1,2}$ can be directly read off from the results in sections \ref{subseq:vac},\,\ref{subseq:singleB} and \ref{subseq:doubleB}. The basis of color operators for these three cases is:
\begin{eqnarray}
&&e^{(0)}=(a_1 a_2)_R, (a_2 a_1)_R,\label{eq:vacbasis}\\
&&e^{(1)}=(a_1 a_2 b)_R, (a_1 b a_2)_R, (b a_1 a_2)_R, (a_2 a_1 b)_R, (a_2 b a_1)_R, (b a_2 a_1)_R,\label{eq:SBbasis}\\
&&e^{(2)}=(a_1 a_2 b_1 b_2)_R, (a_1 b_1 a_2 b_2)_R, (b_1 a_1 a_2 b_2)_R, (a_2 a_1 b_1 b_2)_R, (a_2 b_1 a_1 b_2)_R, (b_1 a_2 a_1 b_2)_R,\nonumber\\
&&\qquad\,\,\,\,\,(a_1 a_2 b_2 b_1)_R, (a_1 b_1 b_2 a_2)_R, (b_1 a_1  b_2 a_2)_R, (a_2 a_1 b_2 b_1)_R, (a_2 b_1 b_2 a_1)_R, (b_1 a_2 b_2 a_1)_R,\nonumber\\
&&\qquad\,\,\,\,\,(a_1 b_2 a_2 b_1)_R, (a_1 b_2 b_1 a_2)_R, (b_1 b_2 a_1 a_2)_R, (a_2 b_2  a_1 b_1)_R, (a_2 b_2 b_1 a_1)_R, (b_1 b_2 a_2 a_1)_R,\nonumber\\
&&\qquad\,\,\,\,\,(b_2 a_1 a_2 b_1)_R, (b_2 a_1  b_1 a_2)_R, (b_2 b_1 a_1 a_2)_R, (b_2 a_2  a_1 b_1)_R, (b_2a_2  b_1 a_1)_R, (b_2 b_1a_2 a_1)_R.\nonumber\\\label{eq:DBbasis}
\end{eqnarray}
For brevity, we have omitted the overall medium color structure of $(b)_{i}$ for the single Born and $(b_{1})_i (b_{2})_j$ for the double Born color basis elements. Combining,  squaring, and averaging over the position of the medium scattering centers we get:
\begin{eqnarray}
&&\sum_{\text{spin, color}}\left\langle \left|\mathcal{M}^{(0)}_{n+2}+\mathcal{M}^{(1)}_{n+2}+\mathcal{M}^{(2c)}_{n+2}   +  \cdots  \right|^2\right\rangle_{\vc{q}_{\perp}}\label{P1to3def}\\
&&={g^4}\,\text{Tr}\left(\frac{\nslash}{2}\bar{n}\mcdot p_3\, J\bar{J}\,\left[\rho_{0}+\frac{1}{2N_c}\frac{N}{A_{\perp}}\int\frac{\text{d}^2\vc{q}_{\perp}}{(2\pi)^2}\left\{\left|\tilde{v}(\vc{q}_{\perp})\right|^2\rho_1+\tilde{v}(\vc{q}_{\perp})\tilde{v}^{*}(-\vc{q}_{\perp})\,\rho_{(2c)}\right\} +  \cdots \right]   \right).  \nonumber
\end{eqnarray}
Note that the term $\propto \Tr   \mathcal{M}^{(0)\dagger}_{n+2}  \mathcal{M}^{(1)}_{n+2}$ vanishes and the term  
$\propto  \Tr   \mathcal{M}^{(2c)\dagger}_{n+2}  \mathcal{M}^{(2c)}_{n+2}$
contributes to higher  order in opacity~~\cite{Gyulassy:2000er,Vitev:2007ve}.
Here $\rho_0$ is given by the vacuum splitting and has been calculated in section \ref{sec:vacuumresult}. The single and double Born terms $\rho_1$ and $\rho_{(2c)}$ are:
\begin{eqnarray}
\rho_1&=&\sum_{j',j}\,\braket{e^{(1)}_{j'}}{ e^{(1)}_j}\,\left(\text{Re}\left[\beta_{1,j'}^{*}\mcdot \beta_{1,j}+\beta_{2,j'}^{*}\mcdot \beta_{2,j}\right]\,\mathbb{I}-\text{Im}\left[\beta_{1,j'}^{*}\mcdot \beta_{2,j}-\beta_{2,j'}^{*}\mcdot \beta_{1,j}\right]\,\Sigma^{3}\right),\nonumber\\
\rho_{(2c)}&=&2\sum_{j',\,j}\,\braket{e^{(0)}_{j'}}{ e^{(2)}_j}\,\left(\text{Re}\left[\alpha_{1,j'}^{*}\mcdot \gamma_{1,j}+\alpha_{2,j'}^{*}\mcdot \gamma_{2,j}\right]\,\mathbb{I}+\text{Im}\left[\alpha_{2,j'}^{*}\mcdot \gamma_{1,j}-\alpha_{1,j'}^{*}\mcdot \gamma_{2,j}\right]\,\Sigma^{3}\right). \qquad  \label{eq:rho12}
\end{eqnarray}
In the equations above for the single Born expression we have used the fact that the  Gram matrix of color basis vectors is symmetric. This is explicitly shown below in this subsection. The dot products between the tensor structures $\alpha, \beta, \gamma$ indicate contractions, for example $\beta_{1,j'}^{*}\mcdot \beta_{1,j}\equiv \sum_{i_1, i_2}\beta_{1,j'}^{*i_1 i_2} \beta_{1,j}^{i_1 i_2}$. The Gram matrices of the color vector basis necessary for the evaluation of the squared matrix element are straightforward to obtain
\begin{eqnarray}
&&\braket{e^{(1)}_{j'}}{ e^{(1)}_j}=T_R
\left[ \begin{array}{cccccc}
c_1 & c_2 & c_3 & c_2 & c_3 & c_4  \\
c_2 & c_1 & c_2 & c_3 & c_4 & c_3  \\
c_3 & c_2 & c_1 & c_4 & c_3 & c_2  \\
c_2 & c_3 & c_4 & c_1 & c_2 & c_3  \\
c_3 & c_4 & c_3 & c_2 & c_1 & c_2  \\
c_4 & c_3 & c_2 & c_3 & c_2 & c_1 \end{array} \right],\label{gram1}\\
&&\braket{e^{(0)}_{j'}}{ e^{(2)}_j}=T_R\left[
\begin{array}{cccccccccccccccccccccccc}
 c_1 & c_2 & c_3 & c_2 & c_3 & c_4 & c_1 & c_1 & c_2 & c_2 & c_2 & c_3 & c_2 & c_1 & c_1 & c_3 & c_2 & c_2 & c_3 & c_2 & c_1 & c_4 & c_3 & c_2 \\
 c_2 & c_3 & c_4 & c_1 & c_2 & c_3 & c_2 & c_2 & c_3 & c_1 & c_1 & c_2 & c_3 & c_2 & c_2 & c_2 & c_1 & c_1 & c_4 & c_3 & c_2 & c_3 & c_2 & c_1
\end{array}
\right].\nonumber\\ \label{gram2}
\end{eqnarray}
Because every element of these two matrices above is a number times a unit matrix in color space, the squared matrix element of both single and double Born amplitudes automatically is a singlet in color space. The color factors as functions of the SU(3) Casimirs are
\begin{eqnarray}
c_1&=&C_F^3, \qquad c_2=C_F^2(C_F-C_A/2),\qquad c_3=C_F(C_F-C_A/2)^2,\nonumber\\ 
c_4&=&C_F(C_F-C_A)(C_F-C_A/2)=2c_3-c_2.
\end{eqnarray}

Unlike in vacuum, the squared two gluon amplitude in medium is in general not a singlet in Dirac space. This was also found for the single gluon probability kernel in Ref. \cite{Ovanesyan:2011xy}. In vacuum $\alpha_{1,j}$ and $\alpha_{2,j}$ are real and, hence, the $\Sigma_3$ piece cancels. This is not the casse in medium, because the longitudinal integrals have a non-zero complex phase. However, if the jet has been created by a pure QCD interaction, the trace $\text{Tr}\left(\frac{\nslash}{2}\,J\bar{J}\,\Sigma^3\right)=0$ and the medium-induced two gluon emission factorizes from the production process, similarly to the single gluon emission \cite{Ovanesyan:2011xy}.

\subsection{Cascade approximation for the two gluon splitting function}
We define the $1\rightarrow 2$ and $1\rightarrow 3$ splittings in the presence of dense QCD matter
\begin{eqnarray}
&&\sum_{\text{spin, color}}\left\langle \left|\mathcal{M}^{(0)}_{n+1}+\mathcal{M}^{(1)}_{n+1}+\mathcal{M}^{(2c)}_{n+1} + \cdots \right|^2\right\rangle_{\vc{q}_{\perp}}=\frac{2g^2}{s_{jk}^2}\langle P_{i\rightarrow jk}[p_j,p_k]\rangle
\sum_{\text{spin,color}}\left|\mathcal{M}^{(0)}_{n}\right|^2 ,  \nonumber  \\
&&\sum_{\text{spin, color}}\left\langle \left|\mathcal{M}^{(0)}_{n+2}+\mathcal{M}^{(1)}_{n+2}+\mathcal{M}^{(2c)}_{n+2}  + \cdots   \right|^2\right\rangle_{\vc{q}_{\perp}}=\frac{4g^4}{s_{jkl}^2}\langle P_{i\rightarrow jkl}[p_j,p_k,p_l]\rangle
\sum_{\text{spin,color}}\left|\mathcal{M}^{(0)}_{n}\right|^2 . \nonumber\\ 
\end{eqnarray}
The full splitting functions become a sum over the opacity series
\begin{eqnarray}
&&\langle P_{i\rightarrow jk}[p_j,p_k]\rangle=\langle P^{(0)}_{i\rightarrow jk}[p_j,p_k]\rangle+\langle P^{(1)}_{i\rightarrow jk}[p_j,p_k]\rangle,\\
&&\langle P_{i\rightarrow jkl}[p_j,p_k,p_l]\rangle=\langle P^{(0)}_{i\rightarrow jkl}[p_j,p_k,p_l]\rangle+\langle P^{(1)}_{i\rightarrow jkl}[p_j,p_k,p_l]\rangle. \qquad
\end{eqnarray}
The first term corresponds to the vacuum splitting function\footnote{It is identical to it due to our normalization} and the second term corresponds to the first order in opacity term, including both single and double Born graphs.
 Note that $p_i, p_j, p_k, p_l$ are momenta of external partons and independent of the medium averaging, which is not shown in the above equations but is present in the second terms. The medium-induced cascade formula, similarly to the vacuum case, is based on the approximation that the probability to emit two gluons can be approximated by a product of single gluon emissions. This approximation is valid up to certain interference terms. The splitting function of the medium-induced ``cascade" is
\begin{eqnarray}
&&\langle P_{q\rightarrow ggq}^{\text{casc}}[p_1,p_2,p_3]\rangle^{(1)}= \nonumber \\
&&\qquad  s_{123}\Bigg(\frac{\langle P^{(0)}_{q\rightarrow gq}[p_2,p_1+p_3]\rangle \langle P^{(1)}_{q\rightarrow gq}[p_1,p_3]\rangle+\langle P^{(1)}_{q\rightarrow gq}[p_2,p_1+p_3]\rangle \langle P^{(0)}_{q\rightarrow gq}[p_1,p_3]\rangle}{s_{13}}\nonumber\\
&& \qquad  +\frac{\langle P^{(0)}_{q\rightarrow gq}[p_1,p_2+p_3]\rangle \langle P^{(1)}_{q\rightarrow gq}[p_2,p_3]\rangle+\langle P^{(1)}_{q\rightarrow gq}[p_1,p_2+p_3]\rangle \langle P^{(0)}_{q\rightarrow gq}[p_2,p_3]\rangle}{s_{23}}\nonumber\\
&& \qquad  +\frac{\langle P^{(0)}_{q\rightarrow gq}[p_1+p_2,p_3]\rangle \langle P^{(1)}_{g\rightarrow gg}[p_1,p_2]\rangle+\langle P^{(1)}_{q\rightarrow gq}[p_1+p_2,p_3]\rangle \langle P^{(0)}_{g\rightarrow gg}[p_1,p_2]\rangle}{s_{12}}\Bigg).  \qquad \label{cascademedium}
\end{eqnarray}
This equation is derived analogously to the one in vacuum, \eq{cascade}, and takes into account that the interaction with the medium can happen either in the first or the second splitting\footnote{There can be medium interactions in both splittings at higher orders in opacity.}. The medium-modified $1\rightarrow 2$ splitting functions are related to the medium-induced splitting kernels $x\,{\rm d} N/{\rm d} x\, {\rm d} \vc{k}_{\perp}$ calculated  in Ref. \cite{Ovanesyan:2011xy,Ovanesyan:2011kn} and are reviewed in Appendix~\ref{sec:appendixSplittingFunctions}.


\section{Angular distributions of  splitting functions}\label{sec:AO}\label{sec:angulardistribs}
In this section we study the angular distributions of the  collinear vacuum and medium-induced splittings. We start with an overview of coherent branching and angular ordering, following closely Ref. \cite{Ellis:1991qj}. Consider an arbitrary hard process with a total of $n$ incoming and outgoing quarks and/or gluons and an exclusive differential cross section $\sigma_n$. In addition, we define $\sigma_{n+1}$ as the lowest order differential cross section to emit an ultrasoft (eikonal) gluon with momentum scaling $(\lambda^2,\lambda^2,\lambda^2)$ from either of the external legs. Using the well known eikonal approximation of QCD we find
\begin{eqnarray}
{\rm d}\sigma_{n+1}={\rm d}\sigma_n\,\frac{{\rm d}\omega}{\omega}\,\frac{{\rm d}\Omega}{2\pi}\,\frac{\alpha_S}{2\pi}\, \sum_{i,j=1}^n C_{ij}\, W_{ij},\label{eq:factorization1}
\end{eqnarray}
where $\omega$ is the energy of the emitted gluon, $C_{ij}$ is a color factor and
\begin{eqnarray}
W_{ij}=\frac{\omega^2\,p_i\mcdot p_j}{p_i\mcdot q\,p_{j}\mcdot q}=\frac{1-\cos \theta_{ij}}{(1-\cos{\theta_{iq}})(1-\cos{\theta_{jq}})}.
\end{eqnarray}
Each term of the sum in \eq{eq:factorization1} corresponds to a different interference term, where the ultrasoft gluon is attached to the leg $i$ in the matrix element and leg $j$ in the complex conjugate of the matrix element. Thus, the ultrasoft branching depends on the global structure of the event. 
We have assumed that all external legs are massless, $p_i$ is the momentum of leg $i$, $q$ is the momentum of the emitted gluon and the angles between legs $i$ and $j$ and leg $i$ and the soft gluon are defined as $\theta_{ij}$ and $\theta_{iq}$, respectively. The function $W_{ij}$ has the well known property of {\it{angular ordering}}. Namely, if one rewrites
\begin{eqnarray}
W_{ij}=W_{ij}^{[i]}+W_{ij}^{[j]},\label{eq:Wij}
\end{eqnarray}
where 
\begin{eqnarray}
W_{ij}^{[i]}=\frac{1}{2}\left(W_{ij}+\frac{1}{1-\cos \theta_{iq}}-\frac{1}{1-\cos \theta_{jq}}\right),
\end{eqnarray}
and a similar definition for $W_{ij}^{[j]}$ with $i\leftrightarrow j$, then $W_{ij}^{[i]}$ has the property
\begin{eqnarray}
\int \frac{{\rm d}\phi_{iq}}{2\pi}\,W_{ij}^{[i]}=\frac{1}{1-\cos \theta_{iq}}\,\Theta(\theta_{ij}-\theta_{iq}).\label{eq:AO_i}
\end{eqnarray}
The integration over the azimuthal part of ${\rm d}\Omega$ in \eq{eq:factorization1} is performed while fixing the $z$ axis along the direction of parton $i$. Thus, the angle $\theta_{iq}$ is kept fixed while $\theta_{jq}$ varies as $\cos\theta_{jq}=\cos\theta_{ij}\,\cos\theta_{iq}+\sin\theta_{ij}\sin\theta_{iq}\cos\phi_{iq}$. \eq{eq:AO_i} means that after azimuthal averaging the interference term $W_{ij}^{[i]}$ emits radiation only inside the cone $R_{ij}^{[i]}$ with opening angle $\theta_{ij}$ centered around parton $i$. $W_{ij}^{[j]}$ obeys an equation analogous to \eq{eq:AO_i} and only radiates soft gluons inside the cone $R_{ij}^{[j]}$.  
In the remainder of this paper we will refer to the cones $R_{ij}^{[i]}$ and $R_{ij}^{[j]}$ as angular ordered cones.
It follows from these properties and equation \eq{eq:factorization1} that {\it{in the eikonal approximation the radiation obeys angular ordering in the sense that the emitted gluons are emitted only inside the angular ordered cones} for all $i$ and $j$}.

Having reviewed the  known properties of coherent branching, we turn to the question of how the collinear branchings behave in terms of angular distributions. In the high energy factorization picture of hard scattering processes there are three widely separated distance scales: the scale of the hard process $ \sim 1/\sqrt{s}$, the scale of collinear splittings or showering $ \sim 1/p_{\perp}$, where $p_{\perp}$ is the scale of the transverse size of the jet, and the scale of soft recombination processes $\sim 1/\Lambda_{\text{QCD}}$. The coherent branchings discussed in the last paragraph which lead to angular ordering correspond to the soft scale. Angular ordering is widely used in the literature on parton showers and is implemented in some of them, for example HERWIG and PYTHIA. To our knowledge, the study of angular distributions of collinear splittings, which are characterized by the intermediate scale of the parton shower, does not exist in the literature. In the following we perform such a study in vacuum and medium using full $1\rightarrow 3$ collinear splittings which include all interferences. For the vacuum we use results from Ref. \cite{Catani:1998nv}, where all $1\rightarrow 3$  parton collinear splittings have been calculated. In section~\ref{sec:vacuumsplitting} we derived one of these splittings, $q\rightarrow ggq$, in SCET and confirmed the result in Ref.~\cite{Catani:1998nv}. In medium we use our new \SCETG result for the $q\rightarrow ggq$ splitting presented in section~\ref{sec:mediumsplitting}

For later use, we define
\begin{eqnarray}
X_{ij}=-\frac{1}{2}\left(W_{ij}-\frac{1}{1-\cos \theta_{iq}}-\frac{1}{1-\cos \theta_{jq}}\right),
\end{eqnarray}
which has the property of {\it{anti-angular ordering}}
\begin{eqnarray}
\int \frac{{\rm d}\phi_{iq}}{2\pi}\,X_{ij}=\frac{1}{1-\cos \theta_{iq}}\,\Theta(\theta_{iq}-\theta_{ij}).
\end{eqnarray}

\subsection{The vacuum case}\label{subsec:AOvac}

We consider the five $1\rightarrow 3$ splitting functions, calculated in Ref. \cite{Catani:1998nv}, and study their angular distributions. Our goal is to clarify if collinear splittings exhibit a feature like angular ordering of soft coherent branching discussed in the previous subsection. It is clear that the notion of angular ordering is only applicable in deterministic parton showers with sequential branching. The obvious example is precisely the $q \rightarrow ggq$ splitting, where the two gluons are indistinguishable.  In order to define a notion of first and second splitting,  we choose the limit when one of the three partons in the final state is much softer than the two others, $z_1\ll z_2,z_3$. One has to be careful in taking this limit, since the collinear power counting breaks down if  parton "1" becomes too soft. So we have to ensure that the energy carried by parton "1" is much smaller than the one carried by partons "2" and "3", but still much larger than  $\Lambda_{\text{QCD}}$. Or, in other words, we take the limit in the collinear branching $1\rightarrow 3$ such that the second branching is at larger distance than the first one, but still in the collinear region, not in the soft recombination regime.
In contrast to the ultrasoft branching in the previous subsection, this limit is process independent because the collinear splitting functions we started with are process independent.

Taking this limit for the five splittings: $q\rightarrow \bar{q}'q'q, q\rightarrow \bar{q}qq, q\rightarrow ggq, g\rightarrow gq\bar{q}, g\rightarrow ggg$ yields
\begin{eqnarray}
&&\langle P_{q_0\rightarrow\bar{q}_1'q_2'q_3}\rangle=  \frac{C_F T_R\,(1-c_{23})}{z_1}\frac{(1-z_2)\left(2(1-z_2)+z_2^2\right)}{z_2}\left(W_{23}^{[2]}+X_{23}\right),\label{eq:approx}\\
&&\langle P_{q_0\rightarrow\bar{q}_1q_2q_3}\rangle=\frac{C_F(1-c_{23})}{z_1}\Bigg[T_R \left(\frac{(1-z_2)(1+(1-z_2)^2)}{z_2}\left(W_{23}^{[2]}+X_{23}\right)
 \right .\nonumber\\
&&\qquad\qquad\qquad\qquad\qquad    \qquad \qquad   +  \left.  \frac{z_2(1+z_2^2)}{(1-z_2)}\left(W_{23}^{[3]}+X_{23}\right)\right)         +2(C_F-C_A/2)\, X_{23}\Bigg],\nonumber\\
&&\langle P_{q_0\rightarrow g_1g_2 q_3}\rangle=\frac{4C_F(1-c_{23})}{z_1^2}z_2(1-z_2)\frac{1-z_2+z_2^2/2}{z_2}\left(C_F \left(W_{23}^{[3]}+X_{23}\right)+ C_A\left(W_{23}^{[2]}\right)\right),\nonumber\\
&&\langle P_{g_0\rightarrow g_1q_2\bar{q}_3}\rangle=\frac{2T_R(1-c_{23}) }{z_1^2}z_2(1-z_2)(z_2^2+(1-z_2)^2)\left(C_F \left(W_{23}^{[2]}+W_{23}^{[3]}\right)+C_A\left(X_{23}\right)\right),\nonumber\\
&&\langle P_{g_0\rightarrow g_1 g_2 g_3}\rangle=\frac{4C_A^2(1-c_{23})}{z_1^2}z_2(1-z_2)
\!\!\left(\!\frac{z_2}{1-z_2}+\frac{1-z_2}{z_2}+z_2(1-z_2)\!\right)
\!\!\left(\! W_{23}^{[2]}+W_{23}^{[3]}+X_{23}\!\right)\!. \nonumber 
\end{eqnarray}

We see from the presence of both terms $W_{ij}^{[i]}$ and $X_{ij}$ that the splittings are neither angular ordered nor anti-angular ordered, however, some individual pieces are. For example, the non-abelian part of the third splitting and the abelian part of the fourth splitting are angular ordered. The identical particle piece of the second splitting proportional to $C_F(C_F-C_A/2)$ and the non-abelian part of the fourth splitting are anti-ordered. All other pieces are neither ordered nor anti-ordered. Note that, as expected, the last three splittings are proportional to the reduced vacuum $1\rightarrow 2$ splitting of the initial parton into partons "2" and "3". Conversely, the first two splittings do not exhibit any similar relation. In addition, the soft behavior of the first two splittings compared to the last three differ. The first two splittings, where the anti-quark is taken to be the softer parton, are proportional to  $1/ \omega\sim 1/z_1$ and thus free of soft singularities after including the phase space factor ${\rm d} z_1\, z_1$. The last three splittings are proportional to $1/z_1^2$ which leads to a soft singularity.

\begin{figure}[H]
\center
\includegraphics[width=185pt]{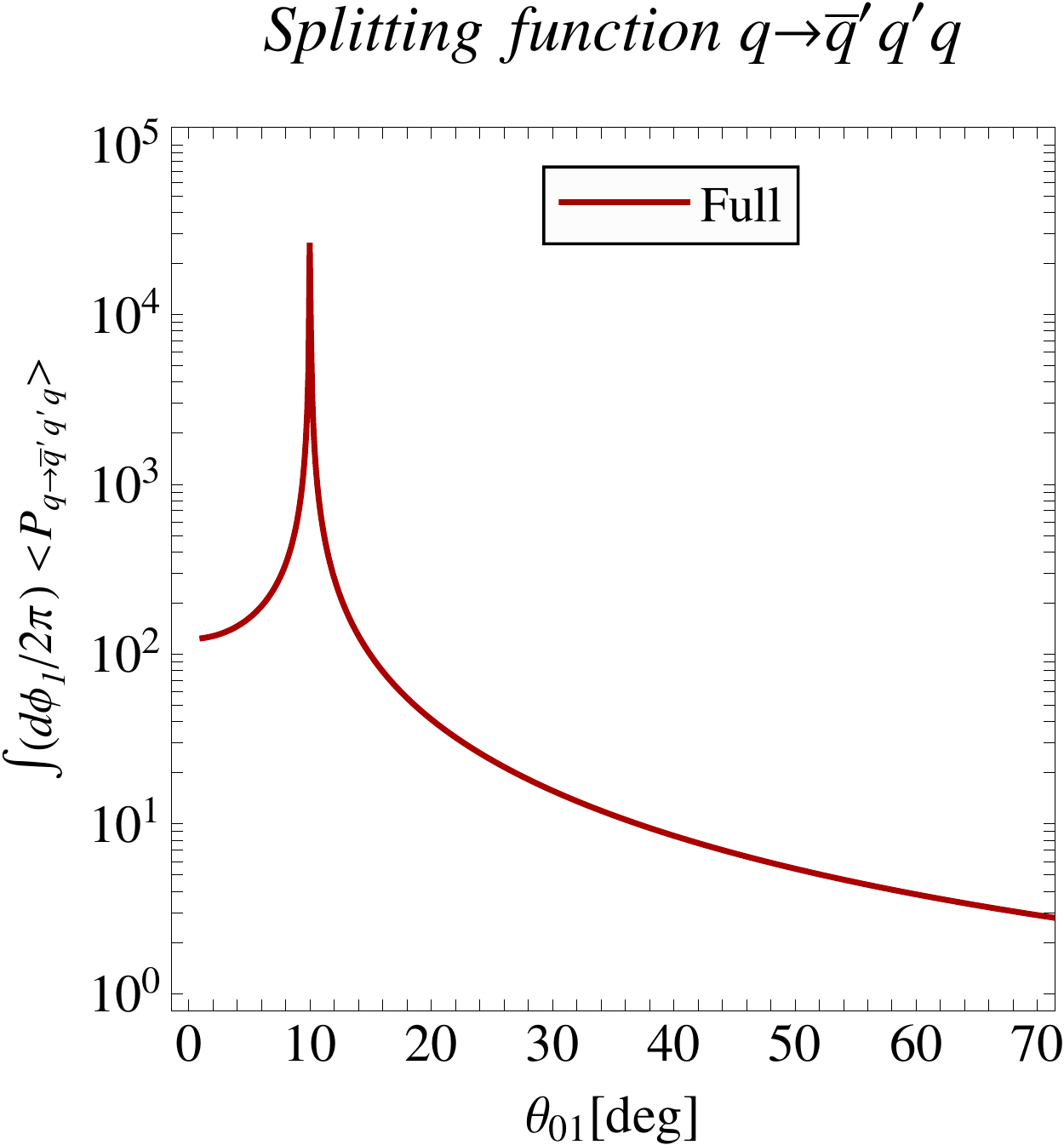}\qquad\includegraphics[width=185pt]{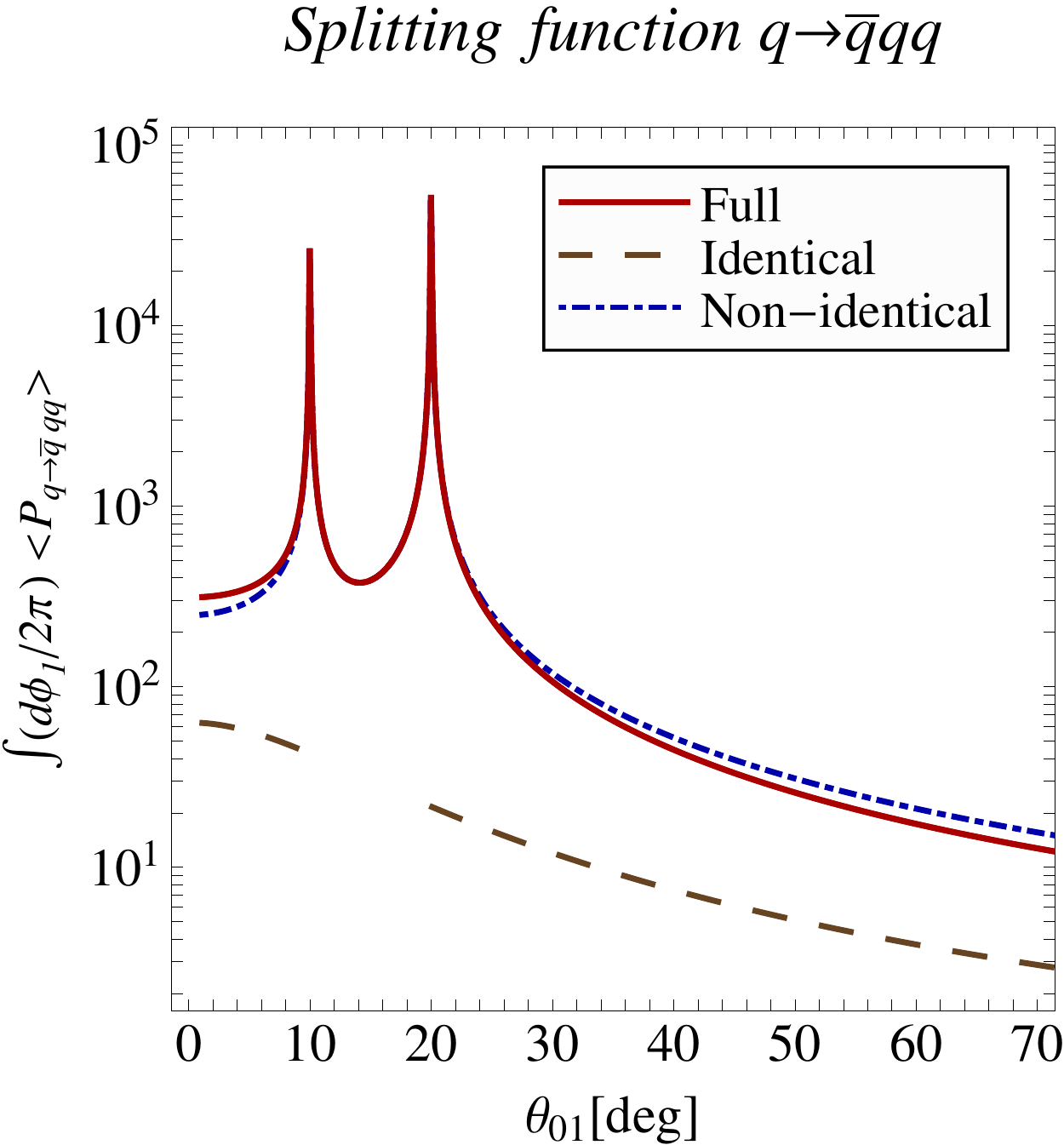}\\
\vspace{0.5cm}
\includegraphics[width=185pt]{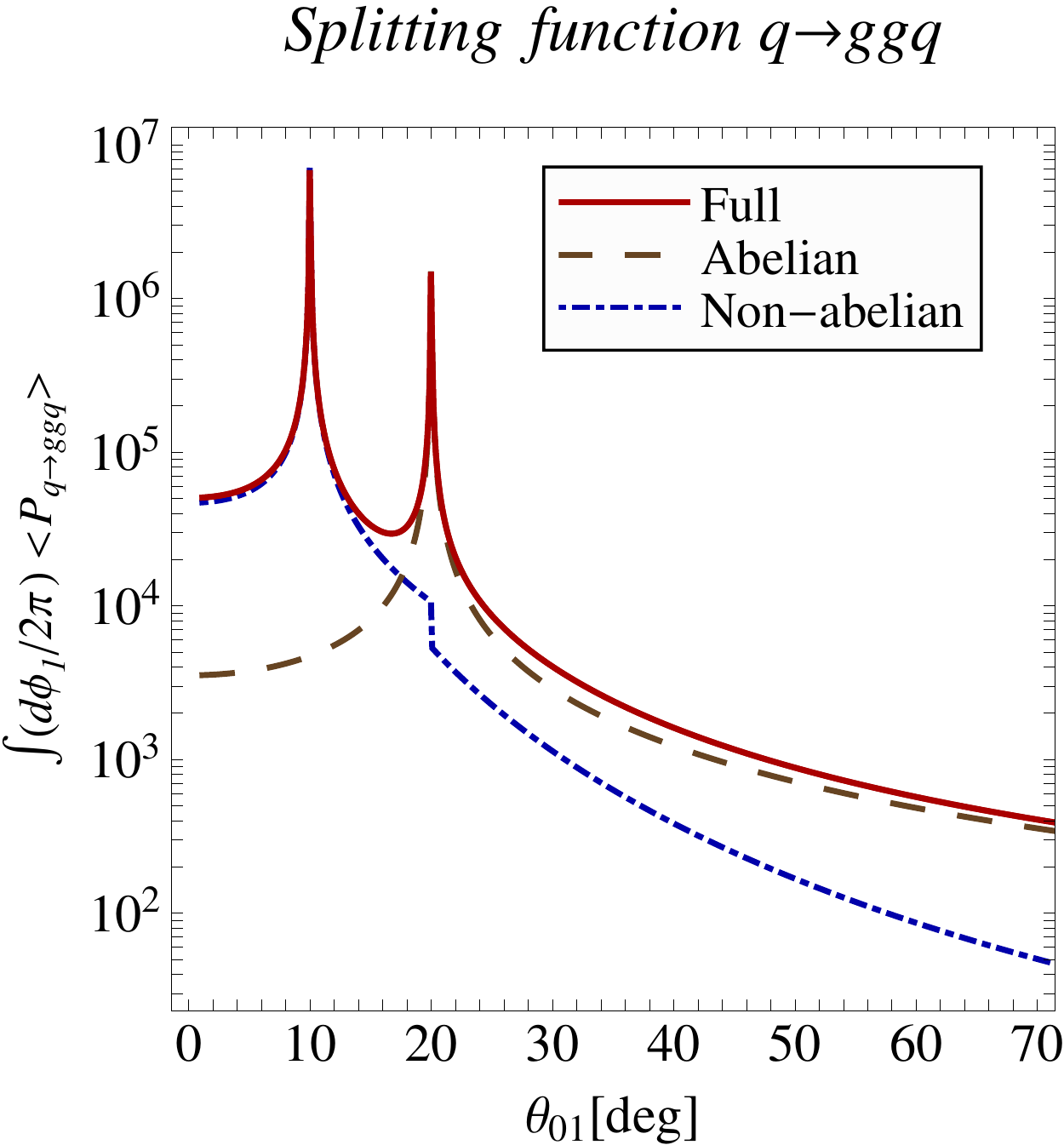}\qquad \includegraphics[width=185pt]{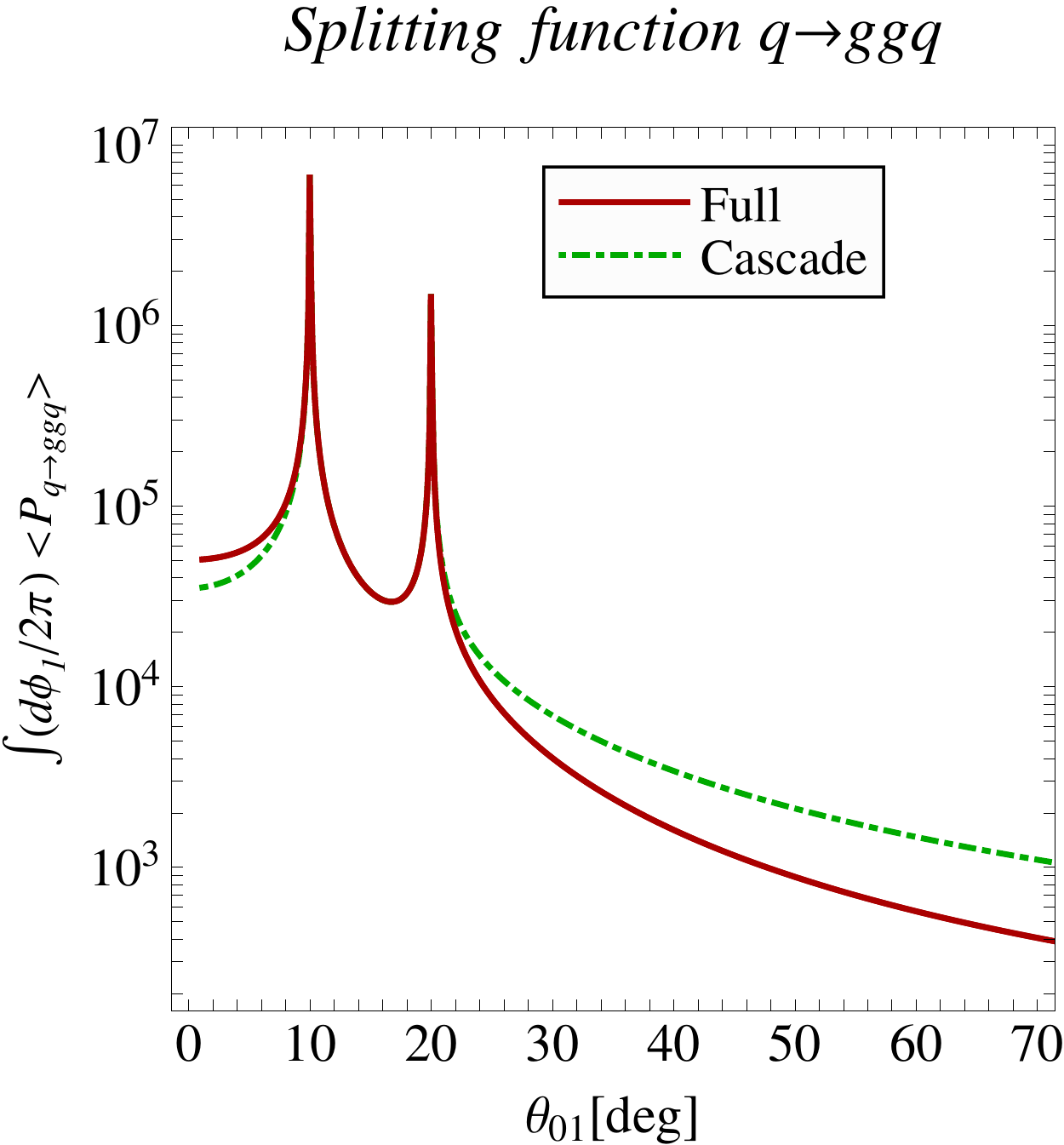}\\
\vspace{0.5cm}
\includegraphics[width=185pt]{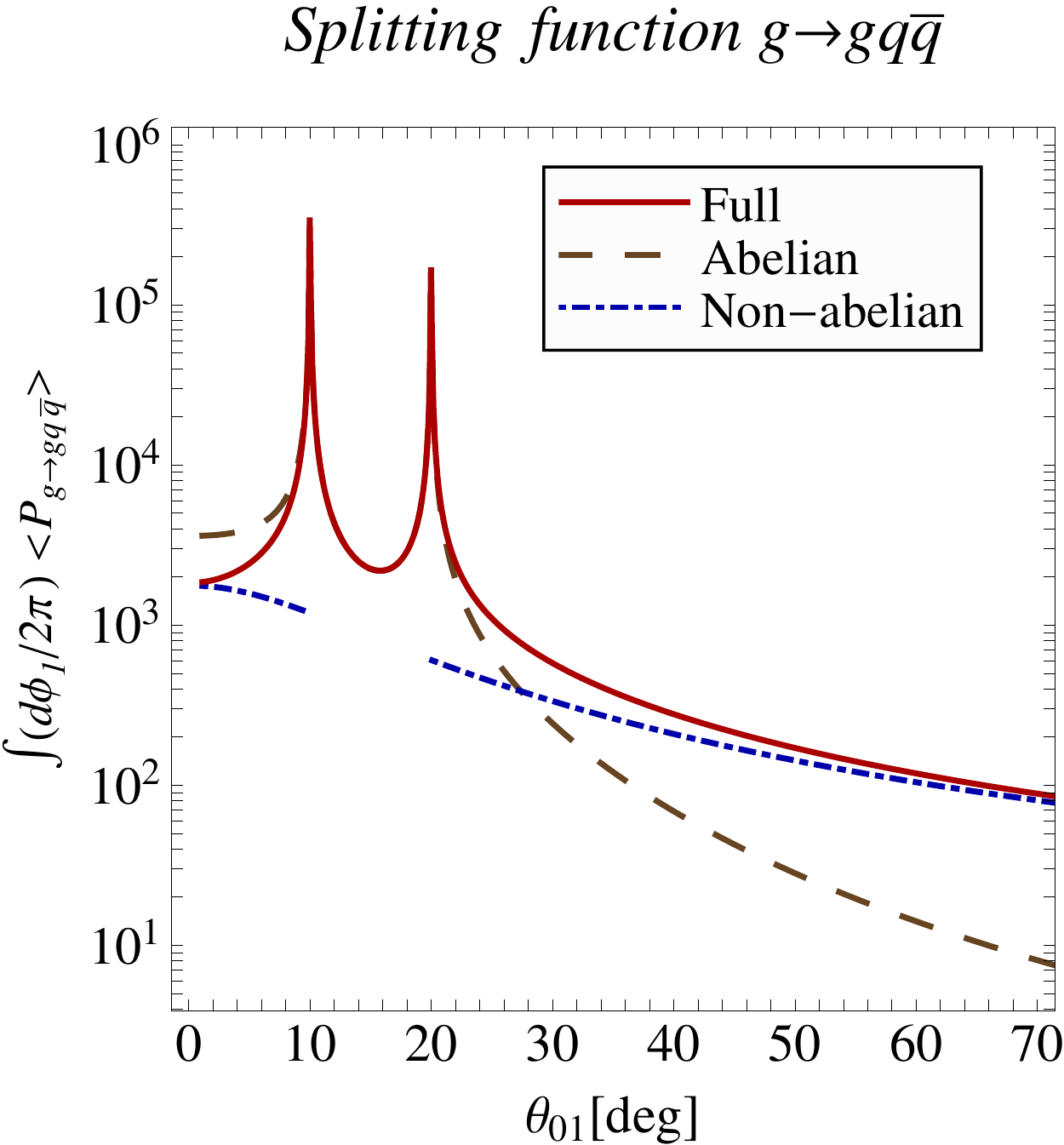}\qquad\includegraphics[width=185pt]{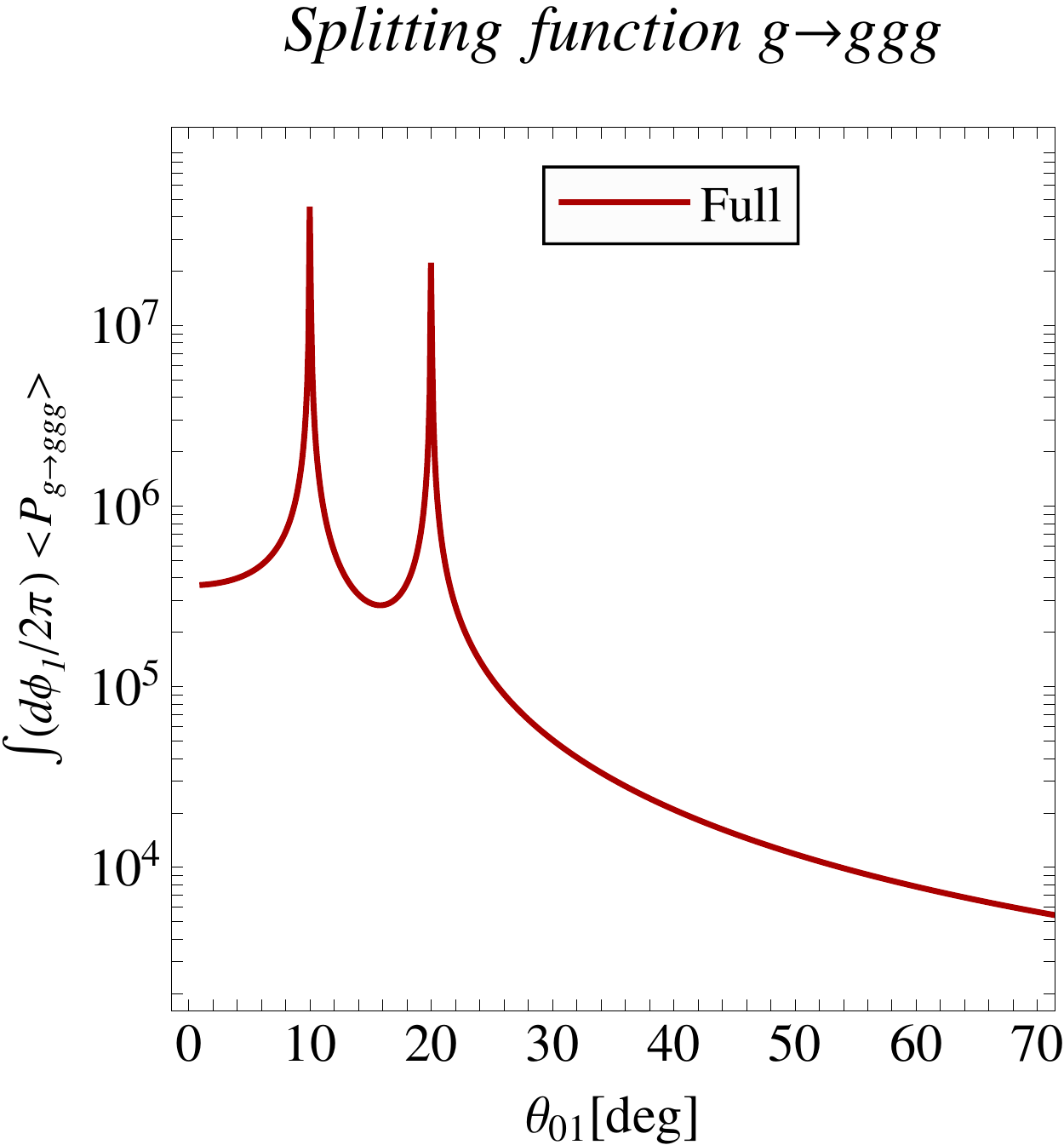}
\caption{Angular distributions of $1\rightarrow 3$ splittings in the vacuum.}
\label{fig:vacuumordering}
\end{figure}

Next we compare the small $z_1$ limit of the full $q\rightarrow ggq$ splitting, given in \eq{eq:approx}, to the small $z_1$ limit of the cascade \eq{cascade},
\begin{eqnarray}
&& \langle P^{\text{cascade}}_{q_0\rightarrow g_1g_2 q_3}\rangle=(1-c_{23})\frac{4C_F}{z_1^2}z_2(1-z_2)\frac{1-z_2+z_2^2/2}{z_2}  \nonumber \\
&&   \qquad \qquad \qquad  \qquad \quad\times  \left(C_F \left(W_{23}^{[3]}+X_{23}\right)+ C_A\left(W_{23}^{[2]}+X_{23}\right)\right).
\end{eqnarray}
The cascade reproduces the abelian part exactly, while only the singular behavior is reproduced for the non-abelian part. The numerical comparison of the full splitting to the cascade is shown in the middle right panel of Figure \ref{fig:vacuumordering}. As expected from the analytical formulas, the cascade reproduces both collinear singularities. 
In the tail of the distribution the cascade is larger by a factor $(C_F+C_A)/C_F$ compared to the equivalent piece ($X_{23}$) in the full splitting.

To visualize the angular distribution we plot the splitting function averaged over the azimuthal angle of the softer parton ($z_1$) with respect to the decaying parton as a function of the  angle between the decaying parton and the softer parton ($z_1$). No other phase space factors are included. We fix the angle between the second (third) and the decaying parton to be $10$ degrees ($20$ degrees). We set the energy of the initial quark $E_0=100$~GeV, $z_1=0.03$ and $z_2=2/3, z_3=1/3$. Note that $z_1$ is much smaller than $z_2$ and $z_3$ but the enrgy of parton "1" is still much larger than $\Lambda_{\text{QCD}}$. Thus, this choice of parameters obeys the desired limit. For the numerics we exploit that partons "2" and "3" have to be approximately back-to-back in the transverse plane in order to balance the total transvers momentum. In Figure~\ref{fig:vacuumordering} we present these plot for all five splittings, which we discussed in the following:
\begin{itemize}
\item{Each splitting has a collinear singularity at angles of $10$ and $20$ degrees as expected.}
\item{The steepness of the angular distributions outside the angular ordered cones (at 40 and 50 degrees with respect to the initial decaying parton) is $\sim 1/\theta_{0q}^4$ in the cases when there is angular ordering and $\sim 1/\theta_{0q}^2$ when there is no ordering, consistent with equation \eq{eq:approx}. Note that in the situations when there is ordering,  radiation is still present outside the angular ordered cones, though it is power suppressed compared to the case of no ordering. The reason for this is that we are averaging over the azimuthal angle with respect to the initial (decaying) parton, not with respect to one of the partons $"2"$ or $"3"$ like in \eq{eq:AO_i}. Since $W_{ij}^{[i]}$ and $W_{ij}^{[j]}$ become non-positive definite outside the angular ordered cones they cannot be interpreted as probabilities anymore. 
Since the solid angle of the emitted gluon is $\sim {\rm d}\theta_{0q}^2$, the amount of radiation outside of the cone depends on the cone size logarithmically for the power law $\sim 1/\theta_{0q}^2$, but by an inverse power for the power law $\sim 1/\theta_{0q}^4$.}
\item{In the middle left panel of Figure~\ref{fig:vacuumordering}, for the $q_0\rightarrow g_1g_2q_3$ splitting, the abilian contribution dominates in the tail, while the non-abilian contribution to the full result is only marginal. This can be understood qualitatively. When a gluon is emitted at large angle, it cannot resolve the small angle between $g_2$ and $q_3$. Thus, the gluon is effectively emitted from an on-shell quark ($q_0$). A similar argument holds for the splitting  $g_0\rightarrow g_1q_2\bar{q}_3$ and leads to the non-abilian radiation dominating over the abilian radiation in the tail, consistent with the bottom left panel of Figure~\ref{fig:vacuumordering}. The same qualitative analysis yields that the tail of the angular distribution for the splitting $g\rightarrow ggg$ is not ordered.}
\item{Note that we plot the absolute value of the angular distribution for the identical-particles term of the splitting $q\rightarrow \bar{q}qq$ and the abelian term of the splitting $g\rightarrow gq\bar{q}$. In both cases, the distribution is not always positive and could not be shown in a logarithmic plot. In the first case the true contribution is positive for $\theta_{01}<10\degree$, zero for $10\degree<\theta_{01}<20\degree$ and negative for $\theta_{01}>20\degree$. In the second case, the true contribution is negative for $\theta_{01}<10\degree$, zero for $10\degree<\theta_{01}<20\degree$ and positive for $\theta_{01}>20\degree$. The difference in sign between first and second case is due to $C_F-C_A/2<0$.}
\end{itemize}

In conclusion, we find that inside the collinear parton shower there is no angular ordering in contrast to the ultrasoft coherent branching. The implications of this result for parton showers remain to be studied phenomenologically by concentrating on observables related to collimated, isolated jets.   This will be done elsewhere.

\subsection{The dense QCD matter case}
In section \ref{sec:mediumsplitting} we calculated the medium-induced splitting $q_0\rightarrow g_1 g_2 q_3$ using \SCETG, which was the most technically demanding part of this paper. See Appendix~\ref{sec:appendixSplittingFunctions} for details on the full calculation and an approximate reduced formula valid in the small $z_1$ limit. In this section, we perform an analysis of the angular distributions in the medium-induced splitting similarly to the one in vacuum. In vacuum, we considered five splittings and were able to analize the angular distributions in the small $z_1$ limit analytically. In medium, we just perform a numerical analysis of the splitting $q\rightarrow ggq$.

\begin{figure}[H]
\center
\includegraphics[width=270pt]{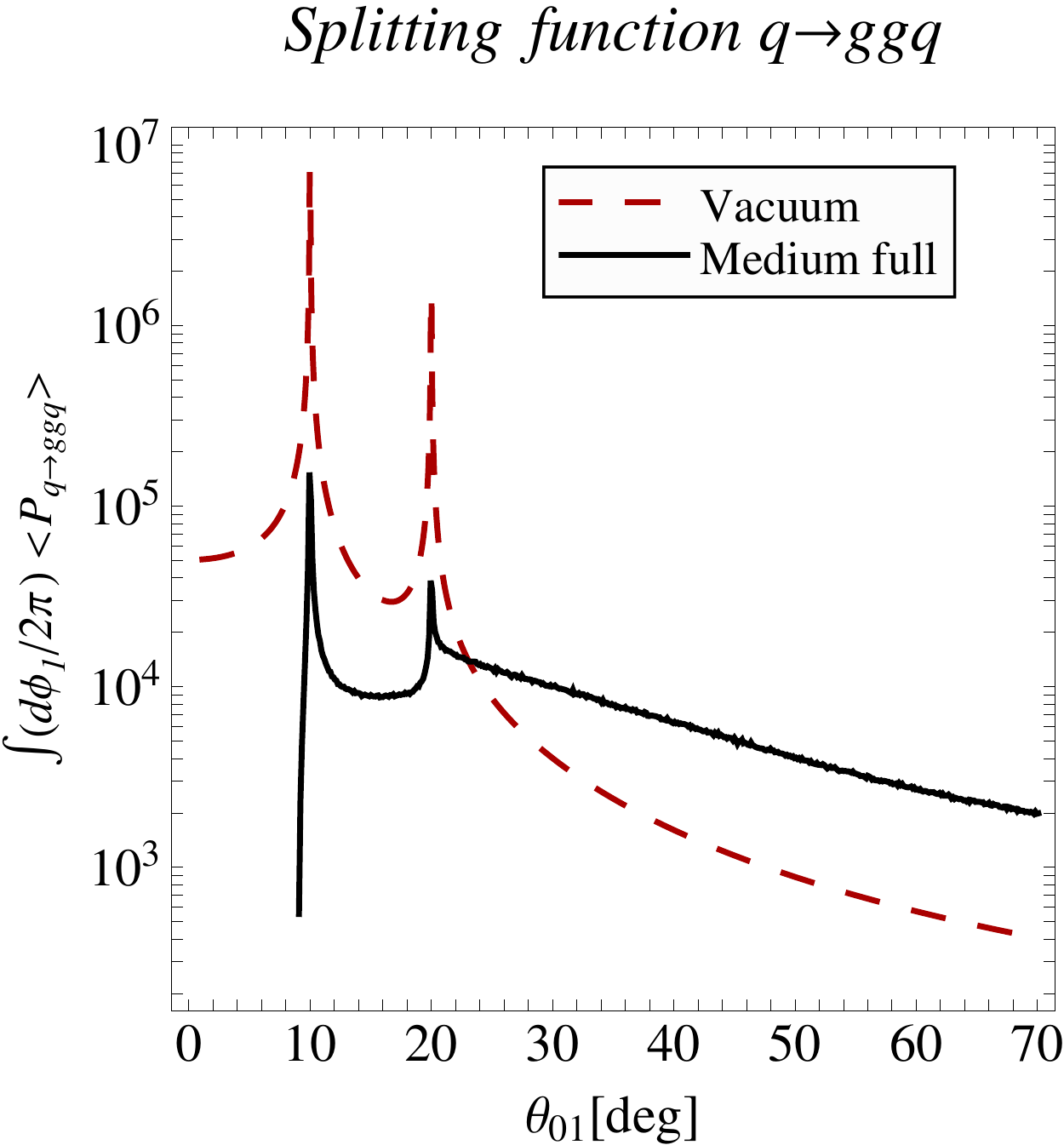}\\
\vspace{0.5cm}
\includegraphics[width=270pt]{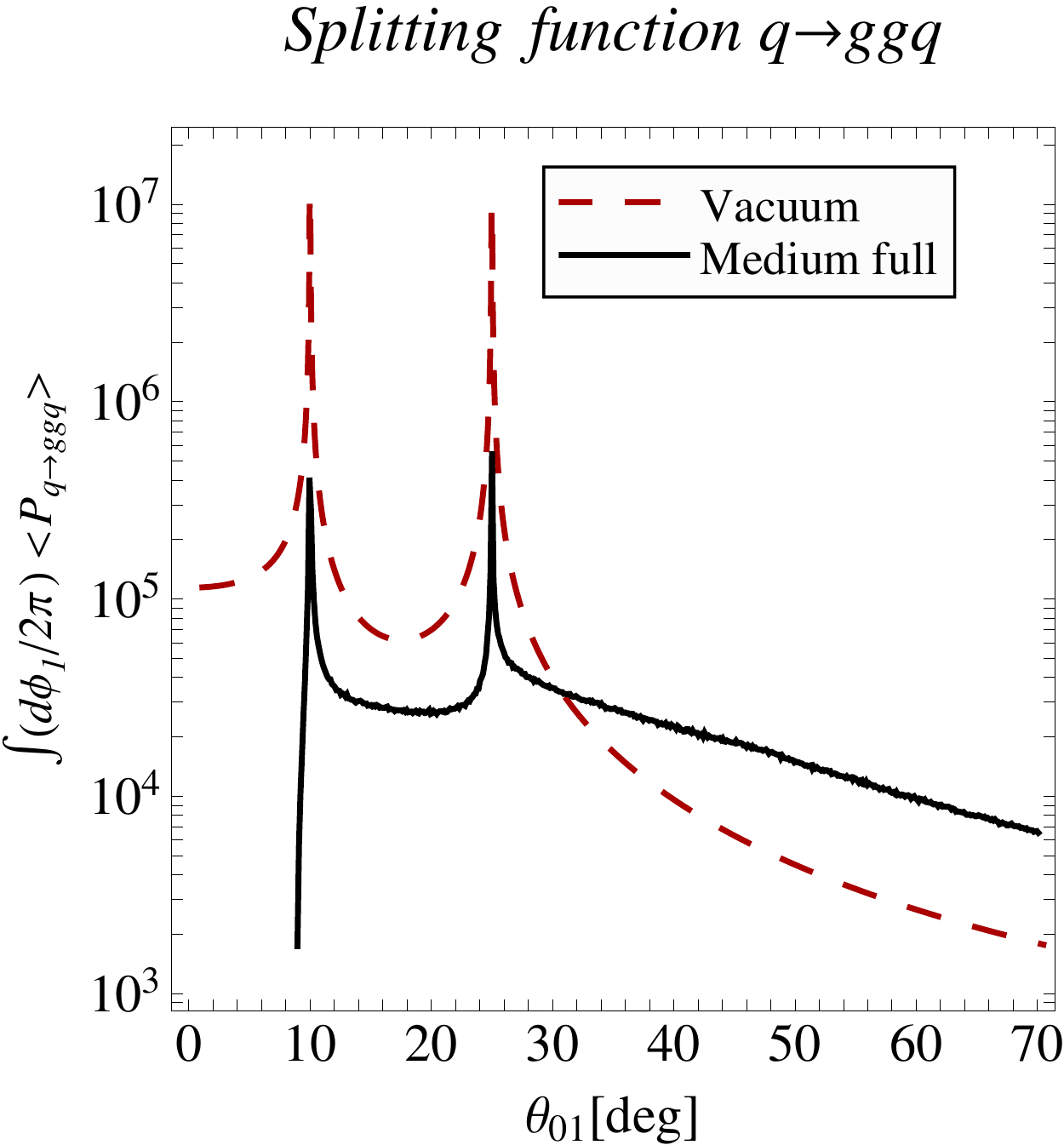}
\caption{Angular distributions of $1\rightarrow 3$ splittings in vacuum and the medium. Color coding is as follows: red$-$vacuum splitting, black$-$medium splitting. The first plot corresponds to scenario 1, the second plot to scenario 2. Further details are given in the text.}
\label{fig:mediumsplittingv1}
\end{figure}

\begin{figure}[H]
\center
\includegraphics[width=210pt]{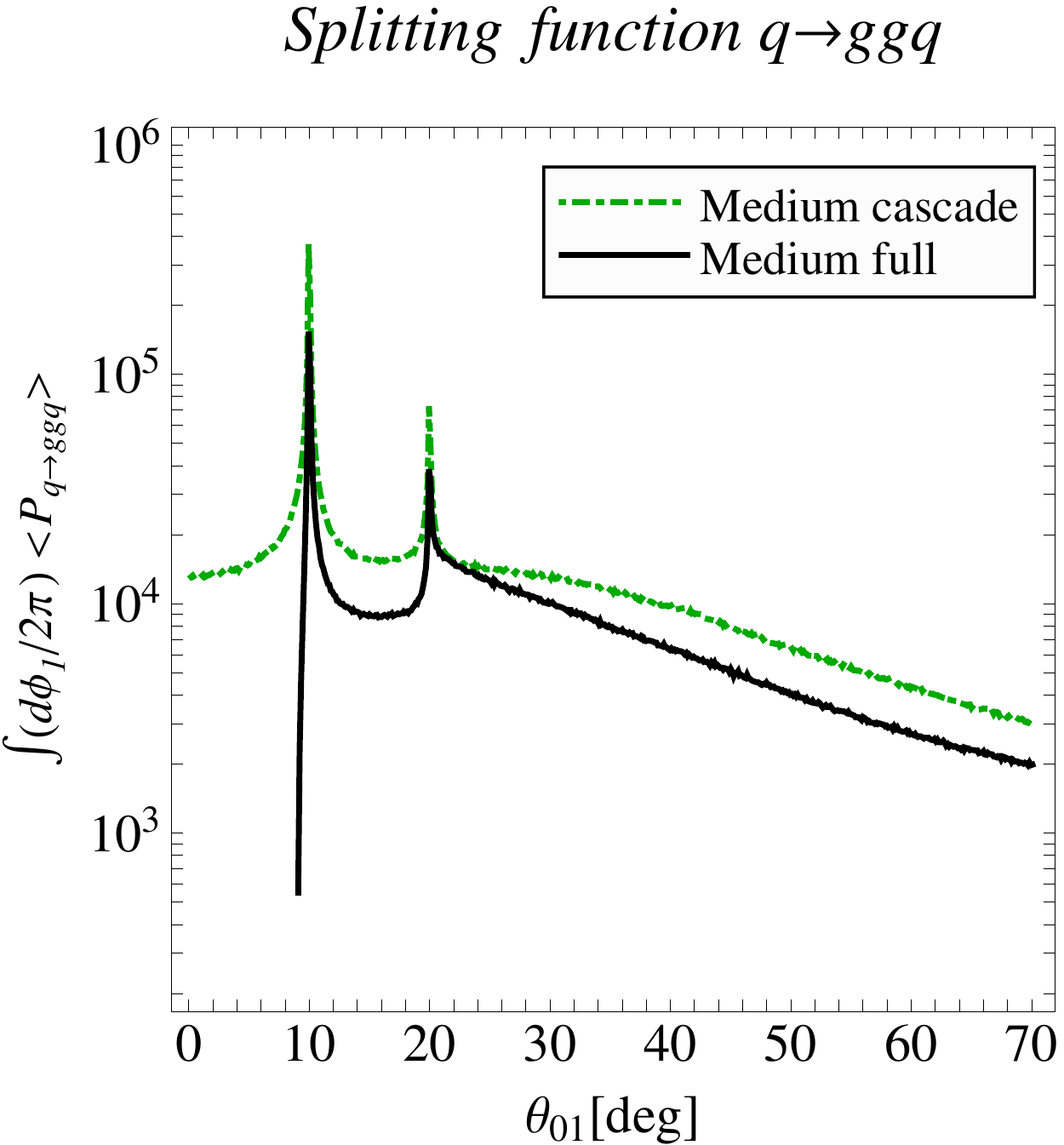}\qquad
\includegraphics[width=210pt]{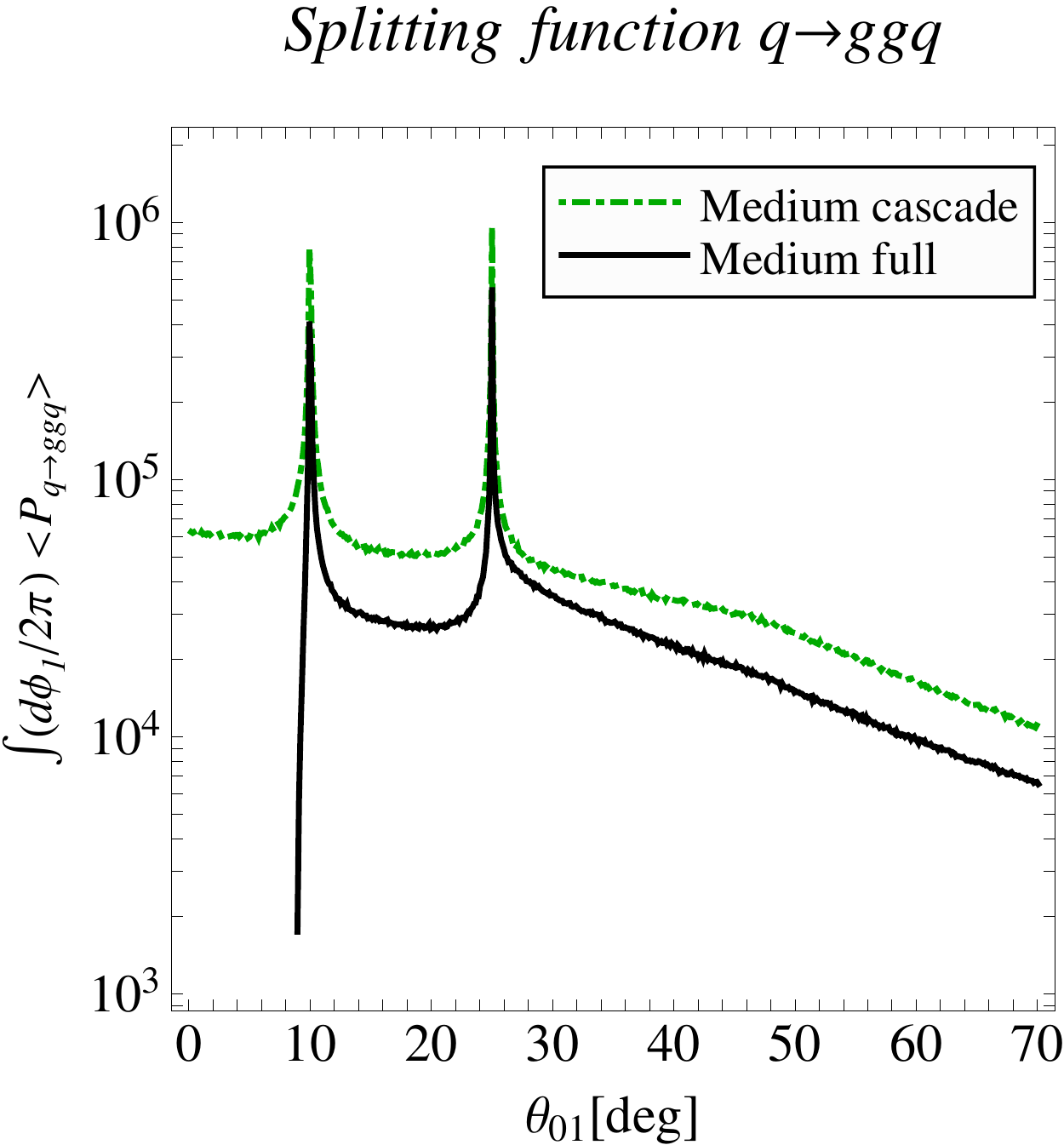}
\caption{Angular distributions of $1\rightarrow 3$ splittings in the medium. Color coding is as follows: black$-$medium splitting, green$-$medium cascade. The first plot corresponds to scenario 1, the second plot to scenario 2. Further details are given in the text.}
\label{fig:mediumsplittingv2}
\end{figure}

\begin{figure}[H]
\center
\includegraphics[width=270pt]{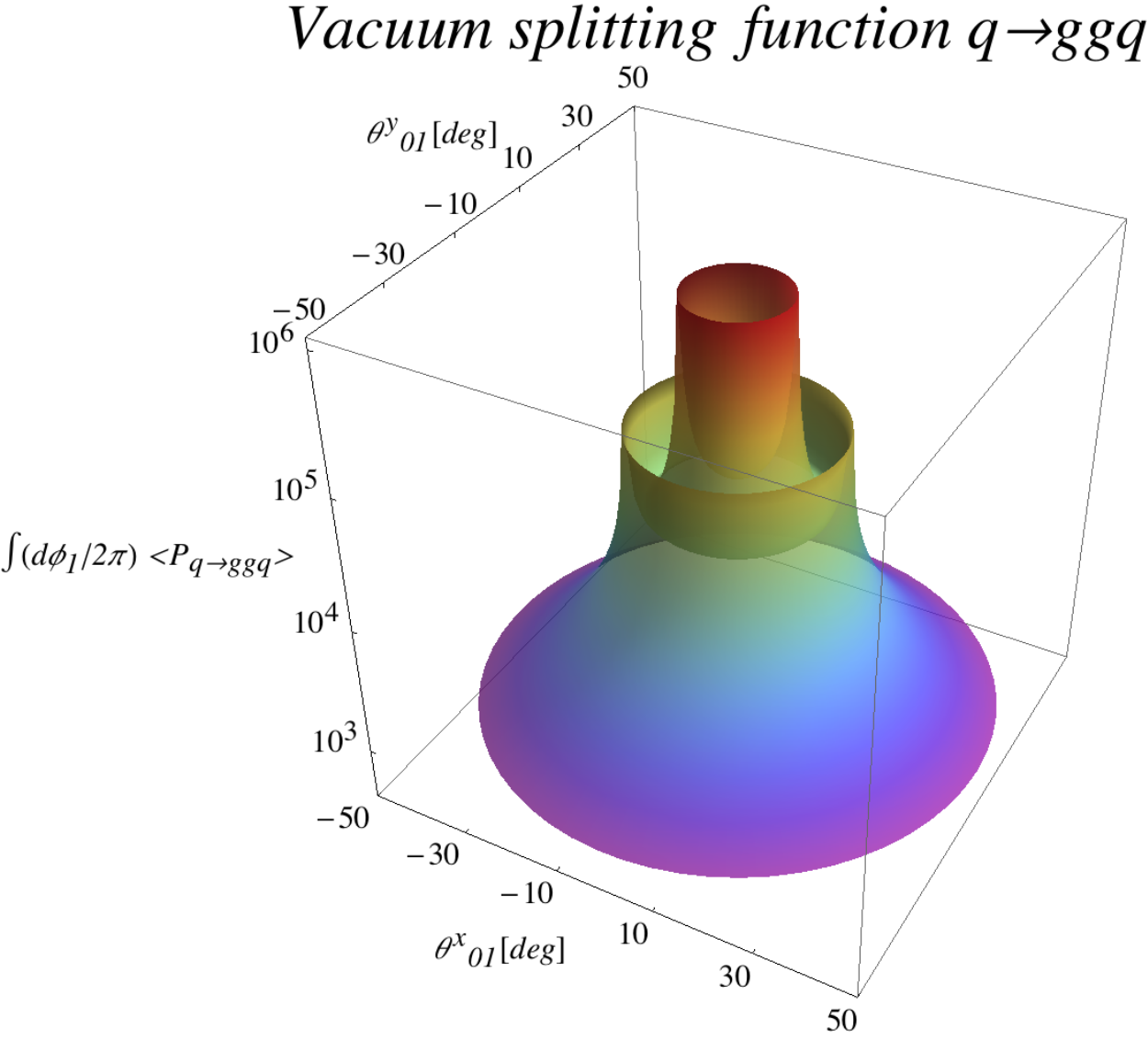}\\
\vspace{0.5cm}
\includegraphics[width=270pt]{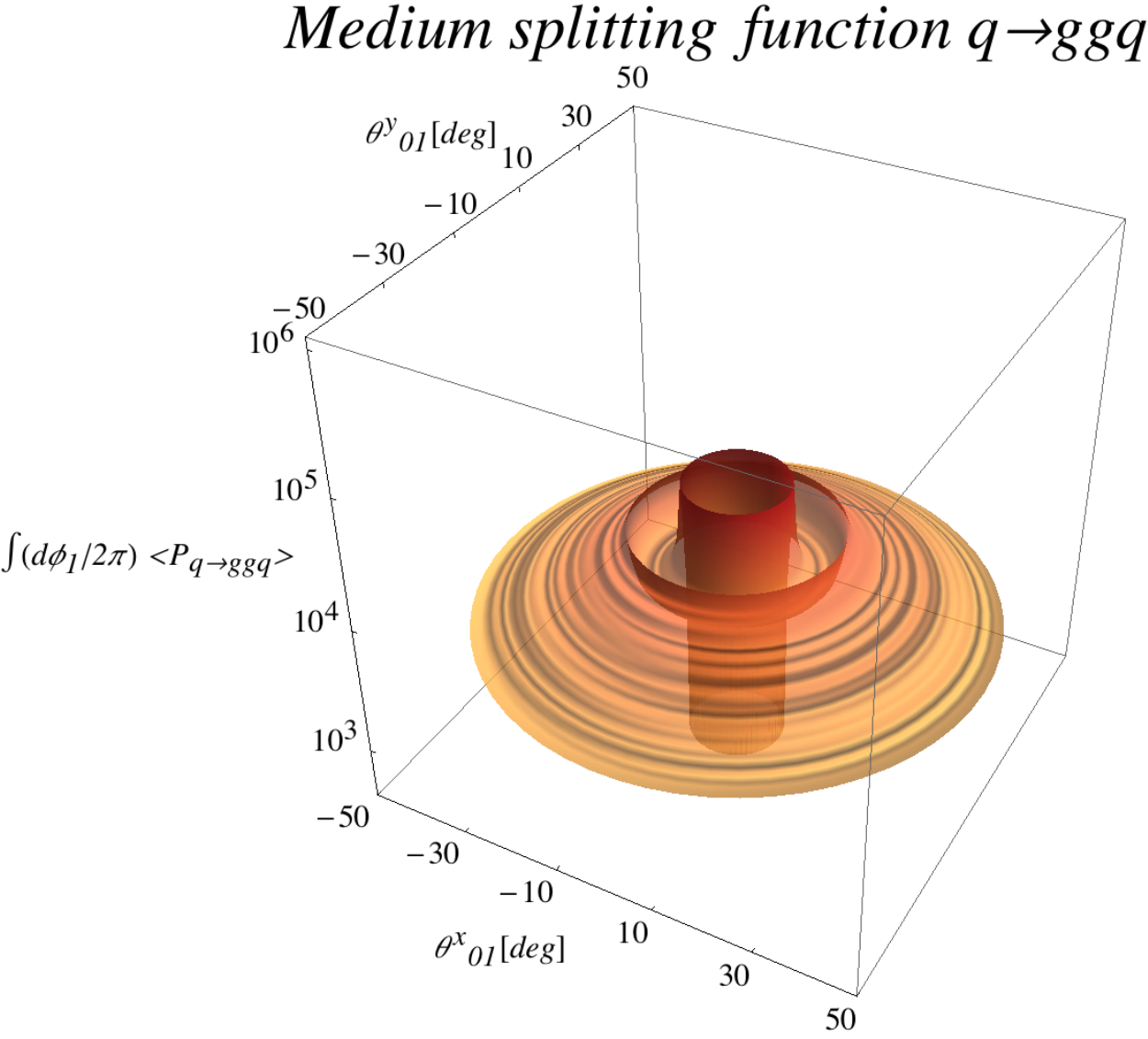}
\caption{Angular distributions of $1\rightarrow 3$ splittings in vacuum and medium. Further details are given in the text.}
\label{fig:mediumsplittingv3}
\end{figure}

To model the QCD medium,  we use the following input parameters: the Debye screening scale in the medium is $\mu=0.75\,\text{GeV}$, the size of the medium $L=5\,\text{fm}$, and the elastic scattering length of gluons in medium $\lambda_g=1\,\text{fm}$. These values have been used in \cite{Ovanesyan:2011kn} and are characteristic of the quark-gluon  plasmas  created at RHIC and LHC. The numerical results are shown in Figures~\ref{fig:mediumsplittingv1},~\ref{fig:mediumsplittingv2},~\ref{fig:mediumsplittingv3}. These plots are the medium equivalent of Figure~\ref{fig:vacuumordering}, but this time we consider two different sets of parameters. In scenario 1 (Figure~\ref{fig:mediumsplittingv3} and top plots in Figures~\ref{fig:mediumsplittingv1},~\ref{fig:mediumsplittingv2}) we use the similar values like in section~\ref{subsec:AOvac} for vacuum: $E_0=100\,\text{GeV}$, $z_1=0.03$, $z_2=0.643, \theta_{20}=10\degree, \theta_{30}=20\degree$. In scenario 2 (bottom plot in Figures~\ref{fig:mediumsplittingv1},~\ref{fig:mediumsplittingv2}) we use: $E_0=100\,\text{GeV}$, $z_1=0.03$, $z_2=0.282, \theta_{20}=25\degree, \theta_{30}=10\degree$.
We present the total medium splitting (solid black curve), medium cascade (dot-dashed green curve) and the vacuum splitting (dashed red curve). In Figure \ref{fig:mediumsplittingv1} we compare medium to vacuum splitting, in Figure \ref{fig:mediumsplittingv2} medium splitting to the medium cascade and in Figure \ref{fig:mediumsplittingv3} medium to vacuum splitting but in three dimensions.
We make the following observations:
\begin{itemize}
\item{The collinear singularities are present in both single and double Born graphs. When combining single and double Born graphs we find large (90\% to 99\%) cancellation for both scenarios. The collinear behavior, corresponding to the gluon $z_1$ being parallel to the quark $z_3$, is in both scenarios significantly reduced.}
\item{For both scenarios the tail of the angular distribution is larger for the medium-induced splitting than for the vacuum splitting. Moreover, in the direction of the parent parton (small $\theta_{01}$) there is significant cancellation in the splitting probability. These features are in agreement with the previously noted features of  medium-induced radiation.  Namely, the ${\cal O}(\alpha_s)$   $q\rightarrow qg$ splitting in dense QCD matter is wider than in the vacuum \cite{Vitev:2005yg}.}
\item{As one can see in Figure \ref{fig:mediumsplittingv2}, the cascade formula for the medium-induced splitting describes the qualitative features of the full splittings for angles between the peaks and in the tail reasonably well. The cancellation of the splitting probability along the direction of the parent quark is not reproduced by the cascade. }
\item{We compared our full medium splitting formula to the approximate formula presented in Appendix~\ref{appendix:subsectionapprox}, which includes only topologies two and four and is valid for small $z_1$. For both scenarios the difference between the exact and approximate formula is smaller than the visible thickness of the lines in our logarithmic plot, Figure \ref{fig:mediumsplittingv1}.  This is a nice cross check on our numerics.}
\item{In Figure \ref{fig:mediumsplittingv3} the spacial distribution of the medium-induced splitting is compared to that of vacuum. The collinear radiation is significantly reduced in the medium, while the medium-induced radiation is larger in the far tail.}
\end{itemize}

In conclusion,  the medium-induced splitting exhibits no angular ordering or angular anti-ordering, similarly to the vacuum splitting. The splitting probability distribution is larger in the tail in comparison to the vacuum splitting. Moreover, there is a cancellation of this probability in the direction of the original parton. These features have been described previously for the lowest order medium-induced parton branchings. 
From a practical point of view, our results imply that in constructing Monte Carlo generators to describe jet physics in heavy ion collisions, an approach without angular ordering would be preferred.


%
%
%

\section{Conclusions}\label{sec:conclusions}
In this paper we studied the the final-state angular distributions of $1\rightarrow 3$ collinear splitting functions at order ${\cal O}(\alpha_s^2)$ in vacuum and in dense QCD matter. We concentrated on the splitting $q\rightarrow ggq$. By comparing the angular distribution of this splitting function to the ones of the other splitting functions, for instance $g\rightarrow ggg$, we showed that $q\rightarrow ggq$ is a representative example. Using SCET, we calculated the vacuum splitting function to demonstrate that collinear modes are sufficient to reproduce the result obtained by Catani and Grazzini~\cite{Catani:1998nv}.
In dense QCD matter we used~\SCETG to derived the  medium-induced  $q\rightarrow ggq$ splitting  to first order in opacity, keeping the full $z_1, z_2$ dependence. 

In vacuum we studied all five ${\cal O}(\alpha_s^2)$ splitting  functions, originally derived in Ref.~\cite{Catani:1998nv}.  In all cases we find no angular ordering. This result is also supported by the fact that the angular distribution of the $q\rightarrow ggq$ splitting can be reproduced well by a parton cascade based on binary branchings. Note that our approach differs from the coherent branching ansatz which yields angular ordering and is widely used in parton shower generators like HERWIG. We study angular distributions of collinear splittings instead of ultrasoft gluon emissions. Thus, the momentum scaling in our approach differs from the one in coherent branching and hence, the different result is not unexpected. Nevertheless, the qualitative argument of color screening used to explain Sudakov effect in QED still works in our case. In large angle gluon emission from a $qg$ antenna, the emitted gluon is only sensitive to the color charge of the initial quark.

Our results indicate that the proper angular distributions inside the collinear parton shower are different from the traditional coherent branching ansatz. Nevertheless, the traditional approach to parton showers, which applies angular ordering to the collinear splitting functions, is claimed to resum large infrared Sudakov logarithms and is phenomenologically successful. One thing which would be interesting to verify is whether the amount of collinear radiation leaking outside of the angular ordered cones leads to a significant correction to parton shower phenomenology. This would have to be checked for intra-jet observables, such as jet shapes of well-isolated jets.  We leave this for future work. The same conclusion holds for the medium induced parton shower. Our detailed analysis  found no evidence of angular ordering or angular anti-ordering. An important feature is that the noticeably broader angular distribution compared to vacuum, found in ${\cal O}(\alpha_s)$  $1\rightarrow 2$ branchings, persists to higher order.   

\appendix
\section{Medium-induced splitting functions}\label{sec:appendixSplittingFunctions}
In this appendix we review the basic formulas for medium-induced splitting functions and how they are related 
to the splitting kernels. 
\subsection{Leading order ${\cal O}(\alpha_s)$  splittings}
The medium-induced splitting kernels have been calculated in Ref.~\cite{Ovanesyan:2011xy,Ovanesyan:2011kn} retaining the full $x$ dependence (beyond the soft 
gluon approximation)  using \SCETG. We rewrite them in a slightly more compact form:
\begin{eqnarray}
\frac{{\rm d} N^{(i)}}{{\rm d} x\,{\rm d}^2\vc{k}_{\perp}}=\frac{\alpha_s}{2\pi^2}\,
P^{(i)}_{\text{vac}}(x)\int\frac{d\Delta z}{\lambda_i(z)}{\rm d}^2\vc{q}_{\perp}
\frac{1}{\sigma_{\text{el}}}\frac{{\rm d}\sigma_{\text{el}}}{{\rm d}^2\vc{q}_{\perp}}\sum_{k=1}^{5}\alpha_k^{(i)}(1-\cos\Phi_k).
\end{eqnarray}
We define the transverse vectors
\begin{eqnarray}
\vc{A}_{\perp}=\vc{k}_{\perp},\,\,\,\,\, \vc{B}_{\perp}=\vc{k}_{\perp}+x\,\vc{q}_{\perp},\,
\,\,\,\,\vc{C}_{\perp}=\vc{k}_{\perp}-(1-x)\vc{q}_{\perp},\,\,\,\,\,\vc{D}_{\perp}=\vc{k}_{\perp}-\vc{q}_{\perp}.
\end{eqnarray}
In terms of these vectors, the five phases $\Phi_k$ are equal to
\begin{eqnarray}
&&\Phi_1=\Psi\vc{B}_{\perp}^2,\,\, \Phi_2=\Psi\vc{C}_{\perp}^2,\,\, \Phi_3=\Psi(\vc{C}_{\perp}^2-\vc{B}^2_{\perp}),\,\, \Phi_4=\Psi\vc{A}_{\perp}^2,\,\, \Phi_5=\Psi(\vc{A}_{\perp}^2-\vc{D}_{\perp}^2),\nonumber\\
&&\text{where}\,\, \Psi=\frac{\Delta z}{x(1-x)\,\bar{n}\mcdot p_0}.
\end{eqnarray}
We recall that $\Delta z$ is the spacial separation between the hard scattering, producing the collinear 
parent parton, and one of the subsequent medium interactions via a Glauber gluon exchange. 
The coefficients $\alpha_k^{(i)}$ are summarized in the following table:
\begin{center}
    \begin{tabular}{ | l | l | l | p{5cm} |}
    \hline
     $k$ &$\alpha_{k}^{(q\rightarrow qg)}$&  $\alpha_{k}^{(g\rightarrow gg)}$ &  $\alpha_{k}^{(g\rightarrow q\bar{q})}$ \\ \hline
    1& $\vc{b}\mcdot\left(\vc{b}-\vc{c}+\frac{\vc{a}-\vc{b}}{N_c^2}\right)$ &  $2\vc{b} \mcdot\left(\vc{b}-\vc{a}-\frac{\vc{c}-\vc{a}}{2}\right)$ & $2\vc{b} \mcdot\left(\vc{b}-\vc{a}+\frac{\vc{c}-\vc{a}}{N_c^2-1}\right)$  \\ \hline
     2&$\vc{c} \mcdot\left(2\vc{c}-\vc{a}-\vc{b}\right)$ & $2\vc{c} \mcdot\left(\vc{c}-\vc{a}-\frac{\vc{b}-\vc{a}}{2}\right)$& $2\vc{c} \mcdot\left(\vc{c}-\vc{a}+\frac{\vc{b}-\vc{a}}{N_c^2-1}\right)$ \\ \hline
  3&  $\vc{b} \mcdot\vc{c}$ & $\vc{b} \mcdot\vc{c}$ & $-2\frac{\vc{b} \,\mcdot\,\vc{c}}{N_c^2-1}$ \\ \hline
  4&   $\vc{a}\mcdot\left(\vc{d}-\vc{a}\right)$ & $\vc{a}\mcdot(\vc{d}-\vc{a})$ & $2\frac{\vc{a}\,\mcdot\,(\vc{a}-\vc{d})}{N_c^2-1}$\\ \hline
   5& $-\vc{a}\mcdot\vc{d}$ & $-\vc{a}\mcdot\vc{d}$ & $2\frac{\vc{a}\,\mcdot\,\vc{d}}{N_c^2-1}$ \\
    \hline
    \end{tabular}
\end{center}
where $\vc{a}=\vc{A}_{\perp}/\vc{A}_{\perp}^2$ and $\vc{b}, \vc{c}, \vc{d}$ are defined similarly. To relate the medium-induced splitting function to the splitting kernel we have to remove the phase space contributions; recall that
\begin{eqnarray}
&&{\rm d}\sigma_{n+1}={\rm d}\sigma_{n}\,\frac{{\rm d} N^{(i)}}{{\rm d} x\,{\rm d}^2\vc{k}_{\perp}}\,{\rm d} x\,{\rm d}^2\vc{k}_{\perp},\\
&&{\rm d}\Phi_{n+1}={\rm d}\Phi_{n}\,\frac{1}{x(1-x)}\frac{1}{16\pi^3}\,{\rm d} x\,{\rm d}^2\vc{k}_{\perp}.
\end{eqnarray}
Using that the splitting function is defined as ratio between the squared matrix element after emission and before emission, gives
\begin{eqnarray}
\langle P^{(1)}\rangle^{(i)}_{1\rightarrow 2}=P^{(i)}_{\text{vac}}(x)\,\vc{k}_{\perp}^2\,\int\frac{d\Delta z}{\lambda_i(z)}{\rm d}^2\vc{q}_{\perp}\frac{1}{\sigma_{\text{el}}}\frac{{\rm d}\sigma_{\text{el}}}{{\rm d}^2\vc{q}_{\perp}}\sum_{k=1}^{5}\alpha_k^{(i)}(1-\cos\Phi_k).
\end{eqnarray}
Note that the $q \rightarrow gq$ kernel is obtained from the $q \rightarrow qg$ kernel via the substitution 
$x\rightarrow 1-x$.
For the normalized elastic scattering cross-section we get
\begin{eqnarray}
\frac{1}{\sigma_{\text{el}}}\frac{{\rm d}\sigma_{\text{el}}}{{\rm d}^2\vc{q}_{\perp}}
=\frac{\mu^2}{\pi\left(\vc{q}_{\perp}^2+\mu^2\right)^2},
\end{eqnarray}
consistent with section~\ref{sec:SCETG}.

\subsection{Next-to-leading order ${\cal O}(\alpha_s^2)$  splittings}\label{appendix:subsectionapprox}
For the $q \rightarrow ggq$ splitting function to first order in opacity 
we find
\begin{eqnarray}
\langle P^{(1)}\rangle_{q\rightarrow ggq}=\frac{1}{N_c}\frac{s_{123}^2}{4}z_3
\int\frac{{\rm d}\Delta z}{\lambda_g(z)}\int{\rm d}^2\vc{q}_{\perp}\,
\frac{1}{\sigma_{\text{el}}}\frac{{\rm d}\sigma_{\text{el}}}{{\rm d}^2\vc{q}_{\perp}}\left(\rho_1+\rho_{(2c)}\right),
\end{eqnarray}
where $\rho_1$ and $\rho_{(2c)}$ are provided in general form in \eq{eq:rho12}. For the special 
case when parton one has softer momentum than the other two partons, $z_1\ll z_2,z_3$, these formula reduces to
\begin{eqnarray}
&&\rho_1\approx 4\left(1-z_2+\frac{z_2^2}{2}\right)
\sum_{k',\,k=1}^{10}\bra{e^{(1)'}_{k'}}\Gamma^{(1)}\ket{e^{(1)'}_{k}}\,C_{k'}^{(1)}C_{k}^{(1)}
\left(\vc{U}_{k'}^{(11)}\,\mcdot\vc{U}_{k}^{(11)}\right)\left(\vc{U}_{k'}^{(12)}\,
\mcdot\vc{U}_{k}^{(12)}\right) \nonumber   \\ 
&&  \hspace{1.9in} \times\text{Re}\,I_{k'}^{(1)*}I_{k}^{(1)},\nonumber\\
&&\rho_{(2c)}\approx 4\left(1-z_2+\frac{z_2^2}{2}\right)\sum_{k'=1,2; k=1,18}^{10}\bra{e^{(0)'}_{k'}}
\Gamma^{(2)}\ket{e^{(2)'}_{k}}\,C_{k'}^{(0)}C_{k}^{(2)}\left(\vc{U}_{k'}^{(21)}\,\mcdot\vc{U}_{k}^{(21)}\right)
\left(\vc{U}_{k'}^{(22)}\,\mcdot\vc{U}_{k}^{(22)}\right)\, \nonumber \\  
&& \hspace{2.4in} \times 2\,\text{Re}\,I_{k}^{(2c)},
\end{eqnarray}
where for $\rho_1$ the sum over $k$ runs over the 10 single Born graphs of topologies 2 and 4; 
for $\rho_{(2c)}$ the sum over $k$ runs over the 18 graphs of topologies 2 and 4; 
as well in vacuum only topologies 2 and 4 matter in this limit. 
Our notation for $\vc{U}^{(11)}, \vc{U}^{(12)}, \vc{U}^{(21)}, \vc{U}^{(22)}$ is as follows. For a given single Born graph, $\vc{U}^{(11)}, \vc{U}^{(12)}$ are the first and second transverse vectors in the corresponding entry in the third raw in the Table \ref{tab:singleborn}. Similarly, $\vc{U}^{(21)}, \vc{U}^{(22)}$ are the first and second transverse vectors in the corresponding entry in the third raw in the Table \ref{tab:doubleborn}. $\Gamma^{(1)}$ and $\Gamma^{(2)}$ 
are defined in \eq{gram1} and \eq{gram2}.
All 
other ingredients follow from the rules in section~\ref{sec:mediumsplitting} for longitudinal integrals, 
coefficients $C_k$ and color operators $e_k$.

\section{Feynman Graphs}\label{sec:FeynmanGraphs}
In section~\ref{sec:mediumsplitting} we presented rules to extract an analytic expression for any 
single and double Born Feynman diagram. \eq{eq:top1} and \eq{eq:top2} are valid for any Feynman graph. In this appendix we give explicit expressions for all coefficients $C_k$, longitudinal integrals $I_k$ and transverse vectors $\vc{U}_{p_{k_1},p_{k_2}}, \vc{U}_{p_{k_3},p_{k_4}}$.

In Figure \ref{fig:singleBdiagrams1} and~\ref{fig:doubleborngraphs} all 19 single Born and 34 double born Diagrams are shown, respectively. The corresponding analytic components can be read off Table~\ref{tab:singleborn} and~\ref{tab:doubleborn}. The frequencies are defined as
\begin{eqnarray}
&&\Omega_0=\Omega(p_1+p_2+p_3,\vc{q}_{\perp}),\nonumber\\
&&\Omega_1=\Omega(p_1,\vc{q}_{\perp}), \qquad \,\,\,\,\,\,\,\,\,\,\Omega_2=\Omega(p_2,\vc{q}_{\perp}) \qquad \,\,\,\,\,\,\,\,\,\,\Omega_3=\Omega(p_3,\vc{q}_{\perp}),\nonumber\\
&&\Omega_4=\Omega(p_2+p_3,\vc{q}_{\perp}), \qquad \Omega_5=\Omega(p_1+p_3,\vc{q}_{\perp}) \qquad \Omega_6=\Omega(p_1+p_2,\vc{q}_{\perp}),\label{Omega1def}
\end{eqnarray}
and
\begin{eqnarray}
&&\bar{\Omega}_{l}=\Omega_l(\vc{q}_{\perp}\rightarrow 0),\qquad \tilde{\Omega}_{l}
=\Omega_l(\vc{q}_{\perp}\rightarrow -\vc{q}_{\perp}),\text{    where  } l=0,1,2,3,4,5,6.
\end{eqnarray}
\begin{center}
\begin{table}[t!]
    \begin{tabular}{ | l | l | l | p{5cm} |}
    \hline
     $k$ &$C_k$&  $\vc{U}^{j_1}_{p_{k_1},p_{k_2}}\vc{U}^{j_2}_{p_{k_3},p_{k_4}}$ &  $I_{k}^{(1)}$ \\ \hline
   1& $1/s_{123}$ &  $-$ & $I_1(\Omega_0)$  \\ \hline
     2&$1/\bar{n}\mcdot p_0$ & $-$&  $I_2(\Omega_0,\Omega_2)$ \\ \hline
       3&$1/\bar{n}\mcdot p_0$  & $-$&  $I_2(\Omega_0, \Omega_1)$ \\ \hline
         4&$1/\bar{n}\mcdot p_0$  & $-$&  $I_2(\Omega_0, \Omega_3)$ \\ \hline
           5&$1/s_{13}s_{123}$ &  $\vc{U}^{j_1}_{p_1,p_3}\vc{U}^{j_2}_{p_2,p_3+p_1}$ &  $I_1(\Omega_0)$ \\ \hline
             6& $1/\bar{n}\mcdot p_0 s_{13}$ & $\vc{U}^{j_1}_{p_1,p_3}\vc{U}^{j_2}_{p_2-q,p_3+p_1}$&  $I_2(\Omega_0, \Omega_2)$ \\ \hline
               7&$1/(\bar{n}\mcdot p_0)^2(z_1+z_3)$ & $\vc{U}^{j_1}_{p_1-q,p_3}\vc{U}^{j_2}_{p_2,p_3+p_1-q}$&  $I_3(\Omega_0,\Omega_1,\Omega_5)$ \\ \hline
                 8&$1/(\bar{n}\mcdot p_0)^2(z_1+z_3)$ & $\vc{U}^{j_1}_{p_1,p_3-q}\vc{U}^{j_2}_{p_2,p_3-q+p_1}$& $I_3(\Omega_0,\Omega_3,\Omega_5)$ \\ \hline
                   9& $1/\bar{n}\mcdot p_0 s_{13}$& $\vc{U}^{j_1}_{p_1,p_3}\vc{U}^{j_2}_{p_2,p_3+p_1-q}$& $I_2(\Omega_0, \Omega_5)$ \\ \hline
                     10&$1/{s_{23}s_{123} }$ & $\vc{U}^{j_1}_{p_2,p_3}\vc{U}^{j_2}_{p_1,p_3+p_2}$&   $I_1(\Omega_0)$ \\ \hline
                       11& $1/\bar{n}\mcdot p_0 s_{23}$ & $\vc{U}^{j_1}_{p_2,p_3}\vc{U}^{j_2}_{p_1-q,p_3+p_2}$& $I_2(\Omega_0, \Omega_1)$ \\ \hline
                         12&$1/(\bar{n}\mcdot p_0)^2(z_2+z_3)$  &  $\vc{U}^{j_1}_{p_2-q,p_3}\vc{U}^{j_2}_{p_1,p_3+p_2-q}$& $I_3(\Omega_0,\Omega_2,\Omega_4)$ \\ \hline
                           13&$1/(\bar{n}\mcdot p_0)^2(z_2+z_3)$  & $\vc{U}^{j_1}_{p_2,p_3-q}\vc{U}^{j_2}_{p_1,p_3-q+p_2}$& $I_3(\Omega_0,\Omega_3,\Omega_4)$ \\ \hline
                             14& $1/\bar{n}\mcdot p_0 s_{23}$ &  $\vc{U}^{j_1}_{p_2,p_3}\vc{U}^{j_2}_{p_1,p_3+p_2-q}$& $I_2(\Omega_0, \Omega_4)$ \\ \hline
                             
                               15&$1/s_{12}s_{123}$ &  $\vc{U}^{j_1}_{p_1,p_2}\vc{U}^{j_2}_{p_1+p_2,p_3}$& $I_1(\Omega_0)$ \\ \hline
                                 16&$1/(\bar{n}\mcdot p_0)^2(z_1+z_2)$ &  $\vc{U}^{j_1}_{p_1,p_2-q}\vc{U}^{j_2}_{p_1+p_2-q,p_3}$& $I_3(\Omega_0,\Omega_2,\Omega_6)$  \\ \hline
                                   17&$1/(\bar{n}\mcdot p_0)^2(z_1+z_2)$  &  $\vc{U}^{j_1}_{p_1-q,p_2}\vc{U}^{j_2}_{p_1-q+p_2,p_3}$& $I_3(\Omega_0,\Omega_1,\Omega_6)$ \\ \hline
                                     18& $1/\bar{n}\mcdot p_0 s_{12}$&  $\vc{U}^{j_1}_{p_1,p_2}\vc{U}^{j_2}_{p_1+p_2,p_3-q}$& $I_2(\Omega_0, \Omega_3)$ \\ \hline
                                       19& $1/\bar{n}\mcdot p_0 s_{12}$ &  $\vc{U}^{j_1}_{p_1,p_2}\vc{U}^{j_2}_{p_1+p_2-q,p_3}$& $I_2(\Omega_0, \Omega_6)$ \\ \hline
    \end{tabular}
      \caption{Entries for single Born graphs.}
      \label{tab:singleborn}
\end{table}
\end{center}

\vspace*{5in}

\begin{center}
\begin{table}[H]
    \begin{tabular}{ | l | l | l | p{5cm} |}
    \hline
     $k$ &$C_k$&  $\vc{U}^{j_1}_{p_{k_1},p_{k_2}}\vc{U}^{j_2}_{p_{k_3},p_{k_4}}$ &   $I_{k}^{(2c)}/(-i)$ \\ \hline
   1& $1/s_{123}$ &  $-$ &  $I_1(\bar{\Omega}_0)/2$  \\ \hline
   2&$1/\bar{n}\mcdot p_0$ & $-$&  $I_2(\bar{\Omega}_0, \bar{\Omega}_2)/2$ \\ \hline
   3&$1/\bar{n}\mcdot p_0$  & $-$&   $I_2(\bar{\Omega}_0, \bar{\Omega}_1)/2$\\ \hline
   4&$1/\bar{n}\mcdot p_0$  & $-$&  $I_2(\bar{\Omega}_0, \bar{\Omega}_3)/2$ \\ \hline
   5&$1/\bar{n}\mcdot p_0$ &  $-$ &  $I_2(\bar{\Omega}_0, \tilde{\Omega}_2+\Omega_3)$ \\ \hline
   6&$1/\bar{n}\mcdot p_0$ & $-$&  $ I_2(\bar{\Omega}_0, \tilde{\Omega}_1+\Omega_3)$ \\ \hline
   7&$1/\bar{n}\mcdot p_0$& $-$&  $I_2(\bar{\Omega}_0, \tilde{\Omega}_2+\Omega_1)$ \\ \hline
   8&$1/s_{13}s_{123}$ & $\vc{U}^{j_1}_{p_1,p_3}\vc{U}^{j_2}_{p_2,p_3+p_1}$& $I_1(\bar{\Omega}_0)/2$  \\ \hline
   9& $1/\bar{n}\mcdot p_0 s_{13}$& $\vc{U}^{j_1}_{p_1,p_3}\vc{U}^{j_2}_{p_2,p_3+p_1}$& $I_2(\bar{\Omega}_0,\bar{\Omega}_2)/2$ \\ \hline
   10&$1/(\bar{n}\mcdot p_0)^2(z_1+z_3)$  & $\vc{U}^{j_1}_{p_1,p_3}\vc{U}^{j_2}_{p_2,p_3+p_1}$&    $I_3(\bar{\Omega}_0,\bar{\Omega}_1, \bar{\Omega}_5)/2$  \\ \hline
   11& $1/(\bar{n}\mcdot p_0)^2(z_1+z_3)$ & $\vc{U}^{j_1}_{p_1,p_3}\vc{U}^{j_2}_{p_2,p_3+p_1}$& $I_3(\bar{\Omega}_0,\bar{\Omega}_3, \bar{\Omega}_5)/2$ \\ \hline
   12&$1/\bar{n}\mcdot p_0 s_{13}$  &  $\vc{U}^{j_1}_{p_1,p_3}\vc{U}^{j_2}_{p_2,p_3+p_1}$&  $I_2(\bar{\Omega}_0,\bar{\Omega}_5)/2$ \\ \hline
   13&$1/(\bar{n}\mcdot p_0)^2(z_1+z_3)$  & $\vc{U}^{j_1}_{p_1,p_3-q}\vc{U}^{j_2}_{p_2+q,p_3-q+p_1}$&$I_3(\bar{\Omega}_0,\tilde{\Omega}_2+\Omega_3,\tilde{\Omega}_2+\Omega_5)$ \\ \hline
   14& $1/(\bar{n}\mcdot p_0)^2(z_1+z_3)$ &  $\vc{U}^{j_1}_{p_1-q,p_3+q}\vc{U}^{j_2}_{p_2,p_3+p_1}$& $I_3(\bar{\Omega}_0,\bar{\Omega}_5, {\Omega}_1+\tilde{\Omega}_3)$ \\ \hline
   15&$1/(\bar{n}\mcdot p_0)^2(z_1+z_3)$ &  $\vc{U}^{j_1}_{p_1-q,p_3}\vc{U}^{j_2}_{p_2+q,p_3+p_1-q}$&  $I_3(\bar{\Omega}_0,\tilde{\Omega}_2+\Omega_1, \tilde{\Omega}_2+\Omega_5)$ \\ \hline
   16&$1/\bar{n}\mcdot p_0 s_{13}$ &  $\vc{U}^{j_1}_{p_1,p_3}\vc{U}^{j_2}_{p_2+q,p_3+p_1-q}$& $I_2(\bar{\Omega}_0, {\Omega}_5+\tilde{\Omega}_2)$  \\ \hline
   17&$1/{s_{23}s_{123} }$  &  $\vc{U}^{j_1}_{p_2,p_3}\vc{U}^{j_2}_{p_1,p_3+p_2}$&  $I_1(\bar{\Omega}_0)/2$ \\ \hline
   18&  $1/\bar{n}\mcdot p_0 s_{23}$&  $\vc{U}^{j_1}_{p_2,p_3}\vc{U}^{j_2}_{p_1,p_3+p_2}$& $I_2(\bar{\Omega}_0, \bar{\Omega}_1)/2$ \\ \hline
   19& $1/(\bar{n}\mcdot p_0)^2(z_2+z_3)$ &  $\vc{U}^{j_1}_{p_2,p_3}\vc{U}^{j_2}_{p_1,p_3+p_2}$& $I_3(\bar{\Omega}_0, \bar{\Omega}_2, \bar{\Omega}_4)/2$ \\ \hline
   20& $1/(\bar{n}\mcdot p_0)^2(z_2+z_3)$ &  $\vc{U}^{j_1}_{p_2,p_3}\vc{U}^{j_2}_{p_1,p_3+p_2}$&     $I_3(\bar{\Omega}_0, \bar{\Omega}_3, \bar{\Omega}_4)/2$  \\ \hline
   21&$1/\bar{n}\mcdot p_0 s_{23}$ &  $\vc{U}^{j_1}_{p_2,p_3}\vc{U}^{j_2}_{p_1,p_3+p_2}$& $I_2(\bar{\Omega}_0, \bar{\Omega}_4)/2$ \\ \hline
   22& $1/(\bar{n}\mcdot p_0)^2(z_2+z_3)$ &  $\vc{U}^{j_1}_{p_2,p_3-q}\vc{U}^{j_2}_{p_1+q,p_3-q+p_2}$& $I_3(\bar{\Omega}_0, \tilde{\Omega}_1+\Omega_4, \tilde{\Omega}_1+\Omega_3)$ \\ \hline
   23& $1/(\bar{n}\mcdot p_0)^2(z_2+z_3)$ &  $\vc{U}^{j_1}_{p_2-q,p_3+q}\vc{U}^{j_2}_{p_1,p_3+p_2}$& $I_3(\bar{\Omega}_0, \bar{\Omega}_4, {\Omega}_2+\tilde{\Omega}_3)$  \\ \hline
   24& $1/(\bar{n}\mcdot p_0)^2(z_2+z_3)$ &  $\vc{U}^{j_1}_{p_2-q,p_3}\vc{U}^{j_2}_{p_1+q,p_3+p_2-q}$& $I_3(\bar{\Omega}_0,{\tilde{\Omega}}_1+{\Omega}_2, \tilde{\Omega}_1+{\Omega}_4)$  \\ \hline
   25& $1/\bar{n}\mcdot p_0 s_{23}$ &  $\vc{U}^{j_1}_{p_2,p_3}\vc{U}^{j_2}_{p_1+q,p_3+p_2-q}$&  $I_2(\bar{\Omega}_0,{\Omega}_4+\tilde{\Omega}_1)$ \\ \hline
   26& $1/s_{12}s_{123}$ &  $\vc{U}^{j_1}_{p_1,p_2}\vc{U}^{j_2}_{p_1+p_2,p_3}$& $I_1(\bar{\Omega}_0)/2$ \\ \hline
   27& $1/(\bar{n}\mcdot p_0)^2(z_1+z_2)$ &  $\vc{U}^{j_1}_{p_1,p_2}\vc{U}^{j_2}_{p_1+p_2,p_3}$&  $I_3(\bar{\Omega}_0, \bar{\Omega}_2, \bar{\Omega}_6)/2$ \\ \hline
   28& $1/(\bar{n}\mcdot p_0)^2(z_1+z_2)$ &  $\vc{U}^{j_1}_{p_1,p_2}\vc{U}^{j_2}_{p_1+p_2,p_3}$&    $I_3(\bar{\Omega}_0, \bar{\Omega}_1, \bar{\Omega}_6)/2$ \\ \hline
   29& $1/\bar{n}\mcdot p_0 s_{12}$ &  $\vc{U}^{j_1}_{p_1,p_2}\vc{U}^{j_2}_{p_1+p_2,p_3}$&$I_2(\bar{\Omega}_0, \bar{\Omega}_3)/2$ \\ \hline
   30& $1/\bar{n}\mcdot p_0 s_{12}$ &  $\vc{U}^{j_1}_{p_1,p_2}\vc{U}^{j_2}_{p_1+p_2,p_3}$& $I_2(\bar{\Omega}_0, \bar{\Omega}_6)/2$ \\ \hline
   31& $1/(\bar{n}\mcdot p_0)^2(z_1+z_2)$ & $\vc{U}^{j_1}_{p_1,p_2-q}\vc{U}^{j_2}_{p_1+p_2-q,p_3+q}$&   $I_3(\bar{\Omega}_0, \tilde{\Omega}_3+\Omega_2, \tilde{\Omega}_3+\Omega_6)$ \\ \hline
   32& $1/(\bar{n}\mcdot p_0)^2(z_1+z_2)$ &  $\vc{U}^{j_1}_{p_1-q,p_2}\vc{U}^{j_2}_{p_1-q+p_2,p_3+q}$&  $I_3(\bar{\Omega}_0, \tilde{\Omega}_3+\Omega_1, \tilde{\Omega}_3+\Omega_6)$ \\ \hline
   33& $1/(\bar{n}\mcdot p_0)^2(z_1+z_2)$ &  $\vc{U}^{j_1}_{p_1-q,p_2+q}\vc{U}^{j_2}_{p_1+p_2,p_3}$& $I_3(\bar{\Omega}_0, \bar{\Omega}_6, \Omega_1+\tilde{\Omega}_2)$ \\ \hline
   34& $1/\bar{n}\mcdot p_0 s_{12}$ &  $\vc{U}^{j_1}_{p_1,p_2}\vc{U}^{j_2}_{p_1+p_2-q,p_3+q}$& $I_2(\bar{\Omega}_0, \tilde{\Omega}_3+{\Omega}_6)$ \\ \hline
    \end{tabular}
      \caption{Entries for double Born graphs.}
      \label{tab:doubleborn}
\end{table}
\end{center}

\section{Longitudinal integrals}\label{appendix:longintegrals}
In this appendix we derive all necessary formulas to calculate the single and double Born 
longitudinal integrals that appear in our paper. All single Born diagrams are of the form
\begin{eqnarray}
I^{(1)}_{n_q}(\alpha_i,\delta z)=\int\frac{{\rm d} q^-}{2\pi}\,\e^{iq^- \delta z}\prod_{i=1}^{n_q}\frac{1}{\alpha_i-q^-},
\end{eqnarray}
where $n_q$ is the total number of $q$-dependent propagators in the graph and 
$\alpha_i=\Omega(Q_i, \vc{q}_{\perp})+i\eps/\bar{n}\mcdot Q_i$. Integration is straightforward using 
Cauchy's theorem
\begin{eqnarray}
&&I^{(1)}_{n_q}(\alpha_i,\delta z>0)=(-i)\sum_{i=1}^{n_q}\,\theta\left(\text{Im}\,\alpha_i\right)\,
\e^{i\alpha_i\delta z}\prod_{l=1,l\ne i}^{n_q}\frac{1}{\alpha_l-\alpha_i},\label{I1eqpositive}\\
&&I^{(1)}_{n_q}(\alpha_i,\delta z=0)=\lim_{\delta z\rightarrow +0}I^{(1)}_{n_q}(\alpha_i,\delta z), 
\text{          if  } n_q>1,\label{I1eqzero}\\
&&I^{(1)}_{n_q}(\alpha_i,\delta z=0)=\frac{1}{2}\lim_{\delta z\rightarrow +0}I^{(1)}_{n_q}(\alpha_i,\delta z), 
\text{  if  } n_q=1.\label{I1eqzero2}
\end{eqnarray}
The first two equations simply follow from Cauchy's theorem when closing the contour above. 
The third case is more subtle since the boundary term at infinity cannot be neglected which causes the factor $1/2$. We explain this in more detail at the end of this section. The single Born diagrams in this paper take values of $n_q=1,2,3$ only. The three corresponding master formulas for these longitudinal integrals are
\begin{eqnarray}
I^{(1)}_{1}(\alpha_1, \delta z)&=&-i\,\e^{i\alpha_1\delta z},\label{I1eq}\\
I^{(1)}_{2}(\alpha_1,\alpha_2, \delta z)&=&i\,\frac{\e^{i\alpha_2\delta z}-\e^{i\alpha_1\delta z}}{\alpha_2-\alpha_1},\label{I2eq}\\
I^{(1)}_{3}(\alpha_1,\alpha_2, \alpha_3,\delta z)&=&i\,
\left(\frac{\e^{i\alpha_2\delta z}-\e^{i\alpha_1\delta z}}{\alpha_2-\alpha_1}
-\frac{\e^{i\alpha_3\delta z}-\e^{i\alpha_1\delta z}}{\alpha_3-\alpha_1}\right)\frac{1}{\alpha_3-\alpha_2}.\label{I3eq}
\end{eqnarray}

Any double Born integral can be written as
\begin{eqnarray}
&&I^{(2)}_{n_{q_1},n_{q_2},n_{q_3}}(\alpha_i; \beta_j; \gamma_k, \delta z_1,\delta z_2)
=\int\frac{{\rm d} q_1^-}{2\pi}\frac{{\rm d} q^-_2}{2\pi}\,\e^{iq_1^- \delta z_1+iq^-_2 \delta z_2}
\left(\prod_{i=1}^{n_{q_1}}\frac{1}{\alpha_i-q^-_1}\right)\times \left(\prod_{j=1}^{n_{q_2}}\frac{1}{\beta_j-q^-_2}\right)\nonumber\\
&&\qquad\qquad\qquad\qquad\qquad\qquad\qquad\qquad\qquad\times\left(\prod_{k=1}^{n_{q_{12}}}\frac{1}{\gamma_k-q^-_1-q_2^-}\right),
\end{eqnarray}
where $\alpha_i, \beta_j, \gamma_k$ are frequencies that appear in the poles 
of $\Delta_g(Q_i,q)$ with Glauber gluon momenta $q_1, q_2$, $q_1+q_2$, respectively. 
Performing the $q_2^-$ integration using Cauchy's theorem, the remaining $q_1^-$ 
integration can be expressed by a single Born integral
\begin{eqnarray}
&&I^{(2)}_{n_{q_1},n_{q_2},n_{q_3}}(\alpha_i; \beta_j; \gamma_k, \delta z_1,\delta z_2)=\label{I2masterformula}\\
&&\qquad\quad(-i)\sum_{j=1}^{n_{q_2}}\,
\theta\left(\text{Im}\,\beta_j\right)\,\e^{i\beta_j\delta z_2}
\prod_{l=1,l\ne j}^{n_{q_2}}\frac{1}{\beta_l-\beta_j}\, I^{(1)}_{n_{q_1}+n_{q_{12}}}(\alpha_i; \gamma_k-\beta_j,\delta z_1)\nonumber\\
&&\qquad\quad+(-i)\sum_{k=1}^{n_{q_{12}}}\,\theta\left(\text{Im}\,\gamma_k\right)\,
\e^{i\gamma_k\delta z_2}(-1)^{n_{q_2}}\prod_{m=1,m\ne k}^{n_{q_{12}}}\frac{1}{\gamma_m-\gamma_k}\, 
I^{(1)}_{n_{q_1}+n_{q_{2}}}(\alpha_i; \gamma_k-\beta_j,\delta z_1-\delta z_2).\nonumber
\end{eqnarray}
This equation solves any longitudinal integral of the double Born graphs in this paper. 
For the first order in opacity calculation we need the contact limit of this integral, i.e. 
$I^{(2)}_{n_{q_1},n_{q_2},n_{q_3}}(\alpha_i; \beta_j; \gamma_k, \delta z,\delta z)$. 
The result can be obtained by applying \eq{I1eqpositive}-\eq{I1eqzero2} to \eq{I2masterformula}.

The longitudinal integral of any double Born diagram in this paper can be relateed to one of seven master integrals by $n_k=(n_{q_1},n_{q_2},n_{q_{12}})$. This master integrals can be calculated to
\begin{eqnarray}
I^{(2c)}_k=(-i)\mcdot\left\{ \begin{array}{c}
I_1(\gamma_1)/2, \text{\,\,\,if\,\,\, } n_k=(1,0,1)  \\
I_2(\gamma_1,\gamma_2)/2, \text{\,\,\,if\,\,\, } n_k=(1,0,2)  \\
I_3(\gamma_1,\gamma_2,\gamma_3)/2, \text{\,\,\,if\,\,\, } n_k=(1,0,3)  \\
I_2(\alpha_1+\beta_1,\gamma_1), \text{\,\,\,if\,\,\, } n_k=(1,1,1)  \\
I_3(\alpha_1+\beta_1, \alpha_2+\beta_1,\gamma_1), \text{\,\,\,if\,\,\, } n_k=(2,1,1)  \\
I_3(\alpha_1+\beta_1,\alpha_1+\beta_2,\gamma_1), \text{\,\,\,if\,\,\, } n_k=(1,2,1)  \\
I_3(\alpha_1+\beta_1,\gamma_1, \gamma_2), \text{\,\,\,if\,\,\, } n_k=(1,1,2)  \\ \end{array} \right\},\label{doublebornI2res}
\end{eqnarray}
where the functions $I_1, I_2, I_3$ are the same as in \eq{I1eq}-\eq{I3eq}. 
There is one technical issue when deriving \eq{doublebornI2res}. By definition $\alpha_i, \beta_j, \gamma_k$ all have positive imaginary parts, due to the $i\epsilon$-prescription. However, one has to be careful when considering the complex number $\gamma_k-\beta_j$ in \eq{I2masterformula}. For all calculation we are concerned with this number has a negative imaginary part. Indeed, we have checked that the results of the double Born integrals do not depend on the sign of the imaginary part of $\gamma_k-\beta_j$. The crucial point is that none of these numbers lie on the real axis.
 
Now let us return to the subtlety in $I^{(1)}_{1}(\alpha_1,\delta z=0)$. 
For $n_q=1$ and $\delta z=0$ the boundary term at infinity does not vanish and the integral 
$I^{(1)}_{1}(\alpha_1,\delta z=0)$ becomes ill-defined. To solve this issue one has to recall that 
the factor $\tilde{v}(q)$ also depends on $q^-$, $\tilde{v}(q^-,\vc{q}_{\perp})\sim \frac{1}{\mu^{2}+\vc{q}_{\perp}^2+(q^-)^2}$, but we dropped this dependence since it is power-suppressed. Re-introducing 
this extra $q^-$ dependence cures the boundary term probem and yields 
$I^{(1)}_{n_q}(\alpha_i,\delta z=0)=-i/2$ and the overall factor $\tilde{v}(0,\vc{q}_{\perp})$. One can convince oneself that in the other 
integrals, like $I^{(1)}_{n_q>1}(\alpha_i,\delta z=0)$ or $I^{(1)}_{n_q}(\alpha_i,\delta z>0)$, re-introducing 
the factor $\tilde{v}(q^-,\vc{q}_{\perp})$ does not change anything to leading power in the EFT.

\section{Color operators}\label{seq:color}
In this appendix we present the color operators of all single and double Born graphs 
in the paper. For the single Born graphs the color operators in terms of the basis vectors given 
in \eq{eq:SBbasis} are
\small
\begin{eqnarray}
\left(
\begin{array}{ccccccccccccccccccccccccccccccccccccccccc}
e_{1a} \\
e_{1b}\\
e_{2a}\\
e_{2b}\\
e_{3a} \\
e_{3b}\\
e_{4a} \\
e_{4b}\\
e_{5}\\
e_{6}\\
e_{7}\\
e_{8}\\
e_{9}\\
e_{10}\\
e_{11}\\
e_{12}\\
e_{13}\\
e_{14}\\
e_{15}\\
e_{16}\\
e_{17}\\
e_{18}\\
e_{19}
\end{array}
\right)=
\left(
\begin{array}{cccccc}
 1 & 0 & 0 & 0 & 0 & 0 \\
 0 & 0 & 0 & 1 & 0 & 0 \\
 1 & -1 & 0 & 0 & 0 & 0 \\
 0 & 0 & 0 & 0 & 1 & -1 \\
 0 & 1 & -1 & 0 & 0 & 0 \\
 0 & 0 & 0 & 1 & -1 & 0 \\
 0 & 0 & 1 & 0 & 0 & 0 \\
 0 & 0 & 0 & 0 & 0 & 1 \\
 1 & 0 & 0 & 0 & 0 & 0 \\
 1 & -1 & 0 & 0 & 0 & 0 \\
 0 & 1 & -1 & 0 & 0 & 0 \\
 0 & 0 & 1 & 0 & 0 & 0 \\
 0 & 1 & 0 & 0 & 0 & 0 \\
 0 & 0 & 0 & 1 & 0 & 0 \\
 0 & 0 & 0 & 1 & -1 & 0 \\
 0 & 0 & 0 & 0 & 1 & -1 \\
 0 & 0 & 0 & 0 & 0 & 1 \\
 0 & 0 & 0 & 0 & 1 & 0 \\
 1 & 0 & 0 & -1 & 0 & 0 \\
 1 & -1 & 0 & 0 & -1 & 1 \\
 0 & 1 & -1 & -1 & 1 & 0 \\
 0 & 0 & 1 & 0 & 0 & -1 \\
 1 & 0 & -1 & -1 & 0 & 1
\end{array}
\right),\label{eq:singleborncoloroperators}
\end{eqnarray}
\normalsize
For the double Born graphs the color operators in terms of the basis given in \eq{eq:DBbasis} are
\footnotesize
\begin{eqnarray}
\left(
\begin{array}{ccccccccccccccccccccccccccccccccccccccccc}
e_{1a} \\
e_{1b}\\
e_{2a}\\
e_{2b}\\
e_{3a} \\
e_{3b}\\
e_{4a} \\
e_{4b}\\
e_{5a} \\
e_{5b}\\
e_{6a} \\
e_{6b}\\
e_{7a} \\
e_{7b}\\
e_{8}\\
e_{9}\\
e_{10}\\
e_{11}\\
e_{12}\\
e_{13}\\
e_{14}\\
e_{15}\\
e_{16}\\
e_{17}\\
e_{18}\\
e_{19}\\
e_{20}\\
e_{21}\\
e_{22}\\
e_{23}\\
e_{24}\\
e_{25}\\
e_{26}\\
e_{27}\\
e_{28}\\
e_{29}\\
e_{30}\\
e_{31}\\
e_{32}\\
e_{33}\\
e_{34}
\end{array}
\right)=
\left(
\begin{array}{cccccccccccccccccccccccc}
 1 & 0 & 0 & 0 & 0 & 0 & 0 & 0 & 0 & 0 & 0 & 0 & 0 & 0 & 0 & 0 & 0 & 0 & 0 & 0 & 0 & 0 & 0 & 0 \\
 0 & 0 & 0 & 1 & 0 & 0 & 0 & 0 & 0 & 0 & 0 & 0 & 0 & 0 & 0 & 0 & 0 & 0 & 0 & 0 & 0 & 0 & 0 & 0 \\
 0 & -1 & 0 & 0 & 0 & 0 & 1 & 1 & 0 & 0 & 0 & 0 & -1 & 0 & 0 & 0 & 0 & 0 & 0 & 0 & 0 & 0 & 0 & 0 \\
 0 & 0 & 0 & 0 & 0 & 0 & 0 & 0 & 0 & 0 & 0 & -1 & 0 & 0 & 0 & 0 & 1 & 1 & 0 & 0 & 0 & 0 & -1 & 0 \\
 0 & 0 & 0 & 0 & 0 & 0 & 0 & 1 & -1 & 0 & 0 & 0 & 0 & 0 & 0 & 0 & 0 & 0 & 0 & -1 & 1 & 0 & 0 & 0 \\
 0 & 0 & 0 & 1 & -1 & 0 & 0 & 0 & 0 & 0 & 0 & 0 & 0 & 0 & 0 & -1 & 1 & 0 & 0 & 0 & 0 & 0 & 0 & 0 \\
 0 & 0 & 0 & 0 & 0 & 0 & 0 & 0 & 0 & 0 & 0 & 0 & 0 & 0 & 1 & 0 & 0 & 0 & 0 & 0 & 0 & 0 & 0 & 0 \\
 0 & 0 & 0 & 0 & 0 & 0 & 0 & 0 & 0 & 0 & 0 & 0 & 0 & 0 & 0 & 0 & 0 & 1 & 0 & 0 & 0 & 0 & 0 & 0 \\
 0 & 0 & 1 & 0 & 0 & 0 & 0 & 0 & -1 & 0 & 0 & 0 & 0 & 0 & 0 & 0 & 0 & 0 & 0 & 0 & 0 & 0 & 0 & 0 \\
 0 & 0 & 0 & 0 & 0 & 0 & 0 & 0 & 0 & 0 & 0 & 1 & 0 & 0 & 0 & 0 & 0 & -1 & 0 & 0 & 0 & 0 & 0 & 0 \\
 0 & 0 & 0 & 0 & 0 & 0 & 0 & 0 & 1 & 0 & 0 & 0 & 0 & 0 & -1 & 0 & 0 & 0 & 0 & 0 & 0 & 0 & 0 & 0 \\
 0 & 0 & 0 & 0 & 0 & 1 & 0 & 0 & 0 & 0 & 0 & -1 & 0 & 0 & 0 & 0 & 0 & 0 & 0 & 0 & 0 & 0 & 0 & 0 \\
 0 & 1 & -1 & 0 & 0 & 0 & 0 & -1 & 1 & 0 & 0 & 0 & 0 & 0 & 0 & 0 & 0 & 0 & 0 & 0 & 0 & 0 & 0 & 0 \\
 0 & 0 & 0 & 0 & 0 & 0 & 0 & 0 & 0 & 0 & 0 & 0 & 0 & 0 & 0 & 1 & -1 & 0 & 0 & 0 & 0 & -1 & 1 & 0 \\
 1 & 0 & 0 & 0 & 0 & 0 & 0 & 0 & 0 & 0 & 0 & 0 & 0 & 0 & 0 & 0 & 0 & 0 & 0 & 0 & 0 & 0 & 0 & 0 \\
 1 & -1 & 0 & 0 & 0 & 0 & 0 & 0 & 0 & 0 & 0 & 0 & -1 & 1 & 0 & 0 & 0 & 0 & 0 & 0 & 0 & 0 & 0 & 0 \\
 0 & 0 & 0 & 0 & 0 & 0 & 0 & 1 & -1 & 0 & 0 & 0 & 0 & 0 & 0 & 0 & 0 & 0 & 0 & -1 & 1 & 0 & 0 & 0 \\
 0 & 0 & 0 & 0 & 0 & 0 & 0 & 0 & 0 & 0 & 0 & 0 & 0 & 0 & 1 & 0 & 0 & 0 & 0 & 0 & 0 & 0 & 0 & 0 \\
 0 & 0 & 0 & 0 & 0 & 0 & 0 & 1 & 0 & 0 & 0 & 0 & 0 & 0 & 0 & 0 & 0 & 0 & 0 & 0 & 0 & 0 & 0 & 0 \\
 0 & 0 & 1 & 0 & 0 & 0 & 0 & 0 & -1 & 0 & 0 & 0 & 0 & 0 & 0 & 0 & 0 & 0 & 0 & 0 & 0 & 0 & 0 & 0 \\
 0 & 0 & 0 & 0 & 0 & 0 & 0 & 0 & 1 & 0 & 0 & 0 & 0 & 0 & -1 & 0 & 0 & 0 & 0 & 0 & 0 & 0 & 0 & 0 \\
 0 & 1 & -1 & 0 & 0 & 0 & 0 & -1 & 1 & 0 & 0 & 0 & 0 & 0 & 0 & 0 & 0 & 0 & 0 & 0 & 0 & 0 & 0 & 0 \\
 0 & 1 & 0 & 0 & 0 & 0 & 0 & -1 & 0 & 0 & 0 & 0 & 0 & 0 & 0 & 0 & 0 & 0 & 0 & 0 & 0 & 0 & 0 & 0 \\
 0 & 0 & 0 & 1 & 0 & 0 & 0 & 0 & 0 & 0 & 0 & 0 & 0 & 0 & 0 & 0 & 0 & 0 & 0 & 0 & 0 & 0 & 0 & 0 \\
 0 & 0 & 0 & 1 & -1 & 0 & 0 & 0 & 0 & 0 & 0 & 0 & 0 & 0 & 0 & -1 & 1 & 0 & 0 & 0 & 0 & 0 & 0 & 0 \\
 0 & 0 & 0 & 0 & 0 & 0 & 0 & 0 & 0 & 0 & 1 & -1 & 0 & 0 & 0 & 0 & 0 & 0 & 0 & 0 & 0 & 0 & -1 & 1 \\
 0 & 0 & 0 & 0 & 0 & 0 & 0 & 0 & 0 & 0 & 0 & 0 & 0 & 0 & 0 & 0 & 0 & 1 & 0 & 0 & 0 & 0 & 0 & 0 \\
 0 & 0 & 0 & 0 & 0 & 0 & 0 & 0 & 0 & 0 & 1 & 0 & 0 & 0 & 0 & 0 & 0 & 0 & 0 & 0 & 0 & 0 & 0 & 0 \\
 0 & 0 & 0 & 0 & 0 & 1 & 0 & 0 & 0 & 0 & 0 & -1 & 0 & 0 & 0 & 0 & 0 & 0 & 0 & 0 & 0 & 0 & 0 & 0 \\
 0 & 0 & 0 & 0 & 0 & 0 & 0 & 0 & 0 & 0 & 0 & 1 & 0 & 0 & 0 & 0 & 0 & -1 & 0 & 0 & 0 & 0 & 0 & 0 \\
 0 & 0 & 0 & 0 & 0 & 0 & 0 & 0 & 0 & 0 & 0 & 0 & 0 & 0 & 0 & 1 & -1 & 0 & 0 & 0 & 0 & -1 & 1 & 0 \\
 0 & 0 & 0 & 0 & 1 & 0 & 0 & 0 & 0 & 0 & -1 & 0 & 0 & 0 & 0 & 0 & 0 & 0 & 0 & 0 & 0 & 0 & 0 & 0 \\
 1 & 0 & 0 & -1 & 0 & 0 & 0 & 0 & 0 & 0 & 0 & 0 & 0 & 0 & 0 & 0 & 0 & 0 & 0 & 0 & 0 & 0 & 0 & 0 \\
 1 & -1 & 0 & 0 & 0 & 0 & 0 & 0 & 0 & 0 & -1 & 1 & -1 & 1 & 0 & 0 & 0 & 0 & 0 & 0 & 0 & 0 & 1 & -1 \\
 0 & 0 & 0 & -1 & 1 & 0 & 0 & 1 & -1 & 0 & 0 & 0 & 0 & 0 & 0 & 1 & -1 & 0 & 0 & -1 & 1 & 0 & 0 & 0 \\
 0 & 0 & 0 & 0 & 0 & 0 & 0 & 0 & 0 & 0 & 0 & 0 & 0 & 0 & 1 & 0 & 0 & -1 & 0 & 0 & 0 & 0 & 0 & 0 \\
 1 & 0 & -1 & -1 & 0 & 1 & 0 & 0 & 0 & 0 & 0 & 0 & 0 & 0 & 0 & 0 & 0 & 0 & -1 & 0 & 1 & 1 & 0 & -1 \\
 0 & 0 & 1 & 0 & 0 & 0 & 0 & 0 & -1 & 0 & 0 & -1 & 0 & 0 & 0 & 0 & 0 & 1 & 0 & 0 & 0 & 0 & 0 & 0 \\
 0 & 0 & 0 & 0 & 0 & -1 & 0 & 0 & 1 & 0 & 0 & 1 & 0 & 0 & -1 & 0 & 0 & 0 & 0 & 0 & 0 & 0 & 0 & 0 \\
 0 & 1 & -1 & 0 & 0 & 0 & 0 & -1 & 1 & 0 & 0 & 0 & 0 & 0 & 0 & -1 & 1 & 0 & 0 & 0 & 0 & 1 & -1 & 0 \\
 0 & 0 & 1 & 0 & 0 & -1 & 0 & 0 & 0 & 0 & 0 & 0 & 0 & 0 & -1 & 0 & 0 & 1 & 0 & 0 & 0 & 0 & 0 & 0
\end{array}
\right)\nonumber\\
\end{eqnarray}
\normalsize

\vspace{9mm}
{\Large\bf Acknowledgements\/}\\\\
We  thank Christian Bauer, Vincenzo Cirigliano, Andrew Hornig, Andrew Larkoski, Christopher Lee, George Sterman, Wouter Waalewijn, and Jon Walsh for useful discussions.  This research is supported by DOE Office of Science, the LDRD program at LANL and in part by the JET Collaboration. The research of M.F.\ is supported in parts by the DOE Office of Science, grants DE-FG02-06ER41449 and DE-FG02-04ER41338, the US National Science Foundation, grant NSF-PHY-0969510 {\em the LHC Theory Initiative}, the Cluster of Excellence {\em Precision Physics, Fundamental Interactions and Structure of Matter\/} (PRISMA -- EXC 1098) and DFG grant NE~398/3-1. M.F.\ thanks the Los Alamos National Laboratory for hospitality.

\bibliographystyle{JHEP}
   \let\oldnewblock=\newblock
    \newcommand\dispatcholdnewblock[1]{\oldnewblock{#1}}
    \renewcommand\newblock{\spaceskip=0.3emplus0.3emminus0.2em\relax
                           \xspaceskip=0.3emplus0.6emminus0.1em\relax
                           \hskip0ptplus0.5emminus0.2em\relax
                           {\catcode`\.=\active
                           \expandafter}\dispatcholdnewblock}

\bibliography{AngOrdSUB}

\providecommand{\href}[2]{#2}\begingroup\raggedright\begin{thebibliography}{10}

\bibitem{Sjostrand:2006za}
T.~Sjostrand, S.~Mrenna, and P.~Z. Skands, {\it {PYTHIA 6.4 Physics and
  Manual}},  {\em JHEP} {\bf 0605} (2006) 026,
  [\href{http://arxiv.org/abs/hep-ph/0603175}{{\tt hep-ph/0603175}}].

\bibitem{Bahr:2008pv}
M.~Bahr, S.~Gieseke, M.~Gigg, D.~Grellscheid, K.~Hamilton, {\em et~al.}, {\it
  {Herwig++ Physics and Manual}},  {\em Eur.Phys.J.} {\bf C58} (2008) 639--707,
  [\href{http://arxiv.org/abs/0803.0883}{{\tt arXiv:0803.0883}}].

\bibitem{Gribov:1972rt}
V.~Gribov and L.~Lipatov, {\it {e+ e- pair annihilation and deep inelastic e p
  scattering in perturbation theory}},  {\em Sov.J.Nucl.Phys.} {\bf 15} (1972)
  675--684.

\bibitem{Dokshitzer:1977sg}
Y.~L. Dokshitzer, {\it {Calculation of the Structure Functions for Deep
  Inelastic Scattering and e+ e- Annihilation by Perturbation Theory in Quantum
  Chromodynamics.}},  {\em Sov.Phys.JETP} {\bf 46} (1977) 641--653.

\bibitem{Altarelli:1977zs}
G.~Altarelli and G.~Parisi, {\it {Asymptotic Freedom in Parton Language}},
  {\em Nucl.Phys.} {\bf B126} (1977) 298.

\bibitem{Marchesini:1983bm}
G.~Marchesini and B.~Webber, {\it {Simulation of QCD Jets Including Soft Gluon
  Interference}},  {\em Nucl.Phys.} {\bf B238} (1984) 1.

\bibitem{Marchesini:1987cf}
G.~Marchesini and B.~Webber, {\it {Monte Carlo Simulation of General Hard
  Processes with Coherent QCD Radiation}},  {\em Nucl.Phys.} {\bf B310} (1988)
  461.

\bibitem{Catani:1998nv}
S.~Catani and M.~Grazzini, {\it {Collinear factorization and splitting
  functions for next-to-next-to-leading order QCD calculations}},  {\em
  Phys.Lett.} {\bf B446} (1999) 143--152,
  [\href{http://arxiv.org/abs/hep-ph/9810389}{{\tt hep-ph/9810389}}].

\bibitem{Catani:1999ss}
S.~Catani and M.~Grazzini, {\it {Infrared factorization of tree level QCD
  amplitudes at the next-to-next-to-leading order and beyond}},  {\em
  Nucl.Phys.} {\bf B570} (2000) 287--325,
  [\href{http://arxiv.org/abs/hep-ph/9908523}{{\tt hep-ph/9908523}}].

\bibitem{Bauer:2000ew}
C.~W. Bauer, S.~Fleming, and M.~E. Luke, {\it {Summing Sudakov logarithms in B
  $\rightarrow X(s \gamma)$ in effective field theory}},  {\em Phys.Rev.} {\bf
  D63} (2000) 014006, [\href{http://arxiv.org/abs/hep-ph/0005275}{{\tt
  hep-ph/0005275}}].

\bibitem{Bauer:2000yr}
C.~W. Bauer, S.~Fleming, D.~Pirjol, and I.~W. Stewart, {\it {An Effective field
  theory for collinear and soft gluons: Heavy to light decays}},  {\em
  Phys.Rev.} {\bf D63} (2001) 114020,
  [\href{http://arxiv.org/abs/hep-ph/0011336}{{\tt hep-ph/0011336}}].

\bibitem{Bauer:2001ct}
C.~W. Bauer and I.~W. Stewart, {\it {Invariant operators in collinear effective
  theory}},  {\em Phys.Lett.} {\bf B516} (2001) 134--142,
  [\href{http://arxiv.org/abs/hep-ph/0107001}{{\tt hep-ph/0107001}}].

\bibitem{Bauer:2001yt}
C.~W. Bauer, D.~Pirjol, and I.~W. Stewart, {\it {Soft collinear factorization
  in effective field theory}},  {\em Phys.Rev.} {\bf D65} (2002) 054022,
  [\href{http://arxiv.org/abs/hep-ph/0109045}{{\tt hep-ph/0109045}}].

\bibitem{Vitev:2005yg}
I.~Vitev, {\it {Large angle hadron correlations from medium-induced gluon
  radiation}},  {\em Phys.Lett.} {\bf B630} (2005) 78--84,
  [\href{http://arxiv.org/abs/hep-ph/0501255}{{\tt hep-ph/0501255}}].

\bibitem{MehtarTani:2010ma}
Y.~Mehtar-Tani, C.~A. Salgado, and K.~Tywoniuk, {\it {Anti-angular ordering of
  gluon radiation in QCD media}},  {\em Phys.Rev.Lett.} {\bf 106} (2011)
  122002, [\href{http://arxiv.org/abs/1009.2965}{{\tt arXiv:1009.2965}}].

\bibitem{MehtarTani:2011jw}
Y.~Mehtar-Tani and K.~Tywoniuk, {\it {Jet coherence in QCD media: the antenna
  radiation spectrum}},  {\em JHEP} {\bf 1301} (2013) 031,
  [\href{http://arxiv.org/abs/1105.1346}{{\tt arXiv:1105.1346}}].

\bibitem{Idilbi:2008vm}
A.~Idilbi and A.~Majumder, {\it {Extending Soft-Collinear-Effective-Theory to
  describe hard jets in dense QCD media}},  {\em Phys.Rev.} {\bf D80} (2009)
  054022, [\href{http://arxiv.org/abs/0808.1087}{{\tt arXiv:0808.1087}}].

\bibitem{D'Eramo:2010ak}
F.~D'Eramo, H.~Liu, and K.~Rajagopal, {\it {Transverse Momentum Broadening and
  the Jet Quenching Parameter, Redux}},  {\em Phys.Rev.} {\bf D84} (2011)
  065015, [\href{http://arxiv.org/abs/1006.1367}{{\tt arXiv:1006.1367}}].

\bibitem{Bauer:2010cc}
C.~W. Bauer, B.~O. Lange, and G.~Ovanesyan, {\it {On Glauber modes in
  Soft-Collinear Effective Theory}},  {\em JHEP} {\bf 1107} (2011) 077,
  [\href{http://arxiv.org/abs/1010.1027}{{\tt arXiv:1010.1027}}].

\bibitem{Ovanesyan:2011xy}
G.~Ovanesyan and I.~Vitev, {\it {An effective theory for jet propagation in
  dense QCD matter: jet broadening and medium-induced bremsstrahlung}},  {\em
  JHEP} {\bf 1106} (2011) 080, [\href{http://arxiv.org/abs/1103.1074}{{\tt
  arXiv:1103.1074}}]. * Temporary entry *.

\bibitem{Ovanesyan:2011kn}
G.~Ovanesyan and I.~Vitev, {\it {Medium-induced parton splitting kernels from
  Soft Collinear Effective Theory with Glauber gluons}},  {\em Phys.Lett.} {\bf
  B706} (2012) 371--378, [\href{http://arxiv.org/abs/1109.5619}{{\tt
  arXiv:1109.5619}}].

\bibitem{Benzke:2012sz}
M.~Benzke, N.~Brambilla, M.~A. Escobedo, and A.~Vairo, {\it {Gauge invariant
  definition of the jet quenching parameter}},
  \href{http://arxiv.org/abs/1208.4253}{{\tt arXiv:1208.4253}}.

\bibitem{Ovanesyan:2012fr}
G.~Ovanesyan, {\it {Medium-induced splitting kernels from $SCET_G$}},
  \href{http://arxiv.org/abs/1210.4945}{{\tt arXiv:1210.4945}}.

\bibitem{Vitev:2009rd}
I.~Vitev and B.-W. Zhang, {\it {Jet tomography of high-energy nucleus-nucleus
  collisions at next-to-leading order}},  {\em Phys.Rev.Lett.} {\bf 104} (2010)
  132001, [\href{http://arxiv.org/abs/0910.1090}{{\tt arXiv:0910.1090}}].

\bibitem{Neufeld:2010fj}
R.~Neufeld, I.~Vitev, and B.-W. Zhang, {\it {The Physics of
  $Z^0/\gamma^*$-tagged jets at the LHC}},  {\em Phys.Rev.} {\bf C83} (2011)
  034902, [\href{http://arxiv.org/abs/1006.2389}{{\tt arXiv:1006.2389}}].

\bibitem{He:2011pd}
Y.~He, I.~Vitev, and B.-W. Zhang, {\it {${\cal O}(\alpha_s^3)$ Analysis of
  Inclusive Jet and di-Jet Production in Heavy Ion Reactions at the Large
  Hadron Collider}},  {\em Phys.Lett.} {\bf B713} (2012) 224--232,
  [\href{http://arxiv.org/abs/1105.2566}{{\tt arXiv:1105.2566}}].

\bibitem{He:2011sg}
Y.~He, B.-W. Zhang, and E.~Wang, {\it {Cold Nuclear Matter Effects on Dijet
  Productions in Relativistic Heavy-ion Reactions at LHC}},  {\em Eur.Phys.J.}
  {\bf C72} (2012) 1904, [\href{http://arxiv.org/abs/1110.6601}{{\tt
  arXiv:1110.6601}}].

\bibitem{Neufeld:2012df}
R.~Neufeld and I.~Vitev, {\it {The $Z^0$-tagged jet event asymmetry in
  heavy-ion collisions at the CERN Large Hadron Collider}},  {\em
  Phys.Rev.Lett.} {\bf 108} (2012) 242001,
  [\href{http://arxiv.org/abs/1202.5556}{{\tt arXiv:1202.5556}}].

\bibitem{Kang:2012zr}
Z.-B. Kang, S.~Mantry, and J.-W. Qiu, {\it {N-Jettiness as a Probe of Nuclear
  Dynamics}},  {\em Phys.Rev.} {\bf D86} (2012) 114011,
  [\href{http://arxiv.org/abs/1204.5469}{{\tt arXiv:1204.5469}}].

\bibitem{Stavreva:2012aa}
T.~Stavreva, F.~Arleo, and I.~Schienbein, {\it {Prompt photon in association
  with a heavy-quark jet in Pb-Pb collisions at the LHC}},  {\em JHEP} {\bf
  1302} (2013) 072, [\href{http://arxiv.org/abs/1211.6744}{{\tt
  arXiv:1211.6744}}].

\bibitem{Dai:2012am}
W.~Dai, I.~Vitev, and B.-W. Zhang, {\it {Momentum imbalance of isolated
  photon-tagged jet production at RHIC and LHC}},
  \href{http://arxiv.org/abs/1207.5177}{{\tt arXiv:1207.5177}}.

\bibitem{Kang:2013wca}
Z.-B. Kang, X.~Liu, S.~Mantry, and J.-W. Qiu, {\it {Probing nuclear dynamics in
  jet production with a global event shape}},
  \href{http://arxiv.org/abs/1303.3063}{{\tt arXiv:1303.3063}}.

\bibitem{ColemanSmith:2012vr}
C.~E. Coleman-Smith and B.~Muller, {\it {Results of a systematic study of dijet
  suppression measured at the BNL Relativistic Heavy Ion Collider}},  {\em
  Phys.Rev.} {\bf C86} (2012) 054901,
  [\href{http://arxiv.org/abs/1205.6781}{{\tt arXiv:1205.6781}}].

\bibitem{Qin:2012gp}
G.-Y. Qin, {\it {Medium modification of photon-tagged jets at the LHC}},
  \href{http://arxiv.org/abs/1210.6610}{{\tt arXiv:1210.6610}}.

\bibitem{Renk:2012ve}
T.~Renk, {\it {Biased Showers - a common conceptual Framework for the
  Interpretation of High $P_T$ Observables in Heavy-Ion Collisions}},
  \href{http://arxiv.org/abs/1212.0646}{{\tt arXiv:1212.0646}}.

\bibitem{Wang:2013cia}
X.-N. Wang and Y.~Zhu, {\it {Medium Modification of $\gamma$-jets in
  High-energy Heavy-ion Collisions}},
  \href{http://arxiv.org/abs/1302.5874}{{\tt arXiv:1302.5874}}.

\bibitem{Ma:2013bia}
G.-L. Ma, {\it {Towards detailed tomography of high energy heavy-ion collisions
  by $\gamma$-jet}},  \href{http://arxiv.org/abs/1302.5873}{{\tt
  arXiv:1302.5873}}.

\bibitem{Aad:2010bu}
{\bf Atlas Collaboration} Collaboration, G.~Aad {\em et~al.}, {\it {Observation
  of a Centrality-Dependent Dijet Asymmetry in Lead-Lead Collisions at
  $\sqrt{s_{NN}}=2.77$ TeV with the ATLAS Detector at the LHC}},  {\em
  Phys.Rev.Lett.} {\bf 105} (2010) 252303,
  [\href{http://arxiv.org/abs/1011.6182}{{\tt arXiv:1011.6182}}].

\bibitem{Chatrchyan:2012vq}
{\bf CMS Collaboration} Collaboration, S.~Chatrchyan {\em et~al.}, {\it
  {Measurement of isolated photon production in $pp$ and PbPb collisions at
  $\sqrt{s_{NN}}=2.76$ TeV}},  {\em Phys.Lett.} {\bf B710} (2012) 256--277,
  [\href{http://arxiv.org/abs/1201.3093}{{\tt arXiv:1201.3093}}].

\bibitem{CMS:2012wxa}
{\bf CMS Collaboration} Collaboration, {\it {Detailed Characterization of Jets
  in Heavy Ion Collisions Using Jet Shapes and Jet Fragmentation Functions}}, .

\bibitem{Aad:2012vca}
{\bf ATLAS Collaboration} Collaboration, G.~Aad {\em et~al.}, {\it {Measurement
  of the jet radius and transverse momentum dependence of inclusive jet
  suppression in lead-lead collisions at $\sqrt{s_{NN}}=2.76$ TeV with the
  ATLAS detector}},  {\em Phys.Lett.} {\bf B719} (2013) 220--241,
  [\href{http://arxiv.org/abs/1208.1967}{{\tt arXiv:1208.1967}}].

\bibitem{Gyulassy:1993hr}
M.~Gyulassy and X.-n. Wang, {\it {Multiple collisions and induced gluon
  Bremsstrahlung in QCD}},  {\em Nucl.Phys.} {\bf B420} (1994) 583--614,
  [\href{http://arxiv.org/abs/nucl-th/9306003}{{\tt nucl-th/9306003}}].

\bibitem{Gyulassy:2000er}
M.~Gyulassy, P.~Levai, and I.~Vitev, {\it {Reaction operator approach to
  nonAbelian energy loss}},  {\em Nucl.Phys.} {\bf B594} (2001) 371--419,
  [\href{http://arxiv.org/abs/nucl-th/0006010}{{\tt nucl-th/0006010}}].

\bibitem{Vitev:2007ve}
I.~Vitev, {\it {Non-Abelian energy loss in cold nuclear matter}},  {\em
  Phys.Rev.} {\bf C75} (2007) 064906,
  [\href{http://arxiv.org/abs/hep-ph/0703002}{{\tt hep-ph/0703002}}].

\bibitem{Ellis:1991qj}
R.~Ellis, W.~Stirling, and B.~Webber, {\it {QCD and collider physics}},  {\em
  Camb.Monogr.Part.Phys.Nucl.Phys.Cosmol.} {\bf 8} (1996) 1--435.

\end{thebibliography}\endgroup

\end{document}